\documentclass[a4paper,11pt]{article}

\pdfoutput=1
\usepackage{amsmath,amssymb,cite,tabularx,graphicx,color,cancel,xspace,flafter,ifpdf,todonotes,subcaption}

\newcommand{\HEJ}{{\tt HEJ}\xspace}
\newcommand{\HIGHEJ}{\emph{High Energy Jets}\xspace}
\newcommand{\gs}{\ensuremath{g}}
\newcommand{\Ca}{\ensuremath{C_{\!A}}\xspace}
\newcommand{\Cf}{\ensuremath{C_{\!F}}\xspace}

\newcommand{\Nc}{\ensuremath{N_{\!C}}\xspace}
\newcommand{\as}{\ensuremath{\alpha_s}\xspace}

\newcommand{\asp}[1]{\langle {#1} \rangle}
\newcommand{\ssp}[1]{[ {#1} ]}
\def\spa#1.#2{\left\langle#1\,#2\right\rangle}
\def\spb#1.#2{\left[#1\,#2\right]}
\def\spaa#1.#2.#3{\langle\mskip-1mu{#1}
                  | #2 | {#3}\mskip-1mu\rangle}
\def\spbb#1.#2.#3{[\mskip-1mu{#1}
                  | #2 | {#3}\mskip-1mu]}
\def\spab#1.#2.#3{\langle\mskip-1mu{#1}
                  | #2 | {#3}\mskip-1mu\rangle}
\def\spba#1.#2.#3{\langle\mskip-1mu{#1}^+
                  | #2 | {#3}^+\mskip-1mu\rangle}
\def\spav#1.#2.#3{\|\mskip-1mu{#1}
                  | #2 | {#3}\mskip-1mu\|^2}
\def\jc#1.#2.#3{j^{#1}_{#2#3}}

\title{\begin{normalsize}
\begin{flushright}
Edinburgh 2017/11, IPPP/17/33, DCPT/17/66, MCnet-17-9
\end{flushright}
\end{normalsize}
\vspace*{1cm} Higgs Boson Plus Dijets: Higher Order Corrections}
\author{Jeppe~R.~Andersen$^{a}$, Tuomas Hapola$^{a}$, Andreas Maier$^{a}$ and
  Jennifer~M.~Smillie$^{b}$\\\mbox{}\\
  $^a$ Institute for Particle Physics Phenomenology,\\University of Durham,
  South Road, Durham DH1 3LE, UK\\
  $^b$ Higgs Centre for Theoretical Physics, University of Edinburgh,\\
  Peter Guthrie Tait Road, Edinburgh EH9 3FD, UK.}  

\begin{document}
\maketitle
\begin{abstract}
  The gluon fusion component of Higgs-boson production in association with
  dijets is of particular interest because it both (a) allows
  for a study of the $CP$-structure of the Higgs-boson couplings to gluons,
  and (b) provides a background to the otherwise clean study of Higgs-boson
  production through vector-boson fusion. The degree to which this background
  can be controlled, and the $CP$-structure of the gluon-Higgs coupling
  extracted, both depend on the successful description of the perturbative
  corrections to the gluon-fusion process. 

  \emph{High Energy Jets} (\HEJ) provides all-order, perturbative predictions for
  multi-jet processes at hadron colliders at a fully exclusive, partonic
  level. We develop the framework of \HEJ to include the
  process of Higgs-boson production in association with at least two jets. We
  discuss the logarithmic accuracy obtained in the underlying all-order
  results, and calculate the first next-to-leading corrections to the
  framework of \HEJ, thereby significantly reducing the
  corrections which arise by matching to and merging fixed-order results.

  Finally, we compare predictions for relevant observables obtained with NLO
  and \HEJ. We observe that the selection criteria commonly used for
  isolating the vector-boson fusion component suppresses the
  gluon-fusion component even further than predicted at NLO.
\end{abstract}

\newpage
\tableofcontents
\section{Introduction}
\label{sec:intro}

Immediately after the observation\cite{Aad:2012tfa,Chatrchyan:2012ufa} at the
CERN LHC of the fundamental Higgs-like boson, attention turned to measuring
the strength and properties of its couplings to other SM particles, and its
intrinsic $CP$-properties. Initially, these measurements were performed by
studying inclusive Higgs boson production in the Higgs boson decay channels
$\gamma\gamma$ and
$ZZ$~\cite{Aad:2013xqa,Aad:2013wqa,Aad:2014aba,Aad:2014eva,Aad:2014eha,Chatrchyan:2012jja,Chatrchyan:2013mxa,Khachatryan:2014iha,Khachatryan:2014ira}.
As the inclusive Higgs boson production is dominated by gluon-fusion Higgs
boson production, any measurement of the strength of the coupling of the
Higgs boson to e.g.~$Z$ will involve a product of this coupling with the
coupling for the production of the Higgs boson through gluon fusion, mediated
by heavy (top and bottom) quark loops.

A precise measurement of the coupling of the Higgs boson to the electroweak
bosons is obviously important to determine if indeed a single fundamental
Higgs boson is fully responsible for the mass-generation of fundamental
particles and electroweak symmetry breaking, as in the Standard Model. In
this respect, it is interesting to study Higgs boson production directly
through weak boson fusion. At the LHC, this process would occur
perturbatively in the process of Higgs boson production in association with
at least two hard jets. This process is of interest then not just as a
perturbative correction (at order $\mathcal{O}(\alpha_s^4)$) to the inclusive
Higgs boson production through gluon fusion, but also as a
$\mathcal{O}(\alpha_w^4)$ Born level process that allows for a direct
measurement of the strength of the coupling between the Higgs boson and the
weak bosons. Since the quantum interference between the two contributing
production channels of so-called vector-boson fusion (VBF) (involving
a direct coupling between the Higgs and the weak bosons) and gluon fusion (GF) is
insignificant\cite{Andersen:2007mp,Bredenstein:2008tm,Dixon:2009uk}, it is
justified to discuss the processes separately. The study of weak boson fusion
production of Higgs bosons then allows for a measurement of the higgs boson
to weak boson coupling without relying on a knowledge of the loop-induced
coupling strength between gluons and the Higgs boson.

The final analyses of data after
Run-I\cite{Aad:2014lwa,Aad:2014tca,Khachatryan:2014ira} allowed for the Higgs
boson production to be studied for small numbers of co-produced jets, in
particular also for the production in association with two or more
jets. These measurements, therefore, start probing directly the VBF
production mechanism, where the Born-level process involves quarks only
scattering by the exchange of a weak boson. This is dominated by
valence quarks, and hence the resulting jets will carry a significantly
larger part of the light-cone proton momenta than what is the case of the
gluon-fusion production mechanism, where the $Hjj$ cross section contribution for
inclusive cuts is dominated by the $gg$-component. The distinctive topology
for VBF allows for event selection cuts on e.g.~a large invariant mass and/or
rapidity separation between the dijets in order to suppress background. This
also suppresses the contribution from the gluon-fusion process relative to
VBF. While the inclusive GF cross section is dominated by the $gg$-component,
the $qg$-component dominates\cite{DelDuca:2001eu,DelDuca:2001fn} after a
large invariant mass between the dijets is required.

Requiring a significant invariant mass between dijets is interesting not just
as a tool to suppress the gluon-fusion contribution over weak-boson fusion,
but for a slightly less restrictive cut on the invariant mass, which allows
more gluon-fusion events in the sample, it is possible to study the
$CP$-structure of the gluon-Higgs
couplings\cite{Klamke:2007cu,Andersen:2010zx}. In particular, such analyses
of the $Hjj$ sample allow for an extraction of mixing parameters in scenarios
with $CP$-violation in the Higgs sector. However, the correct description of
the gluon-fusion contribution in the region of phase space with a significant
invariant mass between the dijets is more challenging than is the case for
weak-boson fusion. The reason is that the gluon-fusion component allows for a
colour-octet exchange between the dijets, whereas the weak-boson fusion
component obviously has no colour exchange between the jets. This leads to a
different radiation pattern for the two processes\cite{Dokshitzer:1991he},
where the gluon-fusion component will radiate more hard, observable jets in
the rapidity region spanned by the colour octet exchange than the weak-boson
fusion process. This again leads to an increase in the expected number of hard jets
in the event as the rapidity span is increased. This behaviour is universal
for all processes allowing for a colour octet exchange between jets, and has
already been observed in both pure
dijet production\cite{Aad:2011jz,Aad:2014pua} and the production of
W+dijets\cite{Abazov:2013gpa}. Not just does the colour-octet exchange
emphasise the contribution from real-emission, higher-order perturbative
corrections, but it is also accompanied by a tower of logarithms from virtual
corrections. Both sources of perturbative corrections are included in the
BFKL-equation\cite{Fadin:1975cb,Kuraev:1976ge,Kuraev:1977fs,Balitsky:1978ic},
which captures the dominant logarithms $(\ln\hat s/|\hat t|)$) which govern the
high-energy limit of the on-shell scattering matrix elements.

However, such logarithms are not systematically included in the standard
perturbative methods for obtaining predictions for LHC observables. Analyses
of e.g.~W production in association with dijets for both
D0\cite{Abazov:2013gpa} (at the 1.96~TeV Tevatron) and ATLAS\cite{Aad:2014qxa}
(at the 7~TeV LHC) consistently reveal a tension between data and a standard
set of predictions in the region of phase space of large dijet invariant mass
or rapidity separation. This is true for the differential cross section
depending on just the Born-level momenta, and for observables describing
additional jet activity. This tension between data and the predictions of
the standard tools is therefore present for the observables and the region of phase
space that is of direct relevance for the study of Higgs boson production in
association with dijets.

The dominant logarithms of $\hat s/|\hat t|$ are, however,
systematically included in the calculations of the on-shell partonic
scattering amplitudes within the framework of \emph{High Energy
  Jets}\cite{Andersen:2009nu,Andersen:2009he,Andersen:2011hs,Andersen:2012gk,Andersen:2016vkp}. The
framework is based on an approximation to the $n$-body on-shell scattering
matrix element. Within this approximation, both real and
virtual corrections are included to all orders in perturbation
theory. The virtual corrections not only cancel the infra-red poles from
the real corrections, but also contribute to the finite part of the
matrix element. In fact, this finite contribution is instrumental in
achieving leading-logarithmic accuracy. This is in contrast to the
standard formulation of a parton shower, where the assumed Sudakov form
of the virtual corrections keeps the shower \emph{unitary}, allowing for
a probabilistic interpretation of emission.

In \emph{High Energy Jets}, the sum over $n$ and the integration over
each $n$-body phase space is performed explicitly using Monte Carlo sampling,
and as such the predictions are made at the partonic level with direct access
to the four-momenta of each of the $n$ particles. The framework merges
fixed-order (currently leading order), high-multiplicity matrix-elements with
an all-order description of the dominant logarithms. The formalism has been
implemented for several processes, and compares favourably to data for
e.g.~dijet (or more)
production\cite{Aad:2011jz,Chatrchyan:2012gwa,Aad:2014pua}, the
production of a W boson in association with two
jets\cite{Abazov:2013gpa,Aad:2014qxa} and the production of a Z-boson or virtual
photon in association with two jets\cite{Andersen:2016vkp}. These studies indicate that in the
large-invariant mass, and the large rapidity difference-region, the
logarithms of \HEJ are important, and their inclusion improves the theoretical
prediction.

The experimental studies of dijets and W+dijets therefore also indicate that
\emph{High Energy Jets} should be relevant for a successful description of
the gluon-fusion production of a Higgs boson in association with dijets, in
particular in the region of interest for the study of $CP$-properties, and
for understanding how to use the radiation pattern to successfully suppress
the gluon-fusion contribution to Higgs boson+dijets when studying weak-boson fusion.

This paper presents the impact on the physics analyses, and the
implementation of \emph{High Energy Jets} for the gluon-fusion contribution
to Higgs-boson production in association with dijets. The earlier application
of \HIGHEJ included the leading logarithms in $\hat s/\hat t$ only. In
Section~\ref{sec:formal-accuracy-high} we discuss the first systematic inclusion of
part of the sub-leading contributions within the framework of \HIGHEJ. The
resulting predictions for several of the observables measured in
Higgs boson+dijet production are presented in Section~\ref{sec:analysis-results}, and the
conclusion discussed in Section~\ref{sec:conclusions}.


\section{The Formal Accuracy of High Energy Jets}
\label{sec:formal-accuracy-high}
In this section we will present the procedure used for obtaining predictions
within \HIGHEJ (\HEJ). \HEJ is concerned with the description of processes involving
a $t$-channel colour exchange between two jets, such as dijet-production, and
QCD production of W+dijets, Z/$\gamma$+dijets (both starting at order
$\alpha_s^2 \alpha_w$), and Higgs boson+dijets (starting at order $\alpha_s^4$).

Underpinning \HEJ is an all-order approximation to the on-shell,
hard-scattering matrix elements, explicit in the momenta of all particles,
and for each multiplicity. The cancellation of IR singularities between real
and virtual corrections is organised with subtraction terms, which are
sufficiently simple to allow the explicit summation over multiplicities, and
the integration over phase space to be performed using Monte Carlo
techniques. The approximation to the hard scattering matrix element ensures a
certain logarithmic accuracy of the predictions, which will be detailed in
Section~\ref{sec:logarithmic-accuracy}. As further discussed in
Section~\ref{sec:match-merg-fixed}, the all-order approximations are
supplemented by corrections using the fixed-order (so far just tree-level)
predictions for several jet multiplicities. As such, \HEJ provides an
alternative procedure for merging fixed-order samples of various jet
multiplicities to that of CKKW-L\cite{Catani:2001cc,Lonnblad:1992tz}, which
is based on the logarithmic accuracy achieved in a parton shower. Instead,
the merging procedure of \HEJ maintains both the logarithmic accuracy at
large invariant mass between jets (as discussed in the next session) and the
fixed-order accuracy of the merged samples.

\subsection{Logarithmic Corrections and Logarithmic Accuracy}
\label{sec:logarithmic-accuracy}
In this section we will first identify the leading contribution to Higgs
boson production in association with dijets when these dijets have a large
invariant mass. We then identify a source of systematic and logarithmically
(in the invariant mass) enhanced perturbative corrections both for real
emissions and virtual corrections, and discuss how these logarithmic
corrections can be summed to all orders using the formalism of \HIGHEJ.

\subsubsection{Leading Contributions at Large Invariant Mass}
\label{sec:leadingcontrlarge}

\begin{figure}[btp]
  \centering
  \includegraphics[width=0.6\linewidth]{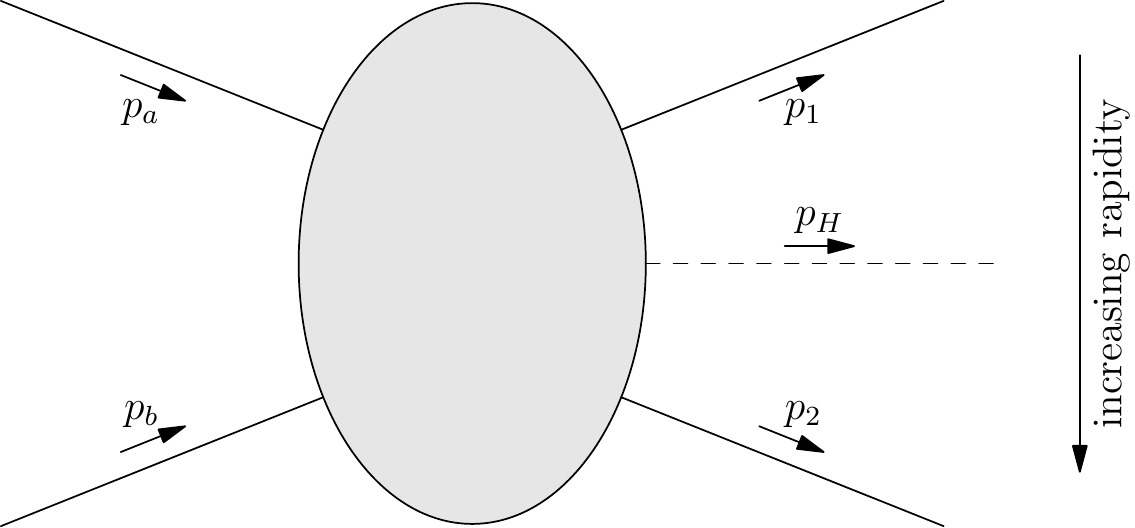}
  \caption{Production of a Higgs boson with momentum $p_H$ in between
    two jets with momenta $p_1, p_2$. Arrows indicate the direction of
    the momentum flow.}
  \label{fig:jhj_amp}
\end{figure}
Consider for illustration the production of a Higgs boson in association
with dijets, with the rapidity of the Higgs boson between that of the
jets.  We label final momenta as shown in Fig.~\ref{fig:jhj_amp}, such
that the rapidities satisfy $y_1 < y_H < y_2$ and the incoming momentum
$p_a$($p_b$) is in the backward (forward) direction. In the following,
we will be frequently interested in amplitudes in the limit of
\emph{Multi-Regge kinematics} (MRK), defined by a large center-of-mass energy
$\sqrt{s_{12}}$, large invariant masses between all outgoing momenta,
and fixed $t$-channel momenta. For our current example, we introduce the
$t$-channel momenta of the system as $t_1 = (p_a - p_1)^2, t_2 = (p_a - p_1
-p_H)^2$ and consider large $s_{1H}, s_{2H}, s_{12}$, keeping $t_1$ and
$t_2$ fixed.  An analysis of the analytic properties of scattering
amplitudes~\cite{Brower:1974yv} (e.g. from Regge theory for
multi-particle production) indicates that in this limit the on-shell
scattering amplitude $\mathcal{M}$ should scale as~\cite{Fadin:2006bj}
\begin{align}
  \label{eq:reggescaling}
  \mathcal{M}\sim\ s_{1H}^{\alpha_1(t_1)}\ s_{2H}^{\alpha_2(t_2)}
\ \gamma\left(t_1, t_2, s_{12}/(s_{1H} s_{2H})\right).
\end{align}
Here, $\alpha_1(t_1)$ is the spin of the particle that can be exchanged in
the $t_1$-channel between the particle of jet 1 and the Higgs boson,
$\alpha_2(t_2)$ is the equivalent for the $t_2$-channel between the Higgs boson
and the particle of
jet 2 and $\gamma$ is a function of transverse scales only. For a
given momentum configuration of the jets and the Higgs boson, the leading
contribution to $Hjj$-production therefore comes from the subprocesses with
a parton flavour assignment to the jets which allows for the particle of the
largest possible spin to be connecting the jets. For QCD this is the
spin-1 gluonic colour-octet exchange. If the flavour assignment of a sub-process is
such that a quark exchange is mandated, then the contribution to the jet cross
section (proportional to the square
of the matrix element) from this subprocess is suppressed by the
invariant
mass of the dijet pair, as compared to the subprocess where a gluon exchange is
possible. For a given momentum configuration of the jets, the flavour
assignments of the incoming states and of the corresponding jets which can proceed through
gluon (colour-octet) exchanges between each jet are called the
Fadin-Kuraev-Lipatov (FKL) configurations. These will form the leading
contribution in $s_{ij}/t_k$ to the given jet configuration.

\begin{figure}[btp]
  \centering
  \begin{tabular}{c@{\qquad}c}
    \begin{subfigure}[t]{0.4\linewidth}
      \includegraphics[width=\linewidth]{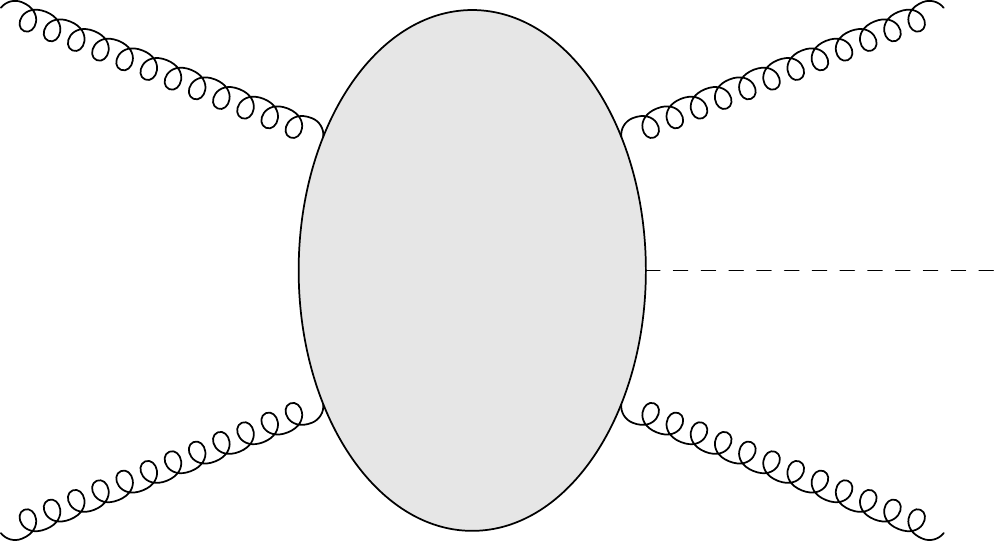}
      \caption{}
      \label{fig:amp_ghg}
    \end{subfigure}
    &
      \begin{subfigure}[t]{0.4\linewidth}
        \includegraphics[width=\linewidth]{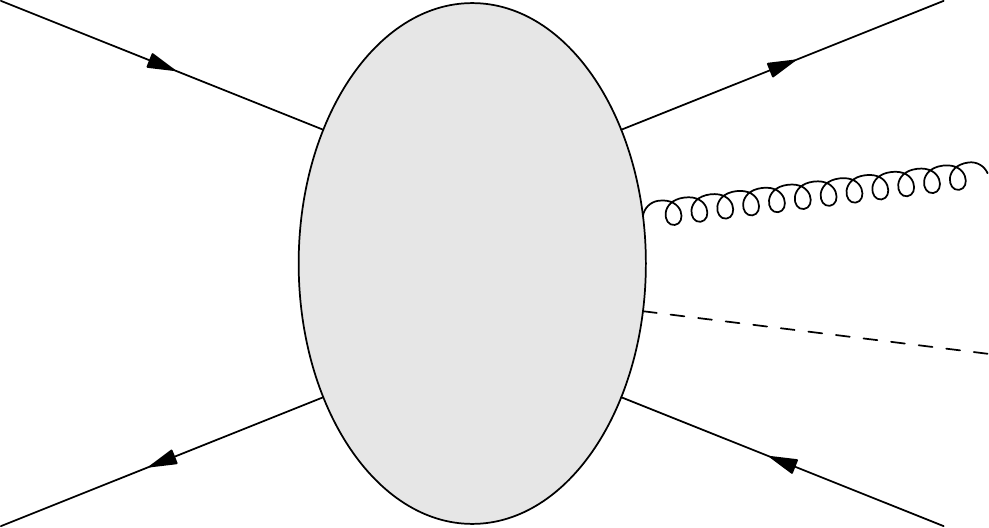}
        \caption{}
        \label{fig:amp_qghQ}
      \end{subfigure}
    \\[2em]
      \begin{subfigure}[t]{0.4\linewidth}
        \includegraphics[width=\linewidth]{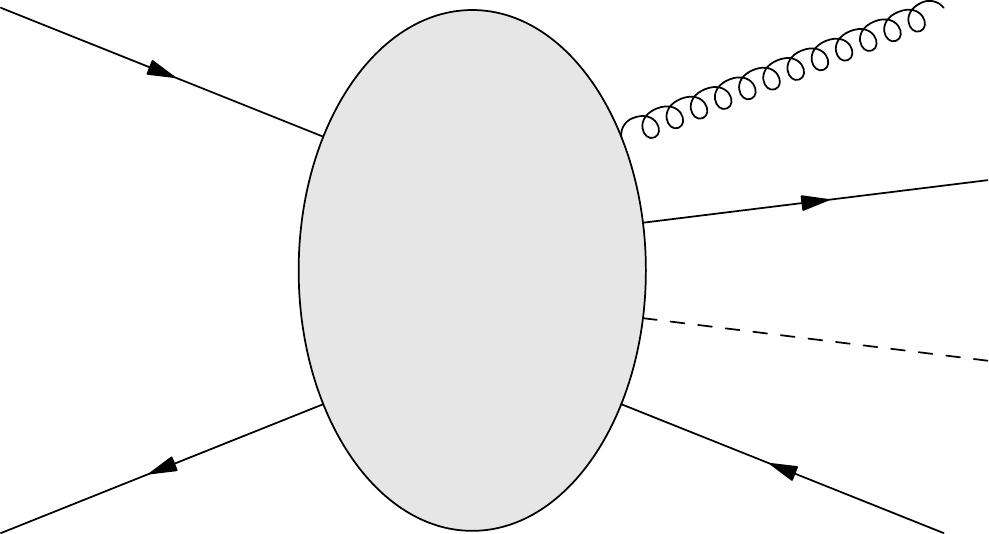}
        \caption{}
        \label{fig:amp_gqhQ}
      \end{subfigure}
    &
      \begin{subfigure}[t]{0.4\linewidth}
        \includegraphics[width=\linewidth]{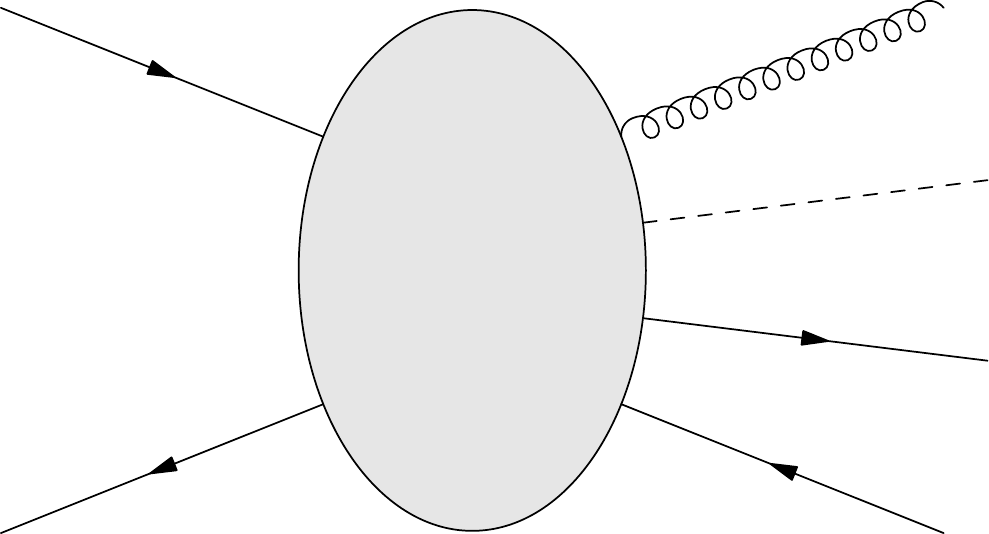}
        \caption{}
        \label{fig:amp_ghqQ}
      \end{subfigure}
    \\
  \end{tabular}

  \caption{Example configurations for Higgs production in association
with jets. Outgoing particles are ordered in increasing rapidity. Both
(a) and (b) can be generated via gluonic $t$-channel exchange between each
pair of adjacent outgoing particles and are therefore FKL
configurations. The non-FKL configurations (c) and (d) require a
quark $t$-channel exchange.}
  \label{fig:ex_FKL_nonFKL}
\end{figure}

\begin{figure}[btp]
  \centering
  \includegraphics[width=.5\textwidth]{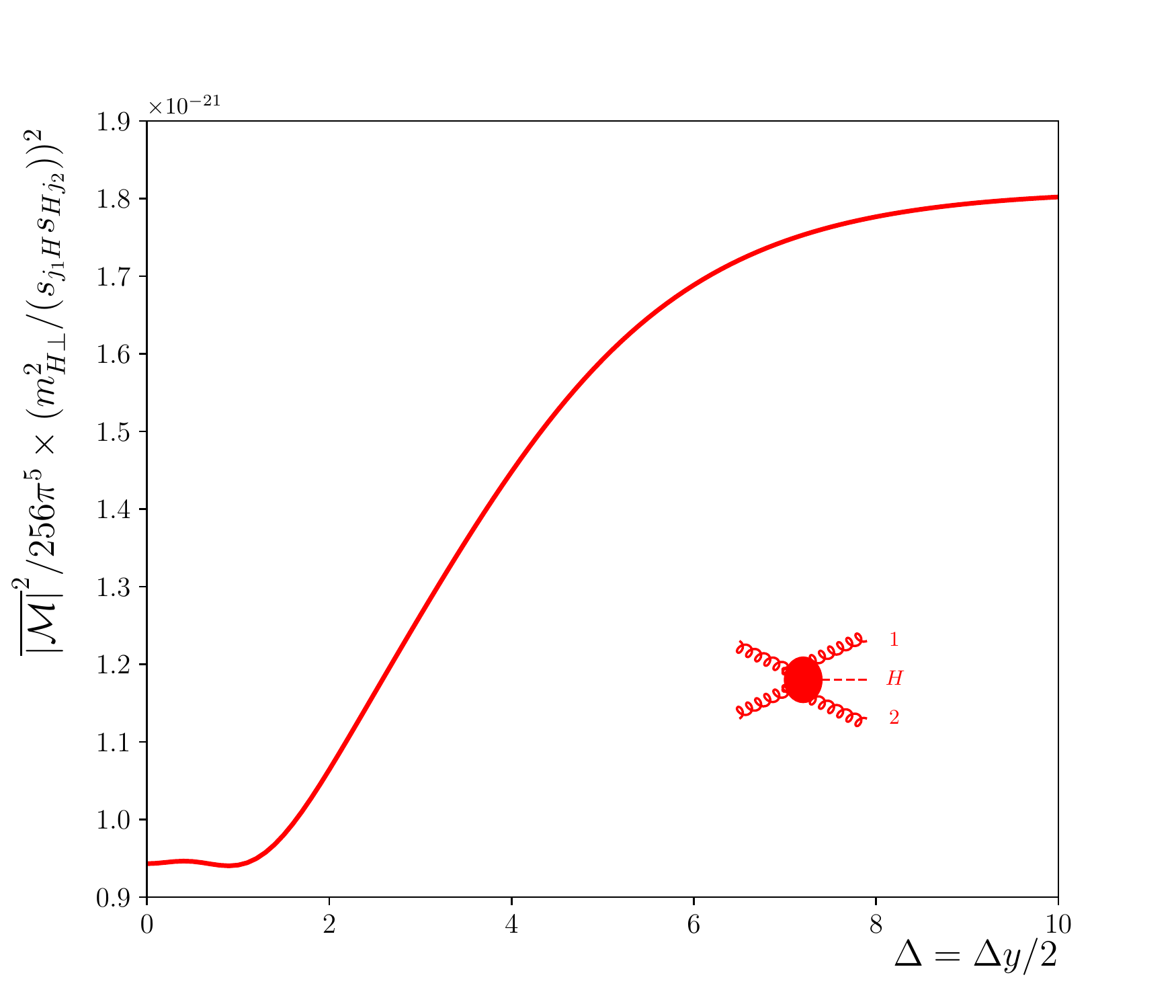}
  \caption{The square of the partonic matrix element for the processes
    $gg\to gHg$ divided by $(s_{j_1H}^2\ s_{j_2H}^2) m_{\perp H}^4$. This is plotted
    for the phase space points parametrised according to
    Eq.~\eqref{eq:ExplorerPHSP}. The square of the matrix element exhibits the
    expected Multi-Regge scaling according to Eq.~\eqref{eq:reggescaling}, for
    spin-1 (gluon) exchanges and $\gamma\propto 1/m_{\perp H}^4$, as the curve
    tends to a constant for increasing $\Delta y$.}
  \label{fig:gghgg}
\end{figure}

We illustrate this by continuing the example above.  Consider first the case
where both the incoming and the outgoing partons making up the jets are
gluons as shown in Fig.~\ref{fig:amp_ghg}. At Born level, the spin of all exchanged particles is 1 (since they are
all gluons), and therefore the amplitude must scale as
$\mathcal{M}\sim s_{1H}\ s_{2H}\ \gamma \left(t_1, t_2, s_{12}/(s_{1H}
  s_{2H})\right)$,
where in the MRK limit
$s_{12}/(s_{1H} s_{2H})\to 1/(m_H^2+p_{\perp H}^2), t_1\to -p^2_{\perp j_1},
t_2\to -p^2_{\perp j_2}$,
such that $\gamma$ depends on transverse scales only. This scaling is indeed
demonstrated in Fig.~\ref{fig:gghgg}. This plot shows
$\overline{|\mathcal{M}|}^2/(s_{1H}^2\ s_{2H}^2) m_{\perp H}^4$, where the
square of the Born level matrix element (extracted from
Madgraph5\_aMC@NLO\cite{Alwall:2014hca}) is evaluated in the phase space configurations of
increasing rapidity separation between all particles.  In particular, the
4-momenta $p = (E, p_x, p_y, p_z)$ of the two jets $p_{j_1}, p_{j_2}$
and the Higgs boson $p_H$ are parametrised in terms of their transverse
momenta, azimuthal angle and rapidity as
\begin{align}
  \label{eq:ExplorerPHSP}
  \begin{split}
    p_{j_1}&=p_{\perp 1} (\cosh y_1,\cos \phi_1 ,\sin \phi_1,\sinh y_1)\\
    p_{j_H}&= (m_{\perp H}\cosh y_H, p_{\perp H}\cos \phi_H ,p_{\perp H}\sin
    \phi_H,m_{\perp H}\sinh y_H)\\
    p_{j_2}&=p_{\perp 2} (\cosh y_2,\cos \phi_2 ,\sin \phi_2,\sinh
    y_2).
  \end{split}
\end{align}
The specific choices for angles and transverse momenta are irrelevant for the
conclusion, but here the phase space points used in the plot were
$p_{\perp 1}=p_{\perp 2}=70$~GeV, $\phi_1=\frac 2 3 \pi$, $y_1=-\Delta$,
$\phi_2=\pi$, $y_2=\Delta$, $y_H=\Delta/3$ and
$p_{\perp H}=-(p_{\perp 1}+p_{\perp 2})$ where $\Delta$ is increasing along the
$x$-axis.  The matrix element exhibits the expected Multi-Regge scaling
according to Eq.~\eqref{eq:reggescaling}, for spin-1 (gluon) exchanges, as
$\overline{|\mathcal{M}|}^2/(s_{1H}^2\ s_{2H}^2) m_{\perp H}^4$ tends to a
constant as $\Delta y$ increases.

We can illustrate the suppression introduced when one requires a quark exchange
in the $t$-channel by considering the squared matrix-elements for non-FKL
configurations versus a corresponding FKL configuration. We will
consider the three
rapidity orderings of the flavour content in the process $pp\to H j_1
j_2 j_3$ shown in panels (b) to (d) of Fig.~\ref{fig:ex_FKL_nonFKL}.
The rapidity-ordering $qQ\to qgHQ$ can proceed through colour-octet exchanges
between each of the jet-pairs $(j_1,j_2)$, and $(j_2,j_3)$ (and the Higgs boson)
and hence is an FKL configuration.  The square of the matrix element for the
cross section then scales as
$|\mathcal{M}_1|^2\propto s_{j_1j_2}^2 s_{j_2H}^2 s_{Hj_3}^2\Gamma_1$ (where
$\Gamma_1$ depends on transverse scales only).  If now the parton content of
$j_1$ and $j_2$ is swapped, the previous possibility of a gluon exchange between
jets 1 and 2 is replaced by a quark exchange. Therefore, the scattering-process
will scale as
$|\mathcal{M}_2|^2\propto s_{j_1j_2} s_{j_2H}^2 s_{Hj_3}^2 \Gamma_2$ (where
$\Gamma_2$ depends on transverse scales only), which is therefore suppressed by
one power of $s_{j_1j_2}$ with respect to the FKL configuration.  The third
configuration we consider is $qQ\to gHqQ$.  Like the second configuration, this
only allows a quark exchange between jets 1 and 2, now with the Higgs boson in
between in rapidity, and hence scales as
$|\mathcal{M}_3|^2\propto s_{j_1H} s_{Hj_2} s_{j_2j_3}^2\Gamma_3$ (where
$\Gamma_3$ depends on transverse scales only).

\begin{figure}[btp]
  \centering
  \includegraphics[width=.49\textwidth]{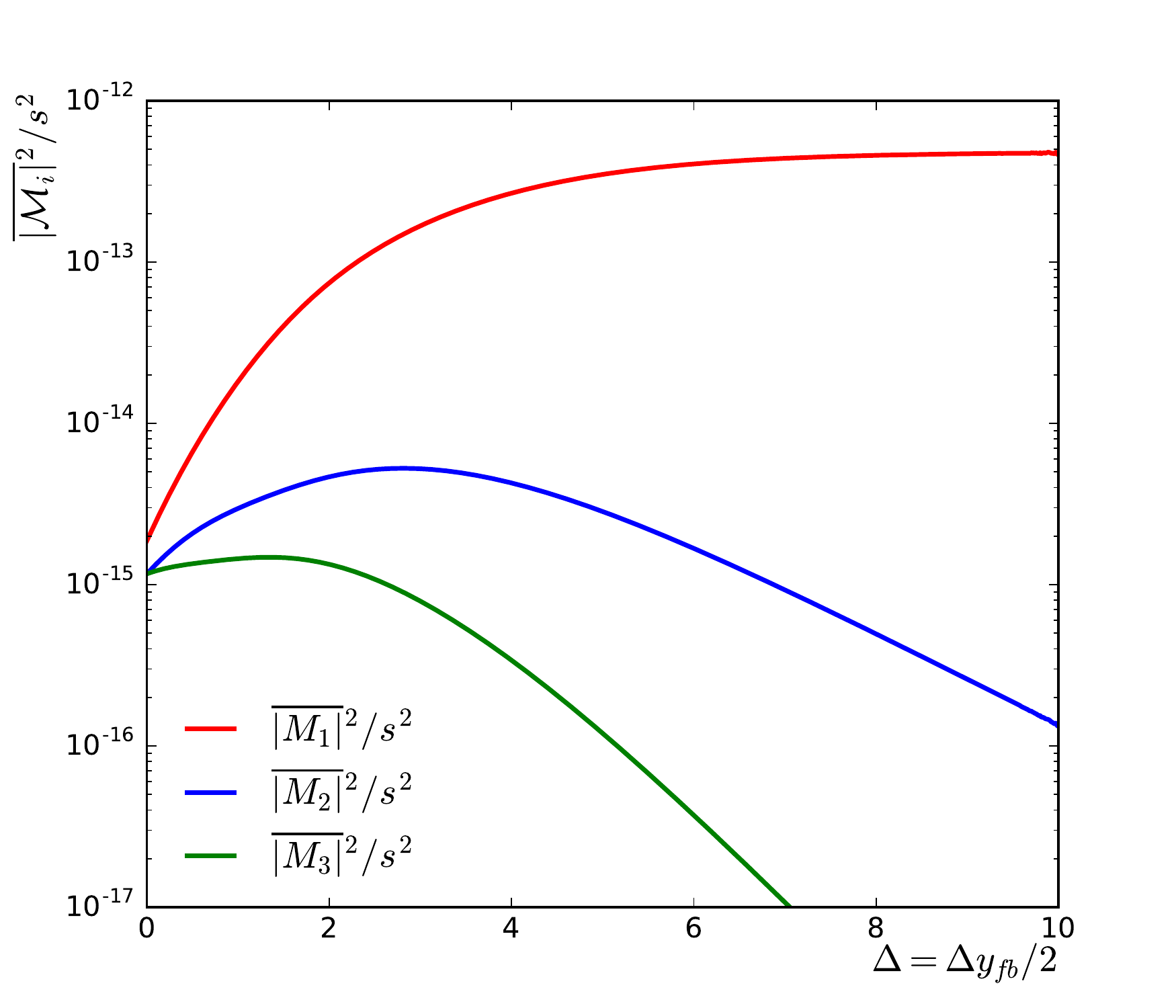}\hfill
  \includegraphics[width=.49\textwidth]{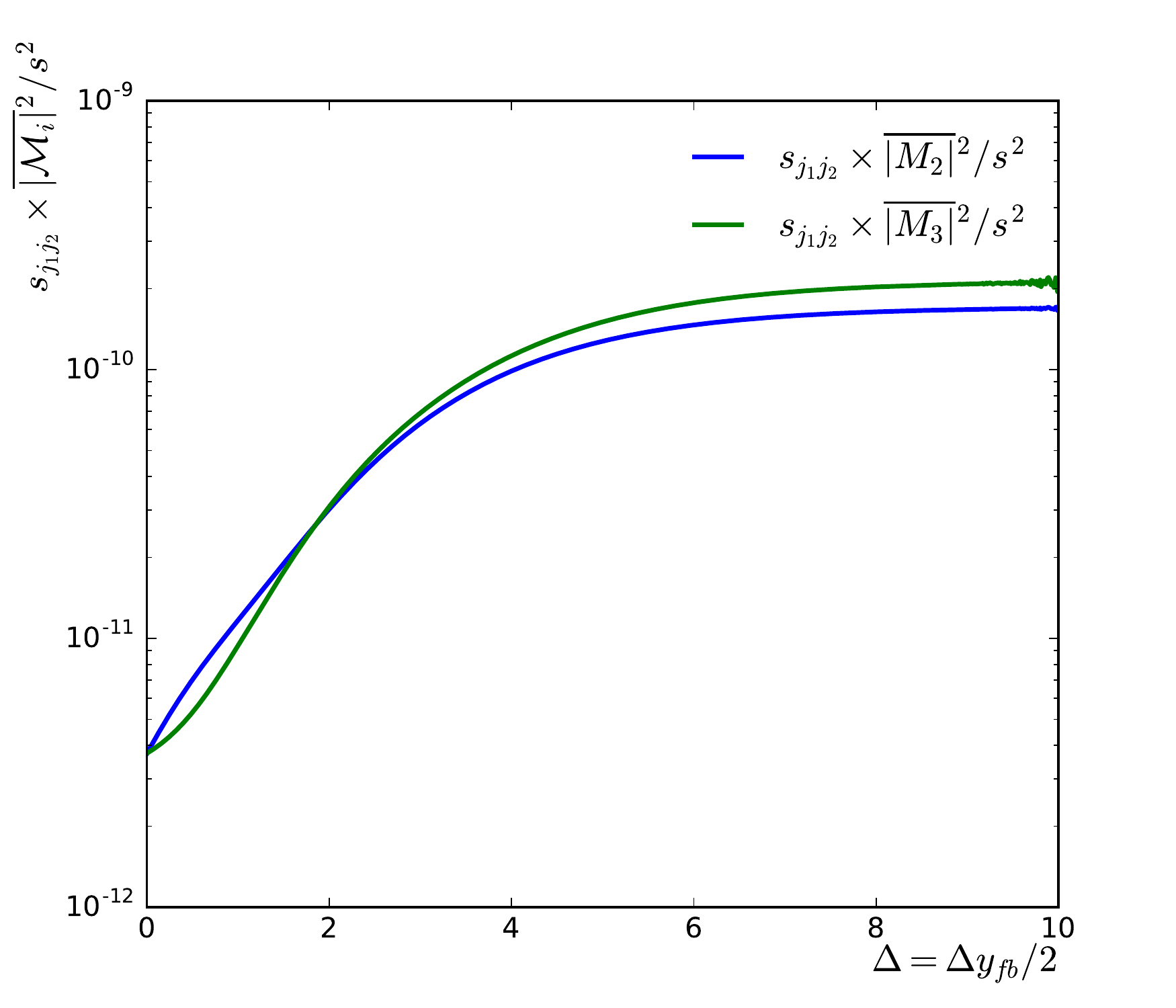}
  \caption{These plots demonstrate that introducing a quark exchange in place of
    a gluon exchange does indeed suppress the amplitude compared to the FKL
    configuration.  In the left plot, the squared matrix elements are shown
    divided by $s^2$ for the three rapidity configurations described in the
    text.  For the FKL configuration, $\overline{|M_1|}^2/s^2$ (red) tends to a
    constant as the rapidity separation increases, while the same quantity for
    the non-FKL configurations $\overline{|M_2|}^2$ (blue) and
    $\overline{|M_3|}^2$ (green) are exponentially suppressed.  In the right
    plot, the suppression is shown to be a factor of $s_{j_1j_2}$ as the same
    quantities multiplied by $s_{j_1j_2}$ now tend to a constant in agreement
    with Eq.~\eqref{eq:reggescaling} .}
  \label{fig:FKLdominance}
\end{figure}

We illustrate the behaviour of these matrix elements in
Fig.~\ref{fig:FKLdominance}.  The left plot clearly shows the resulting
suppression of the square of the matrix elements for the non-FKL configurations
($qQ\to gqHQ$ (blue) and $qQ\to gHqQ$ (green)) compared to the FKL ordering
$qQ\to qgHQ$ (red).  The latter tends to a constant times $s^2$ while the first
two exhibit an exponential suppression for large $\Delta y$ (corresponding to a
power-suppression in $s_{j_1 j_2}$).  The suppression is indeed verified to be
$s_{j_1 j_2}$ on the right-hand plot in
Fig.~\ref{fig:FKLdominance}. Here, the squared matrix elements $|\mathcal{M}|^2$
divided by $s^2$ has been multiplied by $s_{j_1 j_2}$ and tends to a constant
for large $\Delta y$ in both cases.

\subsubsection{Leading Contribution from Perturbative QCD}
\label{sec:LeadColour}

\begin{figure}[btp]
  \centering
  \includegraphics[width=.65\textwidth]{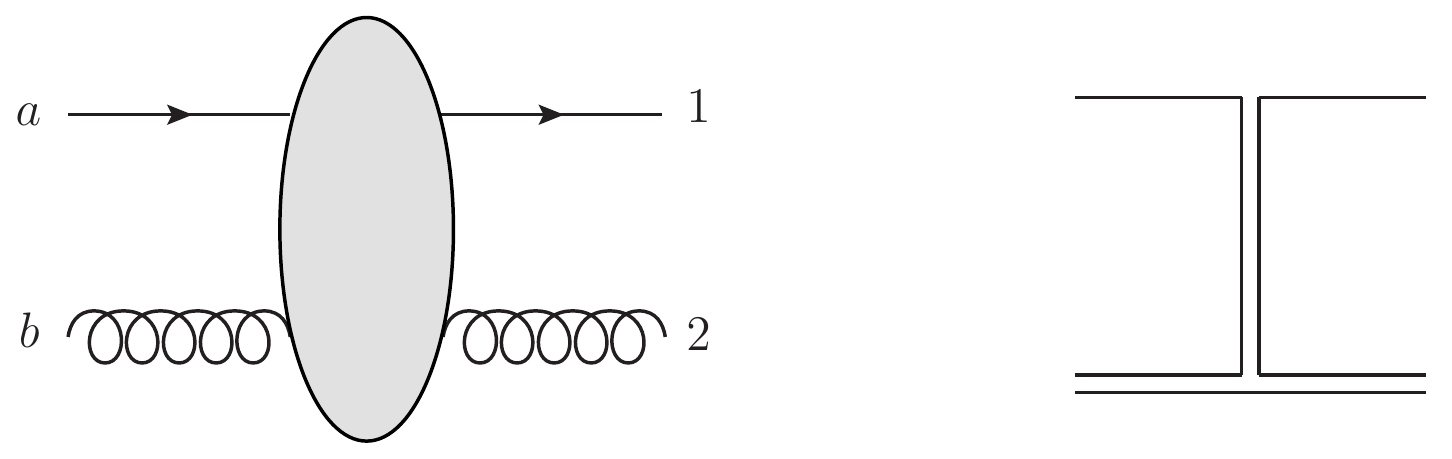}
  \caption{Left: Quark-gluon scattering with rapidity ordering $y_q \ll
    y_g$ and Right: The corresponding leading colour connection in the MRK limit.}
  \label{fig:qgfklT}
\end{figure}

\begin{figure}[btp]
  \centering
  \includegraphics[width=.65\textwidth]{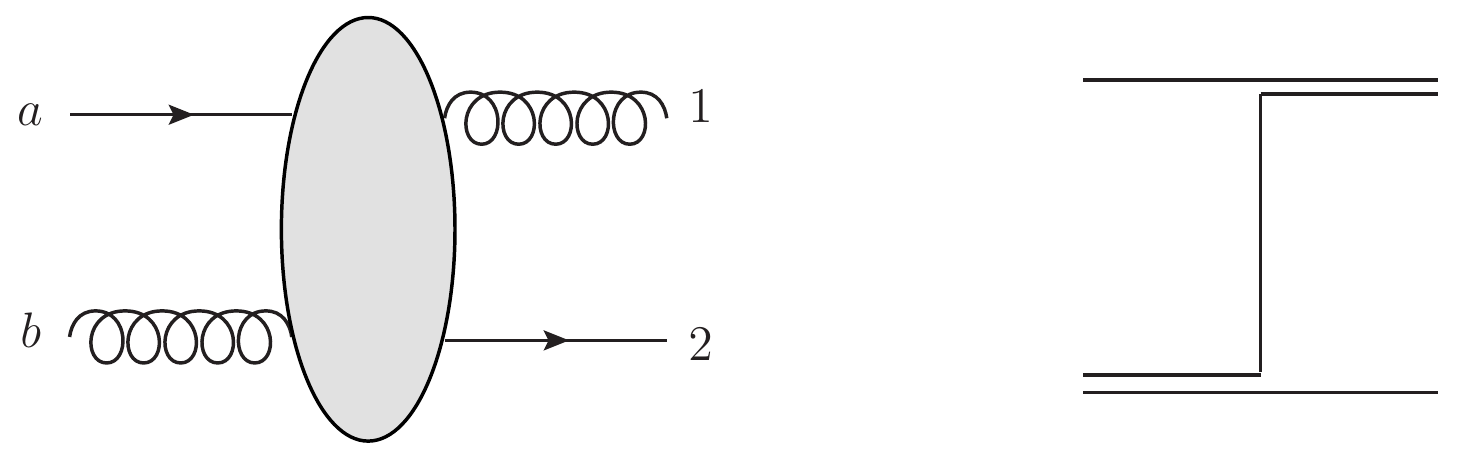}
  \caption{Left: Quark-gluon scattering with rapidity ordering $y_q \gg
    y_g$ and Right: The corresponding leading colour connection in the MRK limit.}
  \label{fig:qgnonfklT}
\end{figure}

An alternative derivation of the dominance of the FKL configurations can be found by considering which
of all the possible colour connections will dominate in the Multi-Regge-Kinematic (MRK) limit.  As the Higgs boson is
colour-neutral and irrelevant for the arguments, we restrict here the
discussion to amplitudes involving just quarks and gluons, and follow the treatment of Ref.~\cite{DelDuca:1995zy}.  We begin by
considering the process $qg\to qg$\footnote{and not $gg\to gg$, since in a pure gluon amplitude the identical
final state particles prevents a clear identification of the $u$ and
$t$ channel, unless of course the scattering is of gluons with different helicities.}.  Without loss of
generality we take the backward incoming parton to be the quark.  For the
outgoing quark and gluon, there are obviously two possible rapidity-orderings
: $y_q < y_g$ and $y_q > y_g$.  These are shown in diagrams with
rapidity-ordered final states in
Figs.~\ref{fig:qgfklT} and \ref{fig:qgnonfklT}, together with the corresponding
planar colour connections.
By explicit calculation one quickly finds (see appendix~\ref{sec:tree-level-ampl}) that the tree-level result for the
initial states fixed as the gluon incoming with positive light-cone momentum
and the quark with negative light-cone momentum, the amplitude for the two
rapidity orderings of the final state in the MRK limit scale as
\begin{align}
  \label{eq:qgqgamps}
  |\mathcal{M}(y_q \ll y_g)| \sim s\ \gamma_1\ \quad {\rm and} \quad |\mathcal{M}(y_q \gg y_g)|
  \sim \sqrt{s}\ \gamma_2,
\end{align}
in agreement with Eq.~\eqref{eq:reggescaling} and hence the dominant
flavour-configuration in the MRK limit is given by the momentum
configuration with $y_q\ll y_g$.  As illustrated in the figures, this is the
configuration where a colour octet (two colour lines) is exchanged, when particles are drawn
ordered in rapidity.

\subsubsection{Dominant Contributions at Arbitrary Multiplicities}
\label{sec:beyondtree}

\begin{figure}[btp]
  \centering
  \includegraphics[width=.7\textwidth]{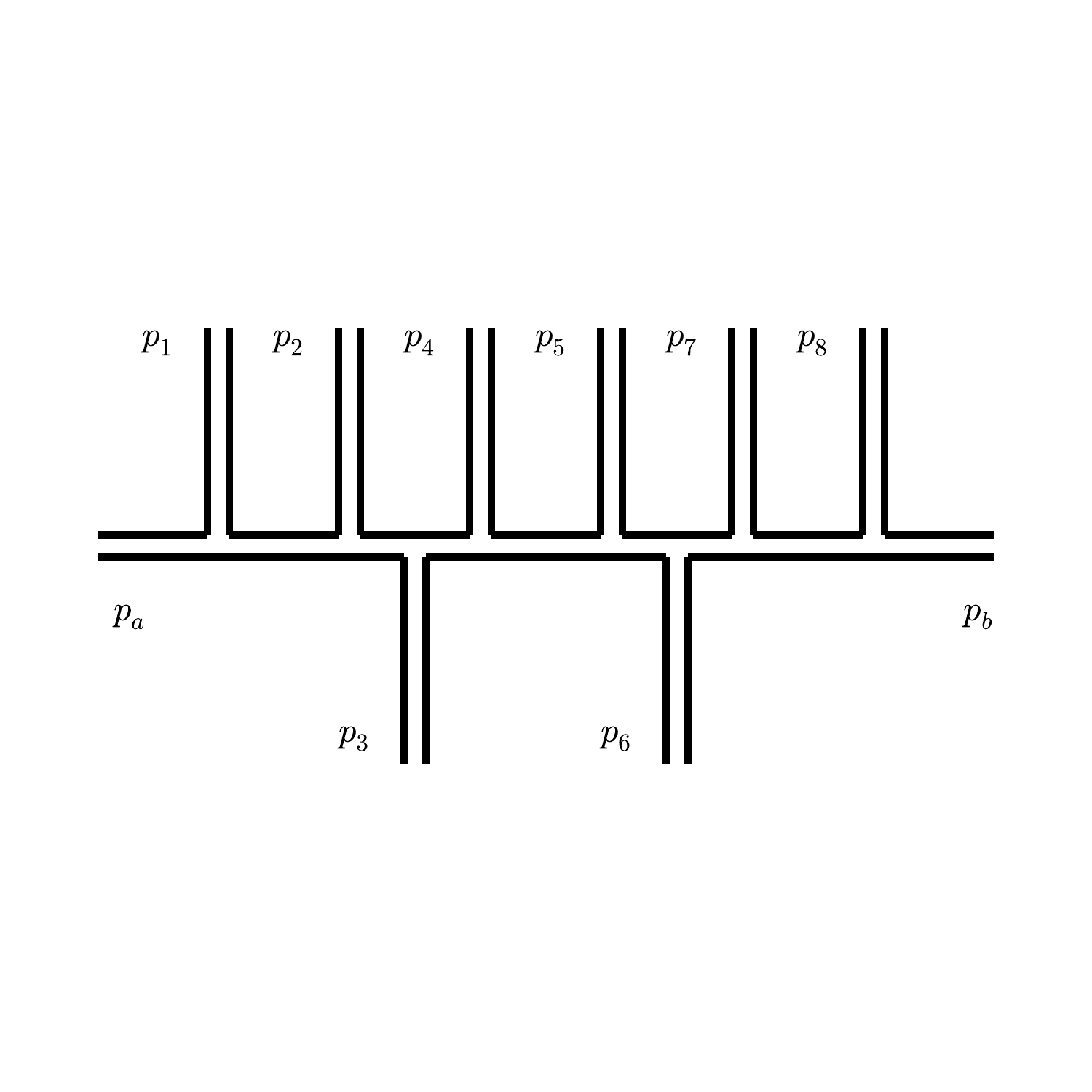}

  \vspace{-3cm}
  \caption{Parke-Taylor amplitude with colour ordering which respects the rapidity ordering $y_1 \ll \cdots \ll y_8$.}
  \label{fig:twosideT}
\end{figure}

The result of the previous section in fact generalises beyond the simple $2\to 2$ process.  In
Ref.~\cite{DelDuca:1995zy}, the compact Parke-Taylor
expression\cite{Parke:1986gb} for the maximally helicity violating (MHV)
amplitudes for all-gluon processes $gg\to g...g$
was used to show that for an arbitrary number of gluons, the
colour connections which dominate kinematically in the MRK limit are those which can be
represented on a so-called two-sided plot.  An example of such a plot is shown
in Fig.~\ref{fig:twosideT}.  The momentum of the incoming particles are
labelled $p_a$ (negative $z$-momentum), and $p_b$ (positive $z$-momentum), and the outgoing particles are ordered in rapidity from left to
right.

The \emph{colour connections} which dominate in the MRK limit are found\cite{DelDuca:1995zy}
to be precisely all those which may be drawn without any crossed
lines. Furthermore, these colour connections all contribute with the same
kinematic factor in the MRK limit.
The colour factor arising from these planar colour connections coincides with
the colour factor from a single diagram with maximal t-channel
gluon exchanges.  In other words, for $2\to n$ gluons, the single colour factor of the FKL amplitude would be
\begin{align}
  \label{eq:fs}
  f^{c_a c_1 d_1}f^{d_1 c_2 d_2}...f^{d_{n-1}c_nc_b},
\end{align}
where $c_a$, $c_1$, ... are the colour indices of the rapidity-ordered external gluons and the
$d_i$ are the repeated indices of t-channel gluons.  All other independent
permutations of the indices multiply kinematic factors which are suppressed in
the MRK limit. The final result for the limit of the colour
summed-and-averaged square of the scattering amplitude agrees with that of
the high-energy limit of QCD derived by Fadin-Kuraev-Lipatov
(FKL)\cite{Kuraev:1976ge}.

The multi-Regge kinematic limit of the \emph{kinematic part} of the
Parke-Taylor amplitudes is found\cite{DelDuca:1993pp} to be such that the full colour summed and
averaged square of the scattering amplitude receives a factor
\begin{align}
  \label{eq:midfactor}
  \frac{4\gs^2 \Ca}{k_{i,\perp}^2}
\end{align}
for each final state gluon beyond the first two. For example, the MRK limit of the colour and spin
summed and averaged matrix element for $gg\to gg$ is
\begin{align}
  \label{eq:ggtoggMRKlimit}
  \overline{\left|\mathcal{M}\right|^2}\longrightarrow \frac {4\hat s^2}{\left(\Nc^2-1\right)}\ \frac{\gs^2\Ca}{k_{1\perp}^2}\ \frac{\gs^2\Ca}{k_{2\perp}^2}.
\end{align}
Similarly, the MRK limit of the colour and spin summed and averaged matrix
element for $gg\to ggg$ is
\begin{align}
  \label{eq:ggtogggMRKlimit}
  \overline{\left|\mathcal{M}\right|^2}\longrightarrow \frac {4\hat
  s^2}{\left(\Nc^2-1\right)}\ \frac{\gs^2\Ca}{k_{1\perp}^2}\
  \frac{4 \gs^2\Ca}{k_{2\perp}^2} \frac{\gs^2\Ca}{k_{3\perp}^2}.
\end{align}
Up to this multiplicity, only MHV configurations contribute to the
amplitude. The above expressions Eqs.~\eqref{eq:ggtoggMRKlimit}
and~\eqref{eq:ggtogggMRKlimit} therefore already cover the most
general case.

In the following, we consider the partons extremal in rapidity (i.e.~partons
1 and 2 for the Born process, 1 and 3 for the $2\to3$-scattering and 1 and
$n$ in the general $2\to n$-scattering) to be hard
in the perturbative sense. Additional partons emitted in-between in rapidity
are then considered part of the radiative corrections to the process.

For a specific choice of rapidities for the extremal partons $p_1, p_3$ in
the limit of the $2\to3$-matrix element of Eq.~\eqref{eq:ggtogggMRKlimit}, the phase space integration of the position
of the middle parton will contribute a factor
\begin{align}
  \int \frac{\mathrm{d^2}k_{2\perp}}{(2\pi)^2}\int_{y_1}^{y_3}\ \frac{d
  y_2}{4\pi} \frac{4 \gs^2\Ca}{k_{2\perp}^2} = \frac{y_3-y_1}{4\pi}  \int
  \frac{\mathrm{d^2}k_{2\perp}}{(2\pi)^2} \frac{4 \gs^2\Ca}{k_{2\perp}^2}
= (y_3-y_1)\ 4 \as\Ca\ \int
  \frac{\mathrm{d^2}k_{2\perp}}{(2\pi)^2} \frac{1}{k_{2\perp}^2}.\label{eq:contribution}
\end{align}
The integral over transverse phase space is IR divergent; the divergence
cancels that introduced by the virtual corrections to the
$2\to2$-scattering. This cancellation is organised by using e.g.~dimensional
regularisation of the integrals, as will be discussed in more detail
later. The point here is that the real (and virtual) corrections to the
Born-level scattering introduce corrections proportional to the rapidity
separation between the extremal (Born-level) partons. In the MRK limit,
$\log \hat s/\hat t\to (y_3-y_1)$, and so we have sketched the appearance of
logarithmic corrections in the perturbative series of the $2\to2$-scattering.

This analysis carries through to any order in \as.
One notes that all dependence on the rapidity of the middle partons is absent in the
factor in Eq.~\eqref{eq:midfactor}, and in the contribution to the
corrections of Eq.~\eqref{eq:contribution} . This
leads to a simple diffential equation for the cross section in $\Delta y=y_n-y_1$; this is called the BFKL
evolution equation\cite{Kuraev:1976ge,Kuraev:1977fs,Balitsky:1978ic}.

Above, we have discussed the
colour connections present in the MRK limit in the tree-level matrix elements
for any number of final-state gluons, i.e.~the real corrections to the Born
level.  The virtual corrections are encoded at all-orders through simple
factors multiplying the $t$-channel poles and hence the colour discussion above
generalises immediately to these cases too.

At higher multiplicities, also non-MHV configurations contribute to the
amplitude. In the MRK limit, the dominant configurations all conserve helicity
between the incoming gluon and the extremal gluon at the respective end (for
MHV configurations, this can be seen directly by considering the numerators in
the Parke-Taylor amplitudes~\cite{DelDuca:1995zy}).  Flipping the helicity of any gluon emitted in-between the extremal gluons only changes the matrix element by a phase in the
MRK limit, so that all helicity configurations which occur in
the MRK limit can be related to the Parke-Taylor formula.

\subsection{Fadin-Kuraev-Lipatov Amplitudes}
\label{sec:FKLalmpl}
In the previous section, we described the behaviour of QCD amplitudes in the
limit of large invariant mass between each particle. Obviously, if the full
amplitude is known, the MRK limit of it can be directly obtained. However,
the limits can also be derived based on the Fadin-Kuraev-Lipatov (FKL)
amplitudes\cite{Fadin:1975cb,Kuraev:1976ge,Kuraev:1977fs}.

\begin{figure}[btp]
  \centering
  \includegraphics[width=.43\textwidth]{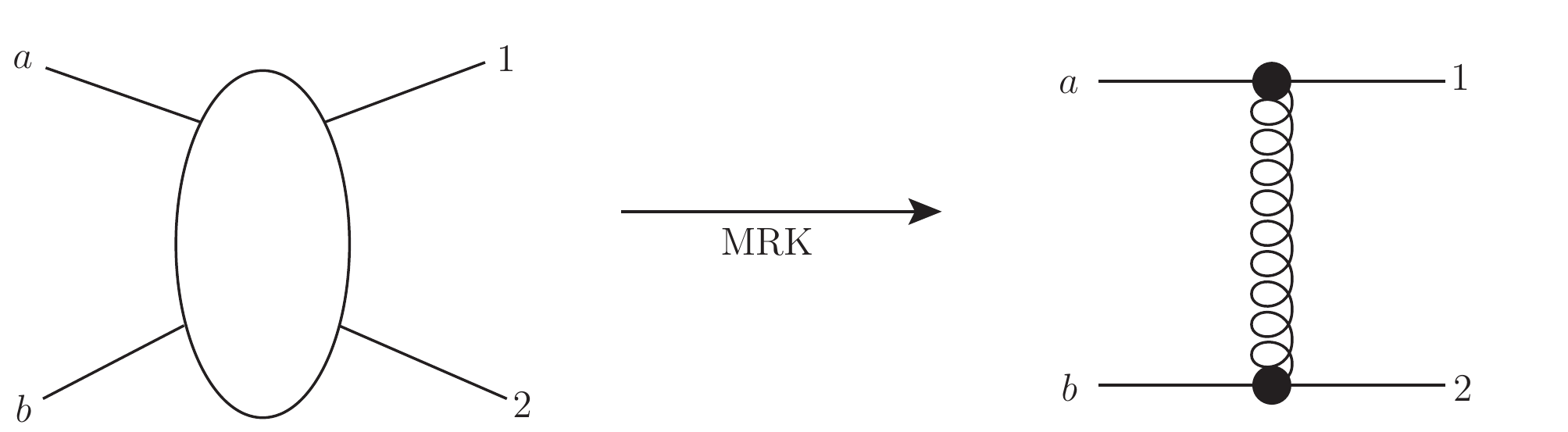}\hspace{1.cm}
  \includegraphics[width=.4\textwidth]{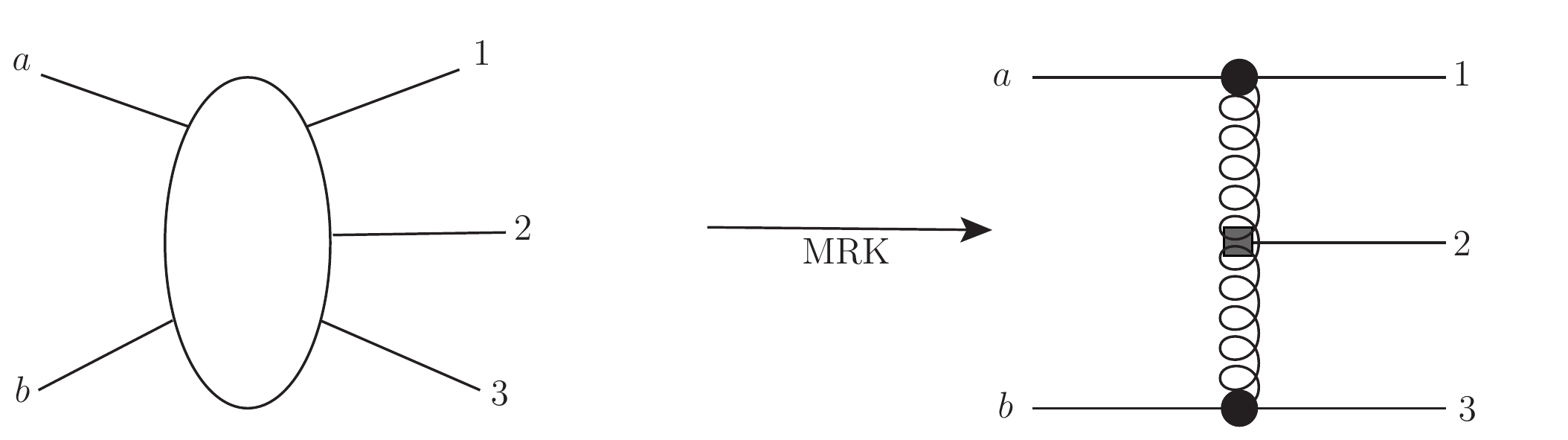}
  \caption{Two simple examples of the factorisation of QCD amplitudes in the MRK
    limit.  Given the process described by the large oval on the left hand side,
    the amplitude in the MRK limit may be written as Left: a product of two
    independent impact factors (black circles) and a gluon exchange and, Right: two independent
    impact factors and a Lipatov emission vertex (grey square) connected with two t-channel
    gluon exchanges.}
  \label{fig:tDominance1}
\end{figure}

QCD scattering amplitudes factorise in the MRK limit into
what in the (B)FKL language are called impact factors and Lipatov vertices,
which are connected by gluon exchanges in the t-channel.  Each of these
components of the amplitude depends only on a much reduced subset of momenta
and is otherwise independent of the rest of the amplitude. This feature
persists after the addition of a Higgs, $W$ or $Z/\gamma^*$ boson to the scattering.  Two simple
examples are shown in Fig.~\ref{fig:tDominance1}.
What is meant by the term ``factorisation of the amplitude'' is that the
correct MRK limit of the amplitude can be obtained from a simple analytic
approximation, which consists of factors, each of which depend only on a
subset of all the momenta of the process. As an example, in the process on
the left-hand-side of Fig.~\ref{fig:tDominance1}, the flavour $f_1,f_2$ of
the external lines may be
quark or gluon and in the MRK limit ($y_1 \ll y_2$), the amplitude may be
expressed in the form:
\begin{align}
  \label{eq:fact1}
  \mathcal{M}_{f_1(p_a) f_2 (p_b)\to f_1(p_1) f_2(p_2)}\to \hat s\ C(p_a,p_1)\ \frac1{(p_a-p_1)^2}\ C(p_b,p_2),
\end{align}
where $C(p_i,p_j)$ indicates an impact factor, which depends on the two momenta along the same
direction on the light-cone only (i.e.~$p_a,p_1$ are the parton momenta
each with the maximum positive light-cone momentum, $p_b,p_2$ have the
largest negative light-cone momentum). The
correct MRK limit of the full amplitude would then be obtained with this
analytic expression, for any configurations of the transverse
momenta. The square of the amplitude is then simply found as
\begin{align}
\label{eq:fact1sq}
  |\mathcal{M}_{f_1(p_a) f_2 (p_b)\to f_1(p_1) f_2(p_2)}|^2\to \hat s^2\
   \frac{|C(p_a,p_1)|^2}{(p_a-p_1)^2}\ \frac{|C(p_b,p_2)|^2}{(p_b-p_2)^2}.
\end{align}
Similarly, the correct MRK limit of the scattering amplitude for the
three-particle final state on the right-hand side may be written
\begin{align}
  \begin{split}
    \label{eq:fact2}
    |\mathcal{M}_{f_1(p_a) f_2 (p_b)\to f_1(p_1) g(p_2)f_2(p_3)}|^2&\to\hat
    s^2 \frac{|C(p_a,p_1)|^2}{(p_a-p_1)^2}
    \frac{|V_L(p_2)|^2}{(p_a-p_1)^2(p_b-p_3)^2} \frac{|C(p_b,p_3)|^2}{(p_b-p_3)^2}\\
    &=\hat s^2 \frac{|C(p_a,p_1)|^2}{t_1} \frac{|V_L(p_2)|^2}{t_1\ t_2}
    \frac{|C(p_b,p_3)|^2}{t_2}
  \end{split}
\end{align}
where $V_L$ is a so-called \emph{Lipatov vertex}. The only difference to the
form of the two-particle final state is the insertion of a vertex and a
propagator in the analytic form of the MRK limit, which has a form suggestive of
the $t$-channel exchange. The $t$-channel interpretation of the analytic form
of the kinematic part of the amplitude is supported by the colour-connections
studied in Sec.~\ref{sec:LeadColour}, but while the contribution from individual $t$-channel
Feynman diagrams are obviously gauge dependent, it is important to realise
that the MRK limit of the scattering amplitude is a gauge-independent
statement. It just happens to have the analytic form expected from a
$t$-channel gluon exchange, as expected from the analysis presented in
Sec.~\ref{sec:leadingcontrlarge}.

For the impact factors one finds $|C(p_a,p_1)|^2=1$, and in the MRK limit
$t_1\to -k_{1\perp}^2,t_2\to -k_{3\perp}^2$, and one finds\cite{Kuraev:1977fs} that the factor
introduced from an additional gluon emission of transverse momentum
$k_{2\perp}$ into the FKL result for the square of the matrix element is
simply
\begin{align}
  \frac{|V_L(p_2)|^2}{t_1\ t_2}\to \frac{4\gs^2\Ca}{k_{2\perp}^2}.
\end{align}
Therefore, the MRK limit of the QCD amplitudes found in
Sec.~\ref{sec:beyondtree} are reproduced by the FKL
amplitudes\cite{DelDuca:1993pp,DelDuca:1995zy}. This is true for an arbitrary
number of gluons emitted, such that the FKL result for the leading-order
contribution to the colour-and-spin summed-and-averaged square of the
scattering amplitude is given by
\begin{align}
  \label{eq:FKLnsummedandaveraged}
  \overline{|\mathcal{M}^\mathrm{FKL}_{gg\to g_1\cdots g_n}|^2}=\frac{2\hat
  s^2}{4\left(\Nc^2-1\right)} \prod_{1}^n \frac{\gs^2\Ca}{k_{i\perp}^2}.
\end{align}

The $t$-channel structure of the FKL amplitudes allows for the inclusion of
the dimensionally regulated virtual corrections (in $D=4-2\varepsilon$
dimensions) through the \emph{Lipatov Ansatz} for the Reggeized $t$-channel
colour-octet exchanges. This is the prescription for including the all-order
virtual corrections to the Born-level colour octet exchange by making the
following substitution in
Eq.~\eqref{eq:fact1sq}:
\begin{align}
  \label{eq:LipatovAnsatz} \frac 1 {t_{i-1}}\ \to\ \frac 1 {t_{i-1}}\ \exp\left[\hat
\alpha (q_i)(y_{i-1}-y_i)\right]
\end{align} where
\begin{align}
  \hat{\alpha}(q_i)&=-\gs^2\ \Ca\
  \frac{\Gamma(1-\varepsilon)}{(4\pi)^{2+\varepsilon}}\frac 2
  \varepsilon\left(q_{i\perp}^2/\mu^2\right)^\varepsilon,\label{eq:ahatdimreg}
\end{align}
with $q_i=p_a-\sum_{j=1}^i p_j$, such that $t_i=q_i^2$. This ansatz for the
exponentiation of the virtual corrections in the appropriate limit of the
$n$-parton scattering amplitude has been proved to even the sub-leading
level\cite{Bogdan:2006af,Fadin:2006bj,Fadin:2003xs,Fadin:2005pt}, which leads
to a perturbative correction to leading-logarithmic results for
$\hat \alpha$, the Lipatov vertex and the impact factors.

The FKL result for the square of the scattering matrix for $2\to n$ obtained
by using the kinematic approximations valid in the multi-regge-kinematic
limit has no dependence on the rapidities of the final-state particles (in
essence because the limit of infinite rapidity-separation has been
applied). The poles in $\epsilon$ in the dimensionally regulated inclusion of
the virtual corrections through the Lipatov ansatz turn out to cancel
order-by-order with the poles from the dimensionally regulated integration
over the soft phase space of additional emissions (intermediate in rapidity
between parton 1 and $n$) included through the FKL result for the square of the
matrix element for $2\to m, m>n$. A finite contribution from the virtual
corrections is left over. If now the contribution to the centre-of-mass
energy $\sqrt{\hat s}$ and therefore also to the longitudinal momentum of the
incoming partons is ignored from all but the most backward and forward
parton, then the sum over the integration over phase space of any parton of
intermediate rapidity can be performed analytically. This leads to the much
celebrated BFKL equation\cite{Balitsky:1978ic}, which captures the leading
(and sub-leading) behaviour in $\log(\hat s/p_t^2)$. It is seen that the
logarithmic behaviour is the same when using the FKL amplitudes of
Eq.~\eqref{eq:FKLnsummedandaveraged} and the limit of the full QCD amplitudes
as discussed in Sec.~\ref{sec:beyondtree}. The large-rapidity behaviour of
the $m$-parton amplitudes of full QCD and FKL is the same in terms of powers
of $\hat s/p_{t i}^2$, which is sufficient to guarantee the same logarithmic
behaviour of the integrated cross section in terms of $\log(\hat s/p_t^2)$.

\subsection{Construction of the Simplest HEJ Amplitude}
\label{sec:constr-hej-ampl}

In the previous two subsections, we have described how the leading behaviour of
scattering amplitudes in QCD arises through the study of $t$-channel poles, and how the simple structure in the MRK limit
is captured to all orders in $\alpha_s$ by the FKL amplitudes. So far with \HEJ, all-order results
have been achieved for such FKL configurations only. All other kinematic
configurations have been included to fixed order only through a matching and
merging procedure described in Sect.~\ref{sec:match-merg-fixed}.  In this
paper, we present for the first time the inclusion of all-order results also
for some sub-leading corrections, namely quark-initiated processes with one
gluon emitted outside the FKL-ordered phase space. These configurations
correspond to the suppressed contributions studied in
Figure~\ref{fig:FKLdominance}. The leading logarithmic corrections to these
processes constitute the first sub-leading logarithmic corrections included
in \HEJ. The configurations constitute the largest part of the sub-leading
cross-section, which previously was included through the na\"ive addition of
fixed-order samples. The inclusion of these sub-leading (and their matching
to fixed-order accuracy) therefore gives a much more satisfactory theoretical
description of the scattering.

The motivation behind the \HEJ framework is to capture the behaviour of
amplitudes at large $\hat s$ \emph{without} applying the full tower of
approximations necessary for obtaining an analytic answer for the cross section
through the BFKL theory. By allowing for numerical integration of multi-particle
amplitudes, we can both allow these to have a more complicated kinematic
dependence than the $1/k_\perp^2$ of the FKL-amplitudes, and account for the
longitudinal momentum-conservation which is invariably lost in any formulation
involving the BFKL equation (at both LL and NLL accuracy).

The simplest of all QCD processes is that of $qQ\to qQ$, proceeding through a
$t$-channel gluon exchange only. The MRK limit of the full QCD result and the
FKL approximation of the square of this amplitude is
\begin{align}
  \label{eq:qQtoqQFKL}
  \overline{|\mathcal{M}^\mathrm{QCD}_{qQ\to qQ}|^2} \to \overline{|\mathcal{M}^\mathrm{FKL}_{qQ\to qQ}|^2}=\frac{2\hat
  s^2}{4\left(\Nc^2-1\right)} \frac{\gs^2\Cf}{k_{1\perp}^2}
  \frac{\gs^2\Cf}{k_{2\perp}^2}=\frac{\left(\gs^2 \Cf\right)^2}{4(\Nc^2-1)}
  \frac{2\hat s^2}{k_{1\perp}^2 k_{2\perp}^2}.
\end{align}
The leading-order QCD result is given by
\begin{align}
  \label{eq:qQQCD}
\overline{|\mathcal{M}^\mathrm{QCD}_{qQ\to qQ}|^2}=\frac{\left(\gs^2
  \Cf\right)^2}{4(\Nc^2-1)}\frac{\hat s^2+\hat u^2}{\hat t^2}.
\end{align}
Two kinematic approximations are necessary to get from the full result to
the approximation of FKL: $\hat s\sim - \hat u$,
$\hat t\sim - k_{1\perp}^2= - k_{2\perp}^2$ (where the last equality holds
for the simple $2\to2$ process). While both of these are valid in the MRK
limit, they are easily off by an order of magnitude within the relevant phase
space of the LHC.

In constructing a Monte Carlo phase space integrator, which is sufficiently
efficient to calculate explicitly the phase space integration over
many-particle (e.g.~up to 30) final state phase space, we can seek to build
an approximation for the matrix elements, which still captures the leading
logarithmic behaviour generated from the $t$-channel poles, but which relies
on fewer kinematic approximations. In particular, we want the
description of the amplitude to be:
\begin{itemize}
\item exact for the simple $2\to2$-process proceeding only through a $t$-channel
  exchange\footnote{We note that the approximation obtained through a
    BFKL-equation cannot improve upon the approximant in
    Eq.~\eqref{eq:qQtoqQFKL}, even through the inclusion of next-to-leading
    logarithmic terms (or higher).};
\item gauge-invariant for any additional gluon emitted, i.e.~the Ward Identity
  is fulfilled (not just asymptotically in the MRK-limit, as for FKL-amplitudes,
  but exactly, everywhere in phase space), $k_n^\mu \mathcal{M}_\mu=0$;
\item such that the soft divergences of the approximant are cancelled by the
  terms generated from the Lipatov Ansatz for the virtual corrections to the
  tree-level results (also for $2\to n$-processes); and
\item sufficiently fast to evaluate such that the numerical integration is
  feasible.
\end{itemize}

Let us first focus on building this simple approximant for the
$2\to2$-processes. The Lipatov Ansatz can most easily be applied if the
analytic structure of the $m$-parton amplitude is factorised into a dependence
on $1$-parton and the $(m-1)$-parton amplitude (obviously evaluated with the
momenta of the $m$-parton phase space). It is therefore important to build a
good approximant to even the simplest processes, since obviously the
multi-particle approximations are built on successive applications of
these. We will see that by using helicity amplitudes, we can build such a
simple structure for approximants of multi-particle amplitudes, which are
valid even before the Multi-Regge-Kinematic limit is applied.

Since we will be evaluating the amplitudes numerically in the Monte-Carlo
integration, there is no problem in keeping the full kinematic dependence on
the $t$-channel propagator-momentum $\hat t$ in Eq.~\eqref{eq:qQQCD} rather
than performing the MRK-approximation $\hat t\to -k_{1\perp}^2$. Clearly, the
$t$-channel poles are described best by maintaining the full dependence on
the $t$-channel momenta.  We now turn to describing the remaining invariants,
$s$ and $u$.  In the full MRK limit, $s=-u$; in practice, there is a large
deviation throughout phase space.  By studying the amplitude for $qQ\to qQ$, we find that
terms proportional to $s^2$ arise from amplitudes where the quarks have
identical helicities; while terms proportional to $u^2$ arise from amplitudes
where the two quark lines have opposite helicities. Explicitly, in terms of currents
$j^{\mu \pm}(p_i,p_j) = \bar{u}^\pm(p_i) \gamma^\mu u^\pm(p_j)$,
one finds that
\begin{align}
  \label{eq:sneu}
  |j^{\mu \pm}(p_1,p_a)\ j_\mu^\pm(p_2,p_b)|^2 = s^2, \quad |j^{\mu \pm}(p_1,p_a)\
  j^\mp_\mu(p_2,p_b)|^2 = u^2.
\end{align}
By working at the helicity amplitude level, we have achieved a description of
the $2\to 2$ amplitude that is exact, and furthermore the analytic form
generalises easily to $2\to n$.  These components then depend on $\{p_a,p_1\}$
and $\{p_b,p_2\}$ separately as in Eq.~\eqref{eq:fact1}.  Hence the product of
two scalar impact factors has been expanded to a contraction of vector currents.

In fact, this factorised form also continues when one moves to $qg\to qg$
with the same quark current as above~\cite{Andersen:2009he}. The gluon
current has an additional scalar factor, but it can still be written in a
form which depends only on the gluon momenta, and can be found in Eq.~(8) of
Ref.\cite{Andersen:2011hs}, with the exact amplitude for $qg\to qg$ written
in terms of the \HEJ building-blocks as
\begin{align}
  \begin{split}
    \label{eq:sqcol}
    |\mathcal{M}_{q^-g^+\to q^- g^+}|^2\ &=  \frac 1 {\Nc^2-1}\ | \spab{b}.\rho.2 \spab1.\rho.a |^2\\
    &\cdot \left( g_s^2\ \Cf\ \frac 1 {t_1}\right)\\
    &\cdot \left( g_s^2\  \left[ \frac 1 2\
       \frac{1+z^2}{z}\ \left(C_A -\frac 1 {C_A}\right)+
       \frac{1}{C_A}\right]\ \frac 1 {t_2}\right),
  \end{split}
\end{align}
where $z=p_2^-/p_b^-$ (and again $t_1=(p_a-p_1)^2=(p_b-p_2)^2=t_2$). This is
written for the case of a backward moving incoming gluon; for a
forward-moving gluon, one would simply define $z=p_2^+/p_b^+$. A similar $t$-channel
factorised form is found for $g^+g^-\to g^+ g^-$ scattering (in the
configuration with scattering of gluons with the same helicity there is of
course no unique concept of the $t$-channel).

We will later discuss how the scattering amplitude can be extended to capture
the all-order leading logarithmic accuracy of the cross section by accounting
for the emissions of additional gluons.

The structure of an amplitude approximated by building blocks, each depending
only on the momenta of a small subset of the particles is obviously appealing
computationally.  Not only though are these factors independent of other
particle momenta, they are completely independent of the rest of the process and
are therefore in that sense, process-independent.  So, if particles $a$ and $1$
are the same flavour in each case (either quark or gluon), the factor of the
FKL formalism
$C(p_a,p_1)$ in Eqs.~\eqref{eq:fact1} and \eqref{eq:fact2} will be identical,
and so will the currents used in \HEJ.


The next building block we need to derive is the Lipatov vertex, $V_L$, for additional FKL-ordered
gluon emissions.  The simplest process to study is $qQ\to qgQ$.
It is necessary to sum the contributions from all five tree-level diagrams.  After
some manipulation in the high-energy limit this yields\cite{Andersen:2009nu}
\begin{align}
  \label{eq:HEJqgQ}
  \begin{split}
    \mathcal{M}^{\rm HEJ}_{\rm tree\ qQ\to qgQ} &= -g_s^2 T^{a_1}_{i_1 i_a}T^{a_2}_{i_3 i_b}\
    \frac{\mathcal{S}_{qQ}(p_1,p_3,p_a,p_b)}{q_1^2 q_2^2} \times ig_s f^{a_{1}b_2a_2}\
    \varepsilon_{\nu_2}(p_2) V^{\nu_2}_L(q_{1},q_{2}),
  \end{split}
\end{align}
where $\mathcal{S}_{qQ}(p_1,p_3,p_a,p_b)$ is still a contraction of currents:
\begin{align}
  \label{eq:Sfunc}
  \mathcal{S}_{qQ}(p_1,p_3,p_a,p_b)=j^{\mu}(p_1,p_a)\ j_\mu(p_3,p_b),
\end{align}
and $V^\nu_L$ is a Lipatov-type vertex for gluon emission, which is given by:
\begin{align}
  \label{eq:GenEmissionV}
  \begin{split}
  V_L^\nu(q_i,q_{i+1})=&-(q_i+q_{i+1})^\nu \\
  &+ \frac{p_a^\nu}{2} \left( \frac{q_i^2}{p_{i+1}\cdot p_a} +
  \frac{p_{i+1}\cdot p_b}{p_a\cdot p_b} + \frac{p_{i+1}\cdot p_n}{p_a\cdot p_n}\right) +
p_a \leftrightarrow p_1 \\
  &- \frac{p_b^\nu}{2} \left( \frac{q_{i+1}^2}{p_{i+1} \cdot p_b} + \frac{p_{i+1}\cdot
      p_a}{p_b\cdot p_a} + \frac{p_{i+1}\cdot p_1}{p_b\cdot p_1} \right) - p_b
  \leftrightarrow p_n.
  \end{split}
\end{align}
This form is slightly more involved than the standard Lipatov (or
Reggeon-Reggeon-particle-) vertex of BFKL\cite{Fadin:1996nw}, since it
maintains the dependence on each of the 4 quark momenta rather than making
the approximation $p_a\sim p_1, p_b\sim p_n$; the two last terms in each
bracket constitutes the eikonal approximation to the emission off each
leg. The difference between the form used in BFKL and in \HEJ is formally
sub-leading, but crucial for obtaining analytic results in BFKL. Conversely,
the full form of Eq.~\eqref{eq:GenEmissionV} unsurprisingly gives a more
accurate description of the sub-asymptotic region of phase space, and thus
leads to smaller matching-corrections. In choosing to perform the phase space
integrations numerically, we are free to choose the numerically more accurate
form.

\begin{figure}[btp]
  \hspace{5.5cm}
  \scalebox{0.6}{\includegraphics{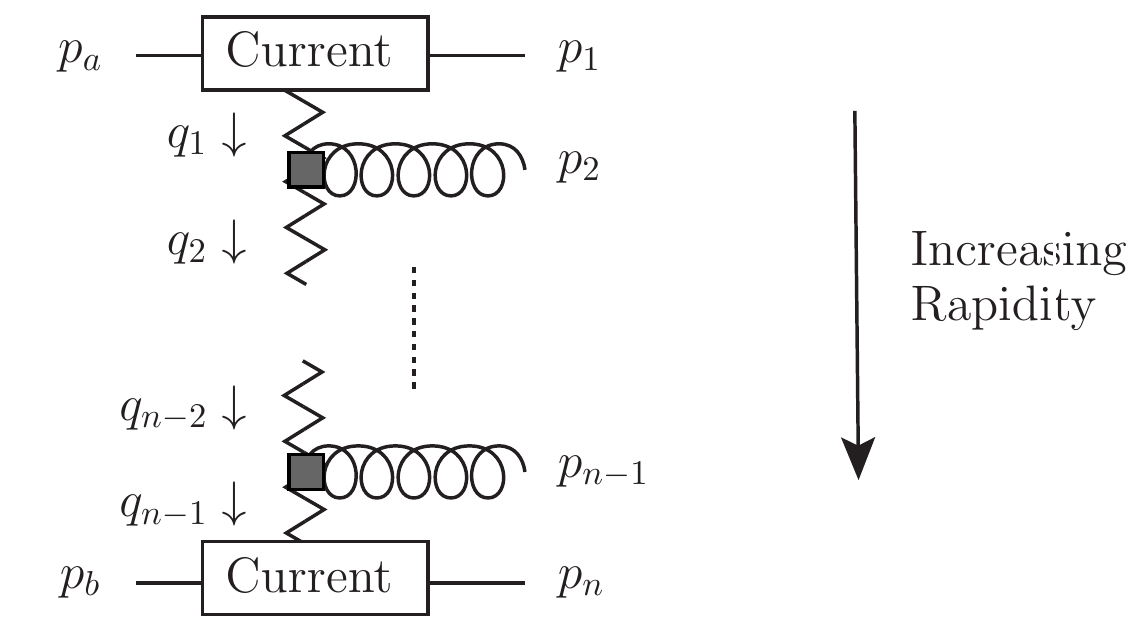}}
  \caption{The analytic structure of the base tree-level scattering amplitude
    for $qQ\to qQ+(n-2)g$ in \emph{High Energy Jets}. Grey boxes denote
    Lipatov vertices.}
  \label{fig:HEJstructure}
\end{figure}

Now the power of the high-energy limit becomes manifest.  With only the building
blocks derived so far, the leading contribution of the scattering
amplitude (in powers of $\hat s/p_t^2$, forming the leading logarithmic
contribution to the integrated cross section) for any number of intermediate gluon emissions is described by
\begin{align}
    \label{eq:MainHEJqQnoH}
  \begin{split}
    \mathcal{M}^{\rm HEJ}_{\rm tree} &= -g_s^2 T^{a_1}_{i_1 i_a}T^{a_{n-1}}_{i_n i_b}\
    \frac{\mathcal{S}_{qQ}(p_1,p_n,p_a,p_b)}{\sqrt{q_1^2 q_j^2
        q_{j+1}^2 q_{n-1}^2}}\times \prod_{k=2}^{n-1} ig_s f^{a_{k-1}b_ka_k}\
    \frac{\varepsilon_{\nu_k}(p_k) V^{\nu_k}_L(q_{k-1},q_{k})}{\sqrt{q_{k-1}^2 q_k^2}}.
  \end{split}
\end{align}
This structure is shown in Fig.~\ref{fig:HEJstructure}, where the Lipatov
vertices are shown as grey boxes.  The amplitude for the
equivalent process with one or two incoming gluons is identical, except for a
minor alteration to the function $\mathcal{S}$.

\subsection{Regularisation and Leading Logarithmic All-Order Cross Sections}
\label{sec:all-order-reg}
In sections~\ref{sec:logarithmic-accuracy}--\ref{sec:constr-hej-ampl} we
identified the leading contributions for jet production in the
multi-Regge-kinematic limit, and showed how to obtain an accurate
approximation to the Born-level matrix elements for such processes for
\emph{any} multiplicity of gluon emissions. The only singularities present in
this approximation are those arising from the $t$-channel propagators in the
colour-octet exchanges of the rapidity-ordered final state, and these
singularities of the Born-level amplitude are outside the physical region. As
discussed in Section~\ref{sec:beyondtree}, logarithmic corrections in
$\hat s/p_t^2$ arise in the region of jets widely separated in rapidity. So
far, we have discussed Born-level results only. In this section, we will
discuss the calculation of the cross section to each order in \as, and the
regularisation of the IR singularities. No UV singularities appear at the
logarithmic order discussed.

The reason for developing an approximation to the $t$-channel poles of the
scattering tree-level matrix elements is that the leading logarithmic
contribution to the loop corrections of these processes can still be obtained
using the \emph{Lipatov ansatz}\cite{Kuraev:1976ge}, just as discussed for
the FKL amplitudes in Section~\ref{sec:FKLalmpl}. This ansatz states that the
leading logarithmic contribution to the virtual corrections for amplitudes in
the MRK limit can be found to all orders in the coupling by replacing each
$t$-channel propagator between the two particles of ordered rapidities
$y_{i}$ and $y_{i+1}$ ($y_{i}<y_{i+1}$) in the amplitudes constructed in
section~\ref{sec:constr-hej-ampl} as follows:-
\begin{align}
  \label{eq:LipatovAnsatz2} \frac 1 {t_i}\ \to\ \frac 1 {t_i}\ \exp\left[\hat
\alpha (q_i)(y_{i+1}-y_{i})\right]
\end{align} with
\begin{align}
  \hat{\alpha}(q_i)&=-\gs^2\ \Ca\
  \frac{\Gamma(1-\varepsilon)}{(4\pi)^{2+\varepsilon}}\frac 2
  \varepsilon\left(q_{i\perp}^2/\mu^2\right)^\varepsilon.
\end{align}
As mentioned earlier, this ansatz for the exponentiation and factorisation of
the virtual corrections in the appropriate limit of the $n$-parton scattering
amplitude has been proved to hold even at the sub-leading
level~\cite{Bogdan:2006af,Fadin:2006bj,Fadin:2003xs,Fadin:2005pt} and
explicitly checked against the two-loop amplitudes for
$qg$-scattering\cite{Bogdan:2002sr}.

As demonstrated in
e.g.~Ref.\cite{Andersen:2011hs} and below, the poles in $\epsilon$ cancel
exactly between the dimensionally regularised (in $D=4-2\epsilon$ dimensions)
virtual and real corrections to processes of any multiplicity, when
calculated with the constructed amplitudes which ensure the correct leading
logarithmic (in $\hat s$) behaviour of the cross section. This allows for the
calculation of the inclusive cross section (for the leading and the included
sub-leading processes) as explicit sums of $n$-body 4-dimensional phase space
integrals of dimensionally regularised $n+2$-particle matrix elements.

The first step in organising the cancellation of the poles in $\varepsilon$ and
obtaining the regularised cross sections is to define for each Born-level
momentum configuration the regions in phase space for which the real corrections
for gluon emissions can be calculated to any order in the coupling. It is the
phase space region in rapidity delimited by the extremal partons. These partons
extremal in rapidity are required to be perturbative (i.e.~of a transverse
momentum similar to the hard jet scale), since these form parts of the
fundamental currents of the formalism, and there is (at LL accuracy) no
accompanying virtual corrections to regulate the divergences present as the
transverse momenta of these extremal partons tend to zero. However, for the
phase space bounded in rapidity by these extremal, hard partons, the soft
singularity from the real emission of additional gluons is regulated by the
singularity from the virtual corrections to all orders in the coupling (i.e.~for
any number of emissions into that region of rapidity).


To illustrate the specifics of this procedure, consider for simplicity the
process $qQ\to jj$. We will now show how the leading logarithmic
perturbative corrections to this process are calculated to all orders through
the explicit construction of regulated, four-dimensional amplitudes, which
can be summed and integrated explicitly using Monte-Carlo techniques.

We will apply dimensional regularisation (working in $D=4-2\varepsilon$
dimensions) in order to facilitate the cancellation of poles from real and
virtual corrections. The colour and spin summed and averaged square of the
scattering matrix element for the process $f_1 f_2\to f_1 \cdot g\cdot f_2$
(where $\cdot g\cdot$ indicates the possibility of any number of gluons),
following from Eq.~\eqref{eq:MainHEJqQnoH} but extended
to $4-2\epsilon$ dimensions, is
\begin{align}
  \label{eq:Mtvepsilon}
  \begin{split}
    \overline{\left|\mathcal{M}^{\rm HEJ}_{\varepsilon\ f_1 f_2 \to f_1\cdot
          g\cdot f_2}\right|}^2 = \ &\frac 1 {4\
       (\Nc^2-1)}\ \left\|S_{f_1 f_2\to f_1 f_2}\right\|^2\\
     &\cdot\ \left(g_s^2\ K_{f_1}\ \frac 1 {t_1}\right) \cdot\ \left(g_s^2\ K_{f_2}\ \frac 1
       {t_{n-1}}\right)\\
     & \cdot \prod_{i=1}^{n-2} \left( \frac{-g_s^2 C_A}{t_it_{i+1}}\
       V^\mu(q_i,q_{i+1})V_\mu(q_i,q_{i+1}) \right)\\
     & \cdot \prod_{j=1}^{n-1} \exp\left[2 \hat
         \alpha (q_j)(y_{j+1}-y_j)\right].
  \end{split}
\end{align}
The colour factors $K_{f_i}$ are $C_F$ if particle $i$ is a quark and $C_A$ if
it is a gluon.  The matrix element above describes the leading-logarithmic
corrections to dijet production at all orders in $\alpha_s$.  These take the
form of additional partons in the final state described by the $V^\mu$ emission
vertices and the corresponding exponential factors arising from the virtual
contributions to the process.  We organise the cancellation of the divergences
by means of a phase-space slicing parameter $\lambda$, which separates the
``hard'' region ($p_\perp>\lambda$) from the ``soft'' region
($p_\perp < \lambda$).  The divergences arising from soft emissions arise from
the singularities of the emission vertices.  Explicitly, in the limit that
$p_k\to 0$,
\begin{align}
  \label{eq:Vsing}
  \frac{V^\mu(q_{k-1},q_k) V_\mu(q_{k-1},q_k)}{t_{k-1}t_k} \to \frac{-4}{p_{k\perp}^2}.
\end{align}
We therefore have
\begin{align}
  \label{eq:Mlim}
  \overline{\left|\mathcal{M}^{\rm HEJ}_{{\rm tree},\varepsilon\ f_1 f_2 \to f_1
  (n-2)gf_2}(\{p_i\}) \right|}^2 \quad \begin{array}{c} \longrightarrow \\
                                         p_k\to 0 \end{array} \quad \left(\frac{4g_s^2 C_A}{p_{k\perp}^2}\right)\
  \overline{\left|\mathcal{M}^{\rm HEJ}_{{\rm tree},\varepsilon\ f_1 f_2 \to f_1
  (n-3)gf_2}(\{p_i\}\backslash p_k)\right|}^2.
\end{align}
The set of particle momenta on the right-hand side (the set of $n-1$ momenta
obtained by removing $p_k$) still satisfies momentum conservation since we
are precisely considering the case of $p_k\to 0$. The divergence in the
$2\to n$-scattering matrix element in the limit $p_k\to 0$ is therefore
identical to that obtained using the simple factor in
Eq.~\eqref{eq:Mlim}. We can therefore organise the cancellation of soft
divergences between real and virtual corrections by first subtracting the
term in brackets from the square of the Lipatov vertices. Since we only need
to regularise the divergence, we will restrict this real-subtraction term to
soft momenta, i.e.~$p_{k\perp}<\lambda$. The integral of the real-emission
subtraction term is then found as
\begin{align}
  \label{eq:softint}
  \begin{split}
    \mu^{-2\varepsilon} \int_{\rm soft}
    \frac{d^{3+2\varepsilon}p_k}{(2\pi)^{3+2\varepsilon} 2 E_k}\
    \frac{4g_s^2C_A}{|p_{k\perp}|^2} &= \mu^{-2\varepsilon}\int_0^\lambda
    \frac{d^{2+2\varepsilon}p_{k\perp}}{(2\pi)^{2+2\varepsilon}}
    \int_{y_{k-1}}^{y_{k+1}} \frac{dy_k}{4\pi}\ \frac{4g_s^2 C_A}{|p_{k\perp}|^2}
    \\
    &= \mu^{-2\varepsilon} \frac{4g_s^2 C_A}{(2\pi)^{2+2\varepsilon} (4\pi)}
    (y_{k+1}-y_{k-1})  \int_0^\lambda
    \frac{d^{2+2\varepsilon}p_{k\perp}}{|p_{k\perp}|^2} \\
    &=\frac{g_s^2 C_A}{\pi(2\pi)^{2+2\varepsilon}}\ (y_{k+1}-y_{k-1})\
    \frac1\varepsilon \frac{\pi^{1+\varepsilon}}{\Gamma(1+\varepsilon)} \left(
      \frac{\lambda^2}{\mu^2} \right)^\varepsilon.
  \end{split}
\end{align}
This contribution will be added to the virtual corrections for the
$n-1$-momenta state. These virtual corrections can be found by expanding the
exponential factor in the last line of Eq.~\eqref{eq:Mtvepsilon} which spans
the rapidity region integrated over in Eq.~\eqref{eq:softint}.  We therefore
find to first order in $\alpha_s$
\begin{align}
  \label{eq:findvirt}
  \begin{split}
   \left( -2(y_{k+1}-y_{k-1})g_s^2 C_A \frac{\Gamma(1-\varepsilon)}{(4\pi)^{2+\varepsilon}}
       \frac2\varepsilon \left( \frac{q_{k\perp}^2}{\mu^2} \right)^\varepsilon
     \right) \overline{\left|\mathcal{M}^{\rm HEJ}_{{\rm tree},\varepsilon\ f_1 f_2 \to f_1
      (n-3)gf_2}(\{p_i\}\backslash p_k)\right|}^2.
  \end{split}
\end{align}
Combining this with the contribution from the integral of the real-emission
subtraction term in Eq.~\eqref{eq:softint} and expanding in $\varepsilon$,
the pole in $\varepsilon$ and (the dependence on $\mu$) cancels exactly. This
is in fact true order-by-order in $\varepsilon$, and the finite correction
which remains can be absorbed into the \emph{regularised trajectory}
\begin{align}
  \label{eq:finite}
  \omega^0(q_{\perp}^2)&=\frac{g_s^2 C_A}{4\pi^{2}}
  \ \log \left(\frac{\lambda^2}{q_{\perp}^2} \right)=
\as \frac {C_A}{\pi} \ \log \left(\frac{\lambda^2}{q_{\perp}^2} \right).
\end{align}
We can repeat this for each real emission between the extremal partons, which
yields the following all-order description of dijet production:
\begin{align}
  \label{eq:finitefinal}
  \begin{split}
        \overline{\left|\mathcal{M}^{\rm HEJ}_{\varepsilon\ f_1 f_2 \to f_1\cdot
          g\cdot f_2}\right|}^2 = \ &\frac 1 {4\
       (\Nc^2-1)}\ \left\|S_{f_1 f_2\to f_1 f_2}\right\|^2\\
     &\cdot\ \left(g_s^2\ K_{f_1}\ \frac 1 {t_1}\right) \cdot\ \left(g_s^2\ K_{f_2}\ \frac 1
       {t_{n-1}}\right)\\
     & \cdot \prod_{i=1}^{n-2} \left( \frac{-g_s^2 C_A}{t_it_{i+1}}\
       V^\mu(q_i,q_{i+1})V_\mu(q_i,q_{i+1}) \right)\\
     & \cdot \prod_{j=1}^{n-1} \exp\left[\omega^0(q_{j\perp})(y_{j+1}-y_j)\right].
  \end{split}
\end{align}
The remaining numerical phase space integration now excludes the soft
region, i.e. we require $p_{k\perp} > \lambda$ for all emitted gluons.

In practice, we find that the contribution from the small, finite integral of
the difference between the Lipatov vertex and the subtraction term is
negligible for transverse momenta less than roughly $\kappa=0.2$~GeV~$=200$~MeV,
but can be relevant if $\lambda$ is larger than that value.  We therefore add the correction
\begin{align}
  \label{eq:ccutcorrection}
  \frac{V^\mu(q_{k-1},q_k) V_\mu(q_{k-1},q_k)}{t_{k-1}t_k} + \frac{4}{p_{k\perp}^2}.
\end{align}
for values of $\kappa<|p_{k\perp}|<\lambda$ and find stable results under
variation of both $\kappa$ and $\lambda$. Numerically stable results can be
obtained with $\kappa$ as low as 0.1~GeV (but we will take $\kappa=0.2$~GeV
since the results are the same, but require less computing time). In fact, if
we choose $\kappa=\lambda$, then the real subtraction term is only applied in
the region of phase space which is integrated over analytically. In the
remaining transverse-momentum phase space, which is integrated over
numerically, the integrand will be positive definite, since the Lipatov
vertex is a space-like 4-vector, and there are no subtraction terms in this
resolved phase space.

The matrix-element squared in Eq.~\eqref{eq:finitefinal} is the basis of the
\HEJ description of dijet production.  In order to generate final cross
sections, this is supplemented with both matching and merging and is then
integrated over the final phase space. However, this procedure will be the same
after the inclusion of the new corrections described in the next section, and we
therefore postpone the discussion of these final aspects until
Section~\ref{sec:match-merg-fixed}.

\subsection{The First Set of Sub-Leading Corrections}
\label{sec:UnorderedandHiggsB}

Section~\ref{sec:logarithmic-accuracy} presented the \emph{FKL
  configurations}, i.e. the flavour and momentum configurations which result
in the leading power behaviour in $s/p_\perp^2$ of the amplitudes. By integrating
these over phase space we find the leading logarithmic contributions in
$s/p_\perp^2$ to the cross section. There is a term of order
$\alpha_s\log(\hat s/t)\approx\alpha_s\Delta y$ contributing for each
additional emission of a gluon between the two quarks in rapidity. These terms
contribute to the cross section as
$\alpha_s^2 (\alpha_s\Delta y)^n$ for all $n$. The
remaining momentum-orderings can be included by simply adding tree-level
predictions for these to the events sample, as done in
Ref.\cite{Andersen:2011hs,Andersen:2012gk,Andersen:2016vkp}. Using this
method, higher-order corrections are included to FKL-orderings only.

In this section, we will describe the inclusion of one set of
\emph{next-to-}leading logarithmic corrections to the cross sections. Such
can arise as sub-leading corrections to processes already included at
leading-logarithmic accuracy (i.e. as a control of a sub-leading behaviour in
the power-expansion of the amplitude), or as the inclusion of processes that
do not contribute at leading logarithmic accuracy. Such processes will also
contribute at sub-leading level in the power-expansion of the amplitude, but
the two contributions to the overall NLL corrections to the amplitude are
physically disconnected. In fact, we will here calculate the leading
logarithmic corrections to flavour and momentum orderings which at
Born-level behave as $\alpha_s^2 \alpha_s$ (i.e. without the $\Delta
y$-enhancement of FKL-orderings). We focus on these, since an investigation
of the non-FKL matching contributions identify these as the largest
contribution. We relax the requirement of an ordering in rapidity of the
emission of exactly one gluon. This means that one gluon is allowed outside
of the rapidity range delimited by the outgoing quarks, e.g.~$qQ\to gqQ$ in
that rapidity order, and we will term these flavour and momentum configuration
\emph{unordered emissions}. Specifically, the approximations for the amplitudes
for these configurations require all terms are kept according to the ordering
$s_{2g}\approx s_{12}$, i.e.~$y_1\approx y_g\ll y_2$.

The discussion in section~\ref{sec:logarithmic-accuracy} tells us that the
square of the amplitude for these \emph{unordered} configurations are
suppressed by one power of $s_{1g}$ compared to the FKL-ordered process; the
leading-logarithmic corrections to this unordered process will then form part of the sub-leading corrections to the cross section.  The
advantage of including an all-order treatment of these processes is two-fold:
firstly we will now be able to apply the resummation of all-order high-energy
logarithms to a greater part of inclusive jet cross sections and secondly, we
reduce our dependence on leading-order matching.  We will explicitly evaluate
their contribution and the impact of their inclusion in section~\ref{sec:match-merg-fixed}; here we describe
their construction.

\begin{figure}[btp]
  \centering
  \includegraphics[width=0.8\textwidth]{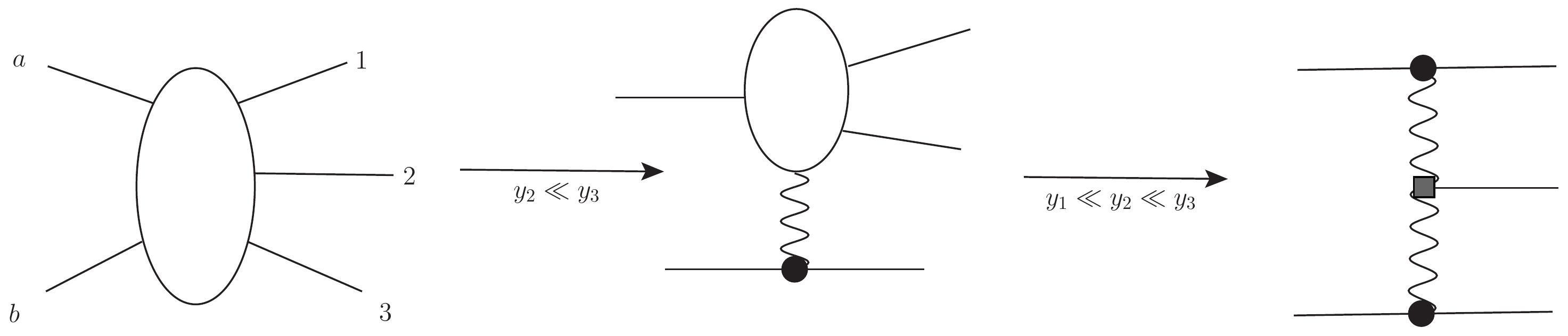}
  \caption{The factorisation property still applies whenever there is a strong
    rapidity order imposed, even if not the full MRK limit.}
  \label{fig:2then3}
\end{figure}

In section~\ref{sec:FKLalmpl}, we described the factorisation of amplitudes in
the MRK limit, illustrated in Fig.~\ref{fig:tDominance1}.  In general, the
factorisation property of the amplitudes is actually stronger still: it holds \emph{whenever}
there is any large rapidity separation between any groups of particles, as illustrated in Fig.~\ref{fig:2then3}.
If the only requirement on the ordered rapidities is a large difference
between $y_{n-1}$ and $y_n$ ($y_{n-1}\ll y_n$, i.e.~$\forall
\{i,j,k,l\}\in\{1,\ldots,n-1\}, i\not=j, k\not=l: s_{ij}\sim s_{kl},
s_{in}\sim s_{jn}$), but
no further requirement on a large difference between any of
$y_1,y_2,...,y_n$, then the leading power of
the amplitude can still be written as a
contraction of the quark current with a sub-amplitude, depending only on the
reduced set of momenta $p_a,p_1,\ldots,p_{n-1}$:
\begin{align}
  \label{eq:fact3}
  \mathcal{M}\sim\mathcal{M}_\text{sub}^\mu(p_a,p_1,...,p_{n-1})\frac 1 {t_{n-1}} j_\mu(p_b,p_n)
\end{align}
In the stricter MRK limit of large rapidity differences between all
$1,\ldots, n-1$, the sub-amplitude
$\mathcal{M}_\text{sub}^\mu(p_a,p_1,...,p_{n-1})$ would factorise further
into another quark current, Lipatov vertices, and $t$-channel
propagators, as indicated on the right-hand
side of Fig.~\ref{fig:2then3}. Clearly, the more complicated sub-amplitude
$\mathcal{M}_\text{sub}^\mu(p_a,p_1,...,p_{n-1})$ includes the
leading-power behaviour in the full MRK limit, and hence one can recover this
fully factorised form starting from Eq.~\eqref{eq:fact3}.

In order to extend the current-based formalism of \HEJ to include
the first next-to-leading logarithmic corrections, we will therefore need to
extract a form $j_\mu^{\rm uno}(p_1,p_g,p_a)$, which takes the place of
$\mathcal{M}_\text{sub}^\mu(p_a,p_1,...,p_{n-1})$ in
Eq.~\eqref{eq:fact3}. Here, we have (without loss of generality) considered
the case of $y_1\sim y_g\ll y_2$. We then
seek an expression for a quantity $j_\mu^{\rm uno}(p_1,p_g,p_a)$ such that
the equation
\begin{align}
  \label{eq:Aunohej}
  \mathcal{M}_{{\rm tree}\ qQ\to gqQ}^{\rm HEJ} =-g_s^3 T^d_{2b}\ \frac{j^{{\rm  uno}\ cd}_\mu(p_1,p_g,p_a)  j^\mu(p_2,p_b)}{t_{b2}}
\end{align}
will contain all the leading-power behaviour of the full tree-level
amplitude. We will give the new current a superscript uno, since it will be used
only for the calculation of unordered emissions, $y_g<y_1$.  Emissions
in-between the quarks, $y_1<y_g$, are already accounted for using the real and virtual
corrections described in the previous section and so we do not apply this
correction there.  The new current now carries colour indices $cd$, where $c$ is
the colour of the emitted gluon, and $d$ is the colour of the gluon exchanged in
the $t$-channel.
\begin{figure}[btp]
  \centering
  \scalebox{0.6}{\includegraphics{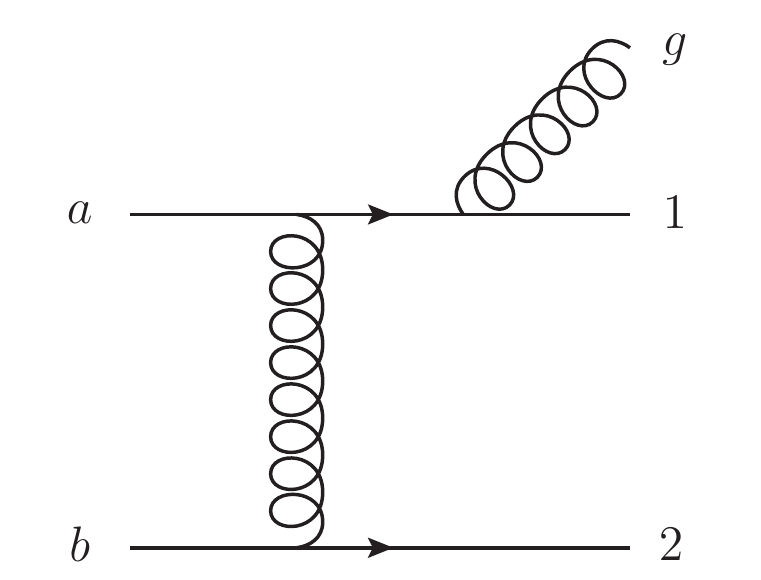}}
  \caption{One of the diagrams which contribute to the process $qQ\to gqQ$,
    illustrating the labelling convention used throughout this section.  We will
    consider the case where $y_g \sim y_1 \ll y_2$. $a$ ($b$) is the incoming
    quark in the backward (forward) direction respectively, which we here assume
    to be of different flavours.}
  \label{fig:tree}
\end{figure}
One of the five Feynman diagrams which contribute to this process is shown in
Fig.~\ref{fig:tree}, which also defines the momentum labelling.  The exact
tree-level expression for the sum of \emph{all five} diagrams is:
\begin{align}
  \label{eq:fulltree}
  \begin{split}
    \mathcal{M}_{\rm tree} =& (ig_s)^3\ T^c_{1i}T^d_{ia}T^d_{2b}\ \varepsilon_{g\nu}
    \frac{\langle 1|\nu|g\rangle \langle g|\mu|a\rangle +
      2p_1^\nu \langle 1|\mu|a \rangle}{s_{1g}t_{b2}} \langle
    2|\mu|b\rangle \\
    +&(ig_s)^3\ T^d_{1i}T^c_{ia}T^d_{2b}\ \varepsilon_{g\nu}
    \frac{2p_a^\nu \langle 1|\mu|a \rangle -
      \langle 1|\mu|g\rangle \langle g|\nu|a\rangle
    }{t_{ag}t_{b2}} \langle 2|\mu|b\rangle \\
    +&(ig_s)^3\ T^c_{2i}T^d_{ib}T^d_{1a}\ \varepsilon_{g\nu} \frac{\langle
      2|\nu|g\rangle \langle g|\mu|b\rangle +2p_2^\nu
      \langle 2|\mu|b \rangle }{s_{2g}t_{a1}} \langle 1|\mu|a\rangle \\
    +&(ig_s)^3\ T^d_{2i}T^c_{ib}T^d_{1a}\ \varepsilon_{g\nu} \frac{
      2p_b^\nu  \langle 2|\mu|b \rangle -\langle
      2|\mu|g\rangle \langle g|\nu|b\rangle}{t_{bg}t_{a1}} \langle 1|\mu|a\rangle \\
    -&g_s^3\ f^{dec} T^d_{1a}T^e_{2b}\ \varepsilon_{g\nu}
    \frac{\langle 1|\rho|a \rangle \langle 2|\mu|b\rangle}{t_{a1}t_{b2}} \left(
      2p_g^\mu g^{\nu\rho}-2p_g^\rho g^{\mu\nu}-(q_1+q_2)^\nu
        g^{\mu\rho} \right),
      \end{split}
\end{align}
where we have used the shorthands
\begin{align}
  \label{eq:defs}
  \langle i | \mu | j \rangle = \bar{u}^-(p_i) \gamma^\mu u^-(p_j), \qquad s_{ij}
  = (p_i+p_j)^2, \qquad t_{ij}=(p_i-p_j)^2,
\end{align}
and the $T^c_{ij}$ are colour matrices.  The external gluon carries the
colour index $c$, and $\{a,b,1,2\}$ in the subscript indicates the colour index
of the relevant external quark.  Repeated indices are summed over.  Of course,
this expression can be considerably simplifed by contracting the Lorentz indices
and re-arranging.  However, we choose not to do so here
as the extended form is particularly convenient for the discussion below.

In the MRK limit, where $y_1\ll y_g \ll y_2$, one term in each of the
first four lines of Eq.~\eqref{eq:fulltree} becomes sub-dominant
(the term with numerator dependence on $p_g$).  This is most easily seen by performing the
Lorentz contractions.  In this case, the expression becomes~\cite{Andersen:2009nu}
\begin{align}
  \label{eq:fklcase}
  \begin{split}
    \mathcal{M}_{\rm tree}^{FKL} &= -g_s^3 f^{dec}T^d_{1a}T^e_{2b} \
    \varepsilon_{g\nu}\ \frac{\langle 1| \mu | a \rangle \langle 2 | \mu | b
      \rangle}{t_{a1}t_{b2}}\ V_L^\nu \\
    &=-g_s^3
    f^{dec}T^d_{1a}T^e_{2b} \   \varepsilon_{g\nu}\
    \frac{j^\mu(p_1,p_a)j_\mu(p_2,p_b)}{t_{a1}t_{b2}}\ V_L^\nu,
    \end{split}
\end{align}
analogously to Eq.~\eqref{eq:HEJqgQ}, and illustrated by the right-hand diagram of
Fig.~\ref{fig:2then3}.  However, for the case at hand, we no longer want to
assume a strong
ordering between $y_g$ and $y_1$.  In this case, the only sub-dominant terms are
the $p_g$-dependent terms in the numerator in lines 3 and 4 of Eq.~\eqref{eq:fulltree}.  We
observe that by discarding these two terms, every other term immediately appears
in the form $j^\mu(p_2,p_b) \times X_\mu$.  The sum of these $X_\mu$-pieces will
therefore become our unordered current.

We now turn our attention to colour factors. The `$b$-$2$' end of the chain
has to
have a single colour matrix of the form $T^d_{2b}$, for a dummy index $d$, in
order to be consistent with the factorised
picture; this is to ensure it can be contracted with either a normal quark- or
gluon-current, or the unordered two-particle $j_\mu^{\rm uno}(p_1,p_g,p_a)$.
This is already the case
for the first, second and fifth lines of Eq.~\eqref{eq:fulltree}.  The MRK limit implies
$p_b\simeq p_2 = p_+$ and the dominant terms in the 3rd and 4th lines are:
\begin{align}
  \label{eq:uno34cont}
  \begin{split}
  & -ig_s^3\ \langle 1|\mu|a \rangle \langle
  2|\mu|b\rangle\ \varepsilon_{g\nu}  \left(
    \frac{2p_2^\nu}{t_{a1}s_{2g}}T^c_{2i}T^d_{ib}T^d_{1a} +
      \frac{2p_b^\nu}{t_{bg}t_{a1}}T^d_{2i}T^c_{ib}T^d_{1a} \right)\\
    \simeq & -ig_s^3\ \langle 1|\mu|a \rangle \langle
  2|\mu|b\rangle\ \varepsilon_{1\nu}\ \frac1{t_{a1}} \frac{p_+^\nu}{p_+\cdot
    p_g}\ T^d_{1a} \left( T^c_{2i}T^d_{ib} - T^d_{2i}T^c_{ib}\right) \\
  = & \quad \;\, g_s^3\ \langle 1|\mu|a \rangle \langle 2|\mu|b\rangle\
    \varepsilon_{1\nu} \  \frac1{t_{a1}} \frac{p_+^\nu}{p_+\cdot
    p_g}\ f^{cde} T^d_{1a}T^e_{2b}\\
  \simeq & \     g_s^3\ \langle 1|\mu|a \rangle \langle 2|\mu|b\rangle\
    \varepsilon_{g\nu}\ \ f^{cde} T^d_{1a}T^e_{2b}\ \ \frac1{2t_{a1}} \left(
      \frac{p_b^\nu}{(p_b\cdot p_g)} + \frac{p_2^\nu}{(p_2\cdot p_g)} \right).
  \end{split}
\end{align}
We have chosen to restore the symmetry of $p_b$ and $p_2$ in the last line (as
we do in $V_L^\nu$ above).  The MRK limit is of course independent of
such choices.  We then arrive at
the following expression for the amplitude for quark-quark scattering with an
additional unordered gluon emission:
\begin{align}
  \label{eq:Afull}
  \mathcal{M}_{{\rm tree}\ qQ\to gqQ}^{HEJ} =  -g_s^3 \frac{\langle 2|\mu|b\rangle
    \varepsilon_{1\nu}}{t_{b2}} T_{2b}^d \left( i T_{1i}^{c}T_{ia}^d\
    U_1^{\mu\nu} + i T_{1i}^{d}T_{ia}^c\ U_2^{\mu\nu} +
    f^{ecd}T_{1a}^e\ L^{\mu\nu} \right).
\end{align}
The tensors $U_1^{\mu\nu}$, $U_2^{\mu\nu}$ and $L^{\mu\nu}$ may then be read off
from Eqs.~\eqref{eq:fulltree} and \eqref{eq:uno34cont} as
\begin{align}
  \label{eq:LU1U2}
  \begin{split}
  U_1^{\mu\nu} &= \frac1{s_{1g}} \left( j_{1g}^\nu j_{ga}^\mu + 2 p_1^\nu
      j_{1a}^\mu \right) \qquad  \qquad U_2^{\mu\nu} = \frac1{t_{ag}} \left( 2
      j_{1a}^\mu p_a^\nu - j_{1g}^\mu  j_{ga}^\nu \right) \\
    L^{\mu\nu} &= \frac1{t_{a1}} \left(-2p_g^\mu j_{1a}^\nu+2p_g.j_{1a}
      g^{\mu\nu} + (q_1+q_2)^\nu j_{1a}^\mu + \frac{t_{b2}}{2} j_{1a}^\mu \left(
      \frac{p_2^\nu}{p_g.p_2} + \frac{p_b^\nu}{p_g.p_b} \right) \right) .
  \end{split}
\end{align}
The three colour factors in Eq.~\eqref{eq:Afull} are not independent and can be
combined to give
\begin{align}
  \label{eq:Afull2}
  \mathcal{A}_{qQ\to gqQ} =  -ig_s^3\ \frac{\langle 2|\mu|b\rangle
    \varepsilon_{g\nu}}{t_{b2}}\ T_{2b}^d \left( T_{1i}^{c}T_{ia}^d\
    \left(U_1^{\mu\nu}-L^{\mu\nu} \right) + T_{1i}^{d}T_{ia}^c\ \left(U_2^{\mu\nu} +
    L^{\mu\nu} \right)\right).
\end{align}
By comparison to Eq.~\eqref{eq:Aunohej}, we extract
\begin{align}
  \label{eq:juno}
  j^{{\rm uno}\; \mu\ cd}(p_1,p_g,p_a) = i \varepsilon_{g\nu} \left(  T_{1i}^{c}T_{ia}^d\
    \left(U_1^{\mu\nu}-L^{\mu\nu} \right) + T_{1i}^{d}T_{ia}^c\ \left(U_2^{\mu\nu} +
    L^{\mu\nu} \right) \right).
\end{align}
Gauge-invariance of this new current is satisfied throughout phase space; it is
easily checked that replacing $\varepsilon_{g\nu}$ with $p_{g\nu}$ gives
identically zero.  One can also check that the use of
Eq.~\eqref{eq:juno} in the MRK limit will result in the BFKL NLO
impact factor derived in Ref.\cite{DelDuca:1999ha}.

After some colour algebra, the final summed and averaged amplitude for $  q(p_a)\ Q(p_b)  \to g(p_g)\
q(p_1)\ Q(p_2)$ is then given by
\begin{align}
  \label{eq:coloursquare}
    \begin{split}
    \left| \overline{\mathcal{M}_{{\rm tree}\ qQ\to gqQ}^{HEJ}} \right|^2
    =& -\frac{g_s^6}{16 t_{b2}^2}\ \sum_{h_a,h_1,h_b,h_2}\ \bigg[ C_F \big( 2
      \mathrm{Re}\big( [(L^{\mu\nu}-U_1^{\mu\nu})\cdot j_{2b\, \mu}]\ [
          (L^{\rho}_{\;\,\nu}+U_{2\;\, \nu}^{\;\rho})\cdot j_{2b\, \rho} ]^* \big)
      \big)  \\
    & \hspace{4cm} + 2\frac{C_F^2}{C_A} \left| (U_1^{\mu\nu}+U_2^{\mu\nu}) \cdot
        j_{2b_\mu} \right|^2 \bigg] \\
      \equiv& -\frac{g_s^6}{16 t_{b2}^2} C_F \left\|S_{f_1 f_2\to gf_1 f_2}^{\rm uno}\right\|^2 ,
  \end{split}
\end{align}
where the sum runs over the helicities of the four quarks and the square in the
second line indicates contraction over the $\nu$-index.  The final line defines
the function $S^{\rm uno}$ and is analogous to the rapidity-ordered case
described in Eq.~\eqref{eq:Sfunc}.

We now follow the formalism for additional
ordered emissions derived in section~\ref{sec:constr-hej-ampl} to arrive at the
following \HEJ matrix element in $4-2\epsilon$ dimensions for $f_1f_2\to gf_1\cdot g\cdot f_2$, where $f_1$
now must be a quark (cf. Eq.~\eqref{eq:Mtvepsilon}:
\begin{align}
  \label{eq:Mtvepsilonuno}
  \begin{split}
    \overline{\left|\mathcal{M}^{\rm HEJ}_{\varepsilon\ q f_2 \to g q\cdot
          g\cdot f_2}\right|}^2 = \ &\frac 1 {4\
       (\Nc^2-1)}\ g_s^2\ \left\|S^{\rm uno}_{q f_2\to gq f_2}\right\|^2\\
     &\cdot\ \left(g_s^2\ C_F\ \frac 1 {t_1}\right) \cdot\ \left(g_s^2\ K_{f_2}\ \frac 1
       {t_{n-1}}\right)\\
     & \cdot \prod_{i=1}^{n-2} \left( \frac{-g_s^2 C_A}{t_it_{i+1}}\
       V^\mu(q_i,q_{i+1})V_\mu(q_i,q_{i+1}) \right)\\
     & \cdot \prod_{j=1}^{n-1} \exp\left[2 \hat
         \alpha (q_j)(y_{j+1}-y_j)\right].
  \end{split}
\end{align}

We are now ready to build the regularised
matrix-element with the appropriate all-order corrections in the manner of
Section~\ref{sec:all-order-reg}.  Such corrections can be added for gluon
emissions in the phase space delimited by the rapidity(ies) of the final
state quark(s). This region will be denoted the \emph{all-order summation
  region}. The momenta of the quarks are still required
to be hard, and form part of the two jets extremal (but one) in rapidity. One
current
includes the \emph{unordered} gluon emission, which allows for a single gluon
to be
emitted outside this all-order summation region. Such an \emph{unordered}
gluon is required to enter a separate hard jet from that of the quark, since the
associated collinear singularity is otherwise unregulated: it would cancel with the singularity associated
with the one-loop correction to the quark production, which form part of the
full NLL corrections, which are not yet included in the formalism. However, in
the all-order summation
  region, the infrared singularities cancel as discussed in the
previous section.  We therefore find
\begin{align}
  \label{eq:finaluno}
  \begin{split}
        \overline{\left|\mathcal{M}^{\rm HEJ}_{q f_2 \to gq\cdot
          g\cdot f_2}\right|}^2 = \ &\frac 1 {4\
       (\Nc^2-1)}\ g_s^2\ \left\|S_{q f_2\to gq f_2}^{\rm uno}\right\|^2\\
     &\cdot\ \left(g_s^2C_F\ \frac 1 {t_1}\right) \cdot\ \left(g_s^2\ K_{f_2}\ \frac 1
       {t_{n-1}}\right)\\
     & \cdot \prod_{i=1}^{n-2} \left( \frac{-g_s^2 C_A}{t_it_{i+1}}\
       V^\mu(q_i,q_{i+1})V_\mu(q_i,q_{i+1}) \right)\\
     & \cdot \prod_{j=1}^{n-1} \exp\left[\omega^0(q_{j\perp})(y_{j+1}-y_j)\right].
  \end{split}
\end{align}
There is a corresponding equation for the gluon emitted instead forward of
the most forward quark, $f_2$.

The currents for the unordered emission therefore enter the
calculation of the all-order, leading corrections to the Born-level three-jet
processes with a gluon jet of larger absolute rapidity than that of the
respective quark
jet.  These three-jet events form part of the
sub-leading logarithmic corrections to inclusive dijet
production.

\subsection{High Energy Corrections to Higgs Boson Production with Jets}
\label{sec:higgs}

In order to develop the formalism for unordered emissions, we have so far worked
with amplitudes purely within QCD.  However, it is straight-forward to extend
the \HEJ description of jet processes to include the production also of a Higgs boson.  In
this paper in particular we are concerned with the production of a Higgs boson
with at least two jets so in this section we briefly review the existing \HEJ
description of this process first developed in \cite{Andersen:2009nu}. This
makes use of the infinite top-mass limit, but this limit not only commutes with the
high-energy limit\cite{DelDuca:2003ba}, but results can be obtained for the
high-energy limit without applying the infinite top-mass limit. We leave such
investigations for a future study, but will here add the possibility of
unordered gluon emissions derived in the previous subsection to the
amplitudes derived in Ref.~\cite{Andersen:2009nu} for Higgs boson production
in association with jets. This will result in different formulae for the
high-energy approximations to the
scattering amplitude for the various rapidity-orderings of particles. We discuss
each here in turn.

\subsubsection{Higgs Boson with Rapidity Between that of Hard Jets}

\begin{figure}[btp]
  \hspace{5.5cm}
  \scalebox{0.6}{\includegraphics{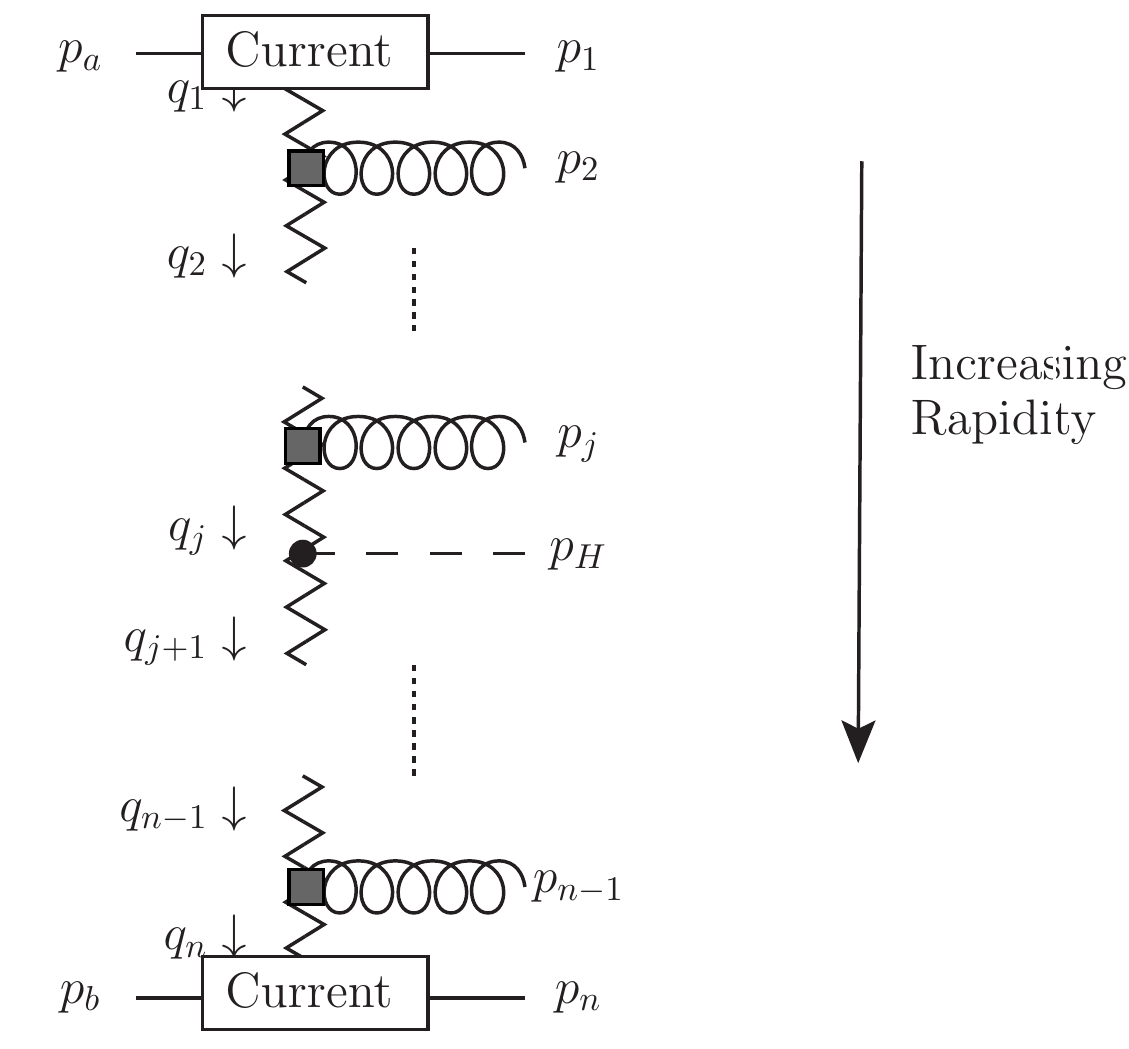}}
  \caption{The analytic structure of the base tree-level scattering amplitude
    for $qQ\to HqQ+(n-2)g$ in \emph{High Energy Jets}.  In this case, the Higgs
    boson is emitted between gluons $j$ and $j+1$ in rapidity.}
  \label{fig:HEJstructureH}
\end{figure}

We begin with the \HEJ approximation to the tree-level amplitude for
$qQ\to HqQ+(n-2)g$.  From the discussion in the previous section, the dominant
momentum configurations in the MRK limit are those where the gluons are all
emitted between the two quarks in rapidity.  We will exploit the factorisation
of the amplitudes discussed in the previous subsection (and which still holds
when a Higgs boson is included) to describe an amplitude as the contraction of
two currents over a $ggH$-vertex, multiplied by a product of vertices for each
additional gluon emission.  This is illustrated schematically in
Fig.~\ref{fig:HEJstructureH} for the case where the Higgs boson is also between
the outer quark jets in rapidity, between gluons $j$ and $j+1$.
This figure also gives the definitions of the momenta
$p_i$ and $q_j$ used in this section.  The vertices for additional gluon
emissions depend on the momenta of that emission
and the momenta of the parton of maximum and minimum rapidity, but \emph{not} on the momenta of any other emissions or
the Higgs boson.  The amplitude can then be
written~\cite{Andersen:2008ue,Andersen:2008gc,Andersen:2009nu}
\begin{align}
  \label{eq:MainHEJ}
  \begin{split}
    \mathcal{M}^{\rm HEJ}_{\rm tree\ qQ\to Hq\cdot g\cdot Q} &= -g_s^2 T^{a_1}_{i_1 i_a}T^{a_{n-1}}_{i_n i_b}\
    \frac{\mathcal{S}_{qQ\to qQH}(p_1,p_n,p_a,p_b,q_j,q_{j+1})}{\sqrt{q_1^2 q_j^2
        q_{j+1}^2 q_{n}^2}}\\ & \hspace{1cm} \times \prod_{k=2}^{j} ig_s f^{a_{k-1}b_ka_k}\
    \frac{\varepsilon_{\nu_k}(p_k) V^{\nu_k}_L(q_{k-1},q_{k})}{\sqrt{q_{k-1}^2
        q_k^2}} \\ & \hspace{1cm} \times \prod_{k=j+1}^{n-1} ig_s f^{a_{k-1}b_ka_k}\
    \frac{\varepsilon_{\nu_k}(p_k) V^{\nu_k}_L(q_{k},q_{k+1})}{\sqrt{q_{k}^2
        q_{k+1}^2}}.
  \end{split}
\end{align}
Here $i_j$ and $b_k$ are the colour indices of the relevant quark and the $k$th
external gluon and the $a_j$ indices are summed over.  The expression
$\mathcal{S}_{qQ\to qQH}(p_1,p_n,p_a,p_b,q_j,q_{j+1})$ represents the contraction of
the two end currents with the $ggH$-vertex in the limit of infinite top mass:
\begin{align}
  \label{eq:Sbit}
  \begin{split}
    &\mathcal{S}_{qQ\to qQH}(p_1,p_n,p_a,p_b,q_j,q_{j+1}) =   j_\mu(p_1,p_a) j_\nu(p_n,p_b)
    V_H^{\mu,\nu}(q_j,q_{j+1}), \\ {\rm where}\
    V_H^{\mu,\nu}&(q_j,q_{j+1}) = \left( \frac{\alpha_s}{3\pi v}
    \right)  \left( g^{\mu \nu} q_j.q_{j+1}-q_j^\nu q_{j+1}^\mu \right), \qquad
    j_{\mu}(p_o,p_i) = \bar{u}(p_o) \gamma_\mu u(p_i).
  \end{split}
\end{align}
The two products which represent the gluon emissions are separated at the point
where the Higgs boson occurs in rapidity in order to correctly assign the
relevant $q_i$.  Our description of processes with incoming gluons follows the
same prescription as that for pure QCD processes.  The relevant quark current(s)
in Eq.~\eqref{eq:Sbit} have the same form multiplied by a scalar factor.

We now wish to include the emission of an unordered gluon in the description of
these Higgs boson processes.  In section~\ref{sec:UnorderedandHiggsB}, the only
modification to the ordered process was in the spinor factor.  Comparing
Eq.~\eqref{eq:finitefinal} and \eqref{eq:finaluno},
\begin{align}
  \label{eq:unoprescrip}
   \left\|S_{f_1 f_2\to f_1 f_2}\right\|^2\ \rightarrow\ \left\|S_{q f_2\to gq f_2}^{\rm uno}\right\|^2.
\end{align}
The factorisation property of the amplitudes implies that we may apply the same
prescription here.  The infrared divergences are regulated here in the same way as in the pure QCD
amplitudes.  The virtual corrections are still given by the Lipatov ansatz with
the prescription given in Eq.~\eqref{eq:LipatovAnsatz2}, which leads to the
following
infrared finite amplitude for a Higgs boson produced between particles $j$
and $j+1$ in rapidity (c.f.~Eq.~\eqref{eq:finitefinal}):
\begin{align}
  \label{eq:HEJHunoreg}
  \begin{split}
    \overline{\left|\mathcal{M}^{\rm HEJ}_{\ f_1 f_2 \to Hgf_1\cdot g\cdot
  f_2}\right|}^2 =& \frac1{4(N_C^2-1)} \left\|\mathcal{S}_{qQ\to
                 qQH}^{\rm uno}(p_1,p_g,p_n,p_a,p_b,q_j,q_{j+1}) \right\|^2 \\
  & \cdot \left(g_s^2 K_{f_1} \frac{1}{t_1} \right) \cdot \left( g_s^2 K_{f_2}
    \frac{1}{t_{n}} \right)
  \\ & \cdot \prod_{k=2}^{j} \left(\frac{-g_s^2 C_A}{t_{k_1} t_k}
       V^{\nu_k}(q_{k-1},q_{k}) V_{\nu_k}(q_{k-1},q_{k}) \right)\\ &
        \cdot\prod_{k=j+1}^{n-1} \left(\frac{-g_s^2 C_A}{t_k t_{k+1}}
           V^{\nu_k}(q_{k},q_{k+1}) V^{\nu_k}(q_{k},q_{k+1}) \right)\\
     & \cdot \prod_{i=1}^{j-1}
     \exp\left[\omega^0(q_{i\perp})(y_{i+1}-y_{i})\right] \cdot \prod_{i=j+2}^n
     \exp\left[\omega^0(q_{i\perp})(y_{i}-y_{i-1})\right] \\ &
     \cdot \exp\left[\omega^0(q_{j\perp})(y_H-y_j) \right] \cdot\exp \left[
       \omega^0(q_{j+1\perp}) (y_{j+1}-y_{H})\right],
   \end{split}
\end{align}
where now
\begin{align}
  \label{eq:jh}
  \mathcal{S}_{qQH}^{\rm  uno}(p_1,p_g,p_n,p_a,p_b,q_i,q_{i+1}) = j^{{\rm
  uno}}_\mu(p_1,p_g,p_a) j_\nu(p_n,p_b) V_{\mu\nu}^H(q_{i-1},q_i),
\end{align}
for a gluon emission most backward in rapidity of all coloured particles.  The
modified current $j^{\rm uno}_\mu(p_1,p_g,p_a)$ is exactly the one given in
Eq.~\eqref{eq:juno}.  If, instead, the unordered emission is forward in rapidity
of all coloured particles, the current pair $j_\mu(p_1,p_g,p_a) j_\nu(p_n,p_b)$ becomes
$j_\mu(p_1,p_a)j_\nu^{\rm uno}(p_n,p_g,p_b)$.

\subsubsection{Higgs Boson with Rapidity Outside that of Hard Jets}
\begin{figure}[btp]
  \centering
  \scalebox{.65}{\includegraphics{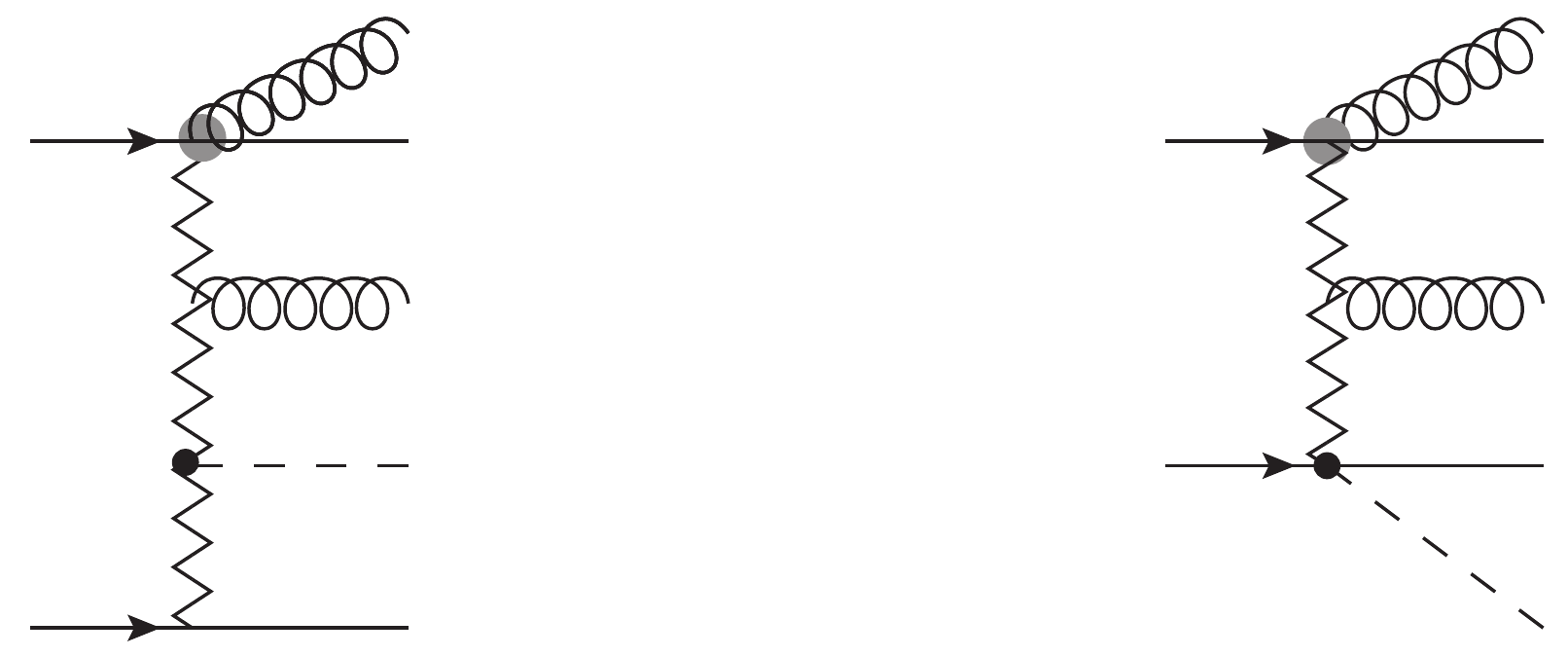}}\\
  (a) \hspace{7cm} (b) \phantom{a}
  \caption{Sample diagrams for a Higgs+3j process including an unordered
    gluon emission: (a) the Higgs is emitted in between the jet system in
    rapidity and (b) the Higgs is emitted outside via an adapted
    current. }
  \label{fig:hdiags}
\end{figure}

The remaining case to consider is the case where the Higgs boson is produced
outside of the coloured particles in rapidity, see e.g.~Fig.~\ref{fig:hdiags}(b).  Motivated by the amplitude for
$qQ\to qQH$ (where of course there is just one $t$-channel amplitude
applied for all momentum configurations), we will apply the leading
factorised amplitude, which is the configuration where the Higgs-boson vertex
is the first (last) vertex in the $t$-channel chain if the rapidity of the
Higgs boson is less (greater) than the rapidity of
the quarks. Therefore, in practice, the two configurations in
Figs.~\ref{fig:hdiags}(a) and (b) have the same description.
When the Higgs boson is produced outside of the coloured particles in rapidity, we will only
include unordered gluon emissions where these occur at the opposite
end of the chain to the Higgs boson (all possibilities could in principle be
included, but these are perturbative corrections to already suppressed
configurations).  The matrix element squared in this case is then given by
Eq.~\eqref{eq:HEJHunoreg} with $j=1$ if the Higgs is most backward in rapidity.
It has $j=n-1$ if the Higgs is most forward in rapidity and
$j_\mu(p_1,p_a)j_\nu^{\rm uno}(p_n,p_g,p_b)$ in place of $j^{\rm
  uno}_\mu(p_1,p_g,p_a)j_\nu(p_n,p_b)$ in Eq.~\eqref{eq:jh}.  For example, the
all-order equation corresponding to an unordered gluon emission as the most
backward outgoing particle and a Higgs boson as the most forward outgoing
particle (the $n$-emission equivalent of Fig.~\ref{fig:hdiags}(b)) is given by:
\begin{align}
  \label{eq:HEJHreg}
  \begin{split}
    \overline{\left|\mathcal{M}^{\rm HEJ}_{\ f_1 f_2 \to gf_1\cdot g\cdot
  f_2H}\right|}^2 =& \frac1{4(N_C^2-1)} \left\|\mathcal{S}_{qQ\to
                 qQH}^{\rm uno}(p_1,p_g,p_n,p_a,p_b,q_{n-1},q_{n}) \right\|^2 \\
  & \cdot \left(g_s^2 K_{f_1} \frac{1}{t_1} \right) \cdot \left( g_s^2 K_{f_2}
    \frac{1}{t_{n}} \right)
  \\ & \cdot \prod_{k=2}^{n-1} \left(\frac{-g_s^2 C_A}{t_{k_1} t_k}
       V^{\nu_k}(q_{k-1},q_{k}) V_{\nu_k}(q_{k-1},q_{k}) \right)\\
     & \cdot \prod_{i=1}^{n-2}
     \exp\left[\omega^0(q_{i\perp})(y_{i+1}-y_{i})\right]  \\ &
     \cdot \exp\left[\omega^0(q_{n-1\perp})(y_H-y_{n-1}) \right] \cdot\exp \left[
       \omega^0(q_{n\perp}) (y_{n}-y_{H})\right],
   \end{split}
\end{align}
where $q_n=q_{n-1}-p_H$ in clear analogy to Eq.~\eqref{eq:HEJHunoreg} with $j=n-1$.

If the flavour $f_1$ (or $f_2$) is a gluon, the amplitude for the emission of
a Higgs boson with more extremal rapidity than the gluons receives
contributions also from top box diagrams, not just the triangle diagrams
implemented in the formalism of the currents. We will use the amplitude
derived for the strict MRK limit in Ref.~\cite{DelDuca:2003ba} for these
kinematic configurations. Their contribution is suppressed for large rapidity
spans, but they are included for completeness.

\subsubsection{Perturbative Validation of the Approximations}
\begin{figure}[btp]
  \centering
  \scalebox{.45}{\includegraphics{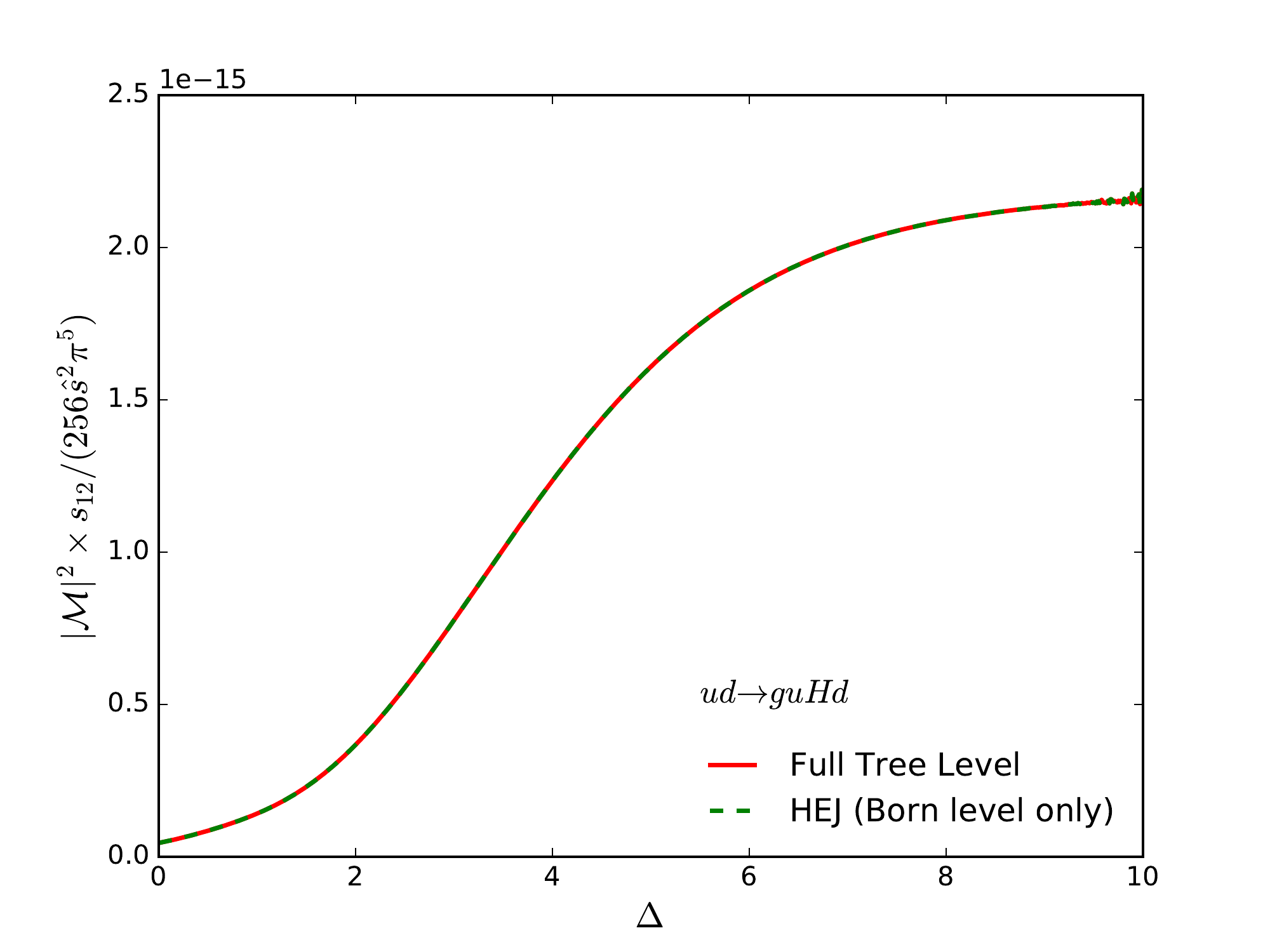}}
  \caption{A slice through phase space of
    $\left| \overline{\cal M} \right|^2\times s_{12}/(256 \pi^5 \hat{s}^2)$ for
    the process $ud\to guHd$.  The rapidities of the final-state particles are
    chosen to be $y_g=-\Delta$, $y_u = -\Delta/3$, $y_H=\Delta/3$ and $y_d=\Delta$.  The new
    \HEJ description of this unordered configuraton  (``HEJ (Born level only)'', green dashed) shows close
    agreement to the full tree level result (red solid) throughout the range.
  }
  \label{fig:uno1}
\end{figure}

We now test the quality of the approximation by comparing this result with the
full matrix element result taken from Madgraph~\cite{Alwall:2014hca}
order-by-order in the strong coupling.  In Fig.~\ref{fig:uno1}, we compare the
matrix element squared for $ud\to guHd$ in a slice through phase space where the
rapidities are chosen to be: $y_g=-\Delta$, $y_u = -\Delta/3$, $y_H=\Delta/3$
and $y_d=\Delta$ for $\Delta\in\{0,10\}$.  The matrix-element squared has been
multiplied by one power of the $gu$ invariant mass, $s_{12}$, to counteract the
suppression discussed in Section~\ref{sec:leadingcontrlarge}.  We observe very
close agreement throughout the rapidity range between the full MadGraph result
(red, solid) and the unordered \HEJ formalism (green, dashed).

\begin{figure}[btp]
  \centering
  \scalebox{.45}{\includegraphics{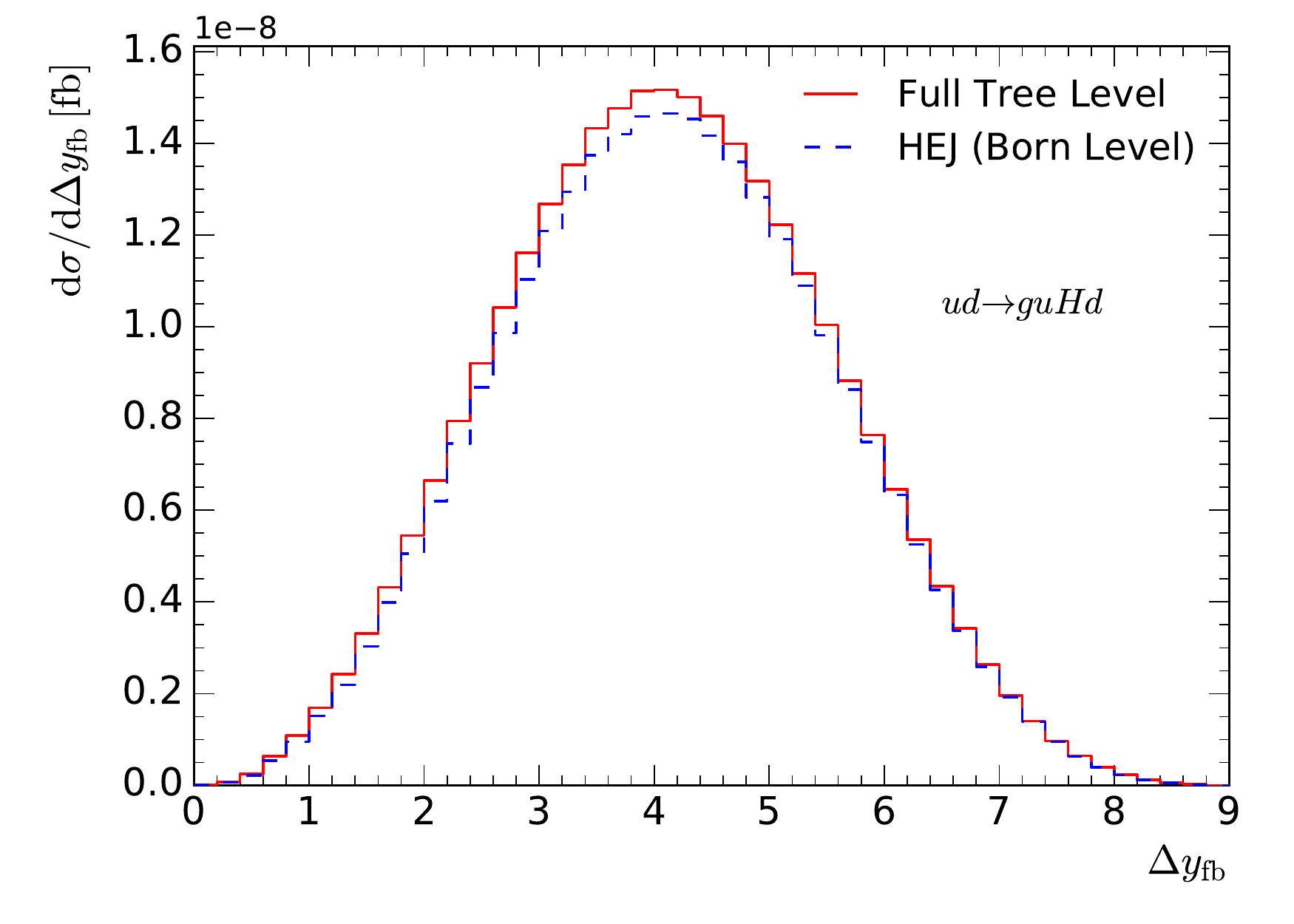}}
  \caption{The rapidity span distribution for $ud\to guHd$ events after
    integration over phase space in the region where the gluon is the most
    backward final state particle: $y_g<y_u<y_H<y_d$ and
    $\Delta y_{fb}=y_d-y_g$.  The approximation (``HEJ Born level'', blue
    dashed) gives an extremely good description of the full tree-level matrix
    element (red, solid). }
  \label{fig:intdydif}
\end{figure}

In Fig.~\ref{fig:intdydif} we show the distribution of the rapidity difference
between the most forward and backward hard jet again for the process
$ud\to guHd$, integrated over the region of phase space where $y_g<y_u<y_H<y_Q$.
We apply modest jet cuts, requiring the partons to form 3 jets with $p_T>30$~GeV
and $|y|<4.4$.  Here, we consider on-shell Higgs-boson production and require
$|y_H|<2.37$.  It is clear that the description from the new impact factor
describing unordered emissions tracks the result from the full matrix element
extremely closely throughout the full range of $\Delta y_{fb}$, becoming
indistinguishable at large $\Delta y_{fb}$.


The square of the matrix elements can be trivially extended to include, for example, the
diphoton decay of the Higgs boson by simply multiplying the square of the matrix
elements of either the FKL-ordered (Eq.~\eqref{eq:finitefinal}) or unordered
configuration (Eq.~\eqref{eq:HEJHunoreg}) by the branching ratio, $BR(H\to
\gamma \gamma)$, and generating the decay products isotropically.  This is
available as an option in the code.


\subsection{Matching and Merging of Fixed Order Samples and Final Results}
\label{sec:match-merg-fixed}

Using the formalism outlined in the previous sections, the all-order summed
contribution to the FKL-ordered plus first unordered cross section for the production of a Higgs
boson which decays to two photons in association with at least two jets can now
be found as
\begin{align}
  \begin{split}
  \label{eq:sigma2jsum}
    \sigma_{H+2j}^\mathrm{resum}=&\sum_{f_a, f_b}\ \sum_{n=2}^\infty\ \left(\int_{p_{1\perp}=p_{\perp\mathrm{ext,min}}}^{\infty}
      \frac{\mathrm{d}^2\mathbf{p}_{1\perp}}{(2\pi)^3}\
      \int_{y_\mathrm{min}}^{y_\mathrm{max}} \frac{\mathrm{d} y_1}{2}
    \right)\ \left(\int_{p_{n\perp}=p_{\perp\mathrm{ext,min}}}^{\infty} \frac{\mathrm{d}^2\mathbf{p}_{n\perp}}{(2\pi)^3}\
      \int_{y_{n-1}}^{y_\mathrm{max}} \frac{\mathrm{d} y_n}{2}\right) \\
&     \times \prod_{i=2}^{n-1}\left(\int_{p_{i\perp}=\kappa}^{\infty}
      \frac{\mathrm{d}^2\mathbf{p}_{i\perp}}{(2\pi)^3}\
      \int_{y_{i-1}}^{y_\mathrm{max}} \frac{\mathrm{d} y_i}{2}
    \right)\
    \int \frac{\mathrm{d}^3p_{\gamma_1}}{(2\pi)^3\ 2 E_{\gamma_1}}\
    \
    \int \frac{\mathrm{d}^3p_{\gamma_2}}{(2\pi)^3\ 2 E_{\gamma_2}}\
    \\
    &\times \frac{\overline{|\mathcal{M}_{\mathrm{HEJ}}^{\mathrm{reg}}(\{ p_i,p_{\gamma_1},p_{\gamma_2}\},\mu_R,\lambda)|}^2}{\hat s^2} \cdot \ x_a f_{f_a}(x_a, Q_a)\
    \cdot x_b  f_{f_b}(x_b, Q_b)\\
    &\times (2\pi)^4\ \delta^2\!\!\left(\sum_{k=1}^n
      \mathbf{p}_{k\perp}+p_{\gamma_1\perp}+p_{\gamma_2\perp}\right )\ \mathcal{O}_{2j}(\{p_i\}),
  \end{split}
\end{align}
where in principle $y_\mathrm{min}=-\infty$ and $y_\mathrm{max}=\infty$ (in
practice, they can both technically be put to $\pm5$ because of the requirement
that the extremal partons form part of the observed extremal jets).  Furthermore, we will
choose by default $\kappa=0.2$~GeV and $\lambda=\kappa$ (see
Section~\ref{sec:all-order-reg} for the definition of these
regulators). $\kappa$ has to be chosen small (as close to 0 as possible), and
setting $\lambda=\kappa$ ensures that events are generated with positive
weight only. While the correct results are obtained in the limit
$\lambda\to0$, the results are stable below $\lambda=2$~GeV. The
factors of $x_if_{i,f_i}(x_i,Q_i)$, $i=a,b$, are the parton density functions
for a parton of flavour $f_i$ evaluated at momentum fraction $x_i$ and
factorisation scale $Q_i$.  In practice, we take both $Q_a$ and $Q_b$ to be
equal to the factorisation scale $\mu_F$ which can be taken to be either fixed
or a number of dynamic scales (including $H_T/2$ or the maximum $p_T$ of any
single jet).  The renormalisation scale $\mu_R$ may also be evaluated at a fixed
or dynamic scale.  The step function, $\mathcal{O}_{2j}(\{p_i\})$, implements
the chosen cuts of the process, which consists of a minimum requirement that
at least two hard jets are observed.

The expression in Eq.~\eqref{eq:sigma2jsum} has leading-logarithmic accuracy by
construction.  We can impose leading-order fixed-order accuracy through matching
to leading-order matrix elements.  Eq.~\eqref{eq:sigma2jsum} only describes FKL
momentum configurations or FKL momentum configurations with one extra unordered
emission, hereafter referred to together as ``\HEJ configurations''.  We therefore
implement matching to full fixed-order in two different ways, depending on
the flavour and
momentum configuration.

Firstly, for the \HEJ configurations covered by the formula above (i.e.~those
where higher-order corrections are systematically summed), we employ
multiplicative matching to leading-order accuracy, where the final state
partons generated by the all-order results is clustered into
two or three jets.  These jets can be formed from a higher number of
partons, which means that they are not necessarily on-shell.  Since the
evaluation of leading-order matrix elements require particles with on-shell momenta, we
reshuffle the jet-momenta to put them on-shell, using an algorithm described in~\cite{Andersen:2011hs}.  After this the matching is
implemented by multiplying the \HEJ matrix-element-squared by the factor
\begin{align}
  \label{eq:weightFKL}
  w_{H+n\rm{-jet}} \equiv \frac{\overline{\left| \mathcal{M}^{LO}_{f_1f_2\to f_1 \cdot g
  \cdot f_2H}\left( \left\{ j_i \right\} \right)\right|}^2}{\overline{\left|
  \mathcal{M}^{HEJ}_{ {\rm tree} \ f_1f_2\to f_1 \cdot g
  \cdot f_2H}\left( \left\{ j_i \right\} \right)\right|}^2},
\end{align}
where $\{j_i\}$ are the on-shell jet-momenta.

An alternative way to think of
this procedure is to view the matching as a merging procedure as used routinely
for parton showers (CKKW-L\cite{Catani:2001cc,Lonnblad:1992tz}) for leading-order matrix
elements at different orders where in place of the logarithms controlled by
a parton shower prescription, the logarithms instead are those which are leading
in the high-energy limit.  This procedure gives
\begin{align}
  \begin{split}
  \label{eq:sigma2jFULL}
    \sigma_{H+2j}^\mathrm{resum,\ match}=&\sum_{f_a, f_b}\ \sum_{n=2}^\infty\
    \left(\int_{p_{1\perp}=p_{\perp\mathrm{ext,min}}}^{\infty}
      \frac{\mathrm{d}^2\mathbf{p}_{1\perp}}{(2\pi)^3}\
      \int_{y_\mathrm{min}}^{y_\mathrm{max}} \frac{\mathrm{d} y_1}{2}
    \right)\ \left(\int_{p_{n\perp}=p_{\perp\mathrm{ext,min}}}^{\infty} \frac{\mathrm{d}^2\mathbf{p}_{n\perp}}{(2\pi)^3}\
      \int_{y_{n-1}}^{y_\mathrm{max}} \frac{\mathrm{d} y_n}{2}\right) \\
&     \prod_{i=2}^{n-1}\left(\int_{p_{i\perp}=\kappa}^{\infty}
      \frac{\mathrm{d}^2\mathbf{p}_{i\perp}}{(2\pi)^3}\
      \int_{y_{i-1}}^{y_\mathrm{max}} \frac{\mathrm{d} y_i}{2}
    \right)\
    \int \frac{\mathrm{d}^3p_{\gamma_1}}{(2\pi)^3\ 2 E_{\gamma_1}}\
    \
    \int \frac{\mathrm{d}^3p_{\gamma_2}}{(2\pi)^3\ 2 E_{\gamma_2}}\
    \\
    &\frac{\overline{|\mathcal{M}_{\mathrm{HEJ}}^{\mathrm{reg}}(\{
        p_i,p_{\gamma_1},p_{\gamma_2}\},\mu_R,\lambda)|}^2}{\hat s^2} \times
    \left( \sum_{m=1}^\infty \mathcal{O}^e_{mj}(\{p_i\})\ w_{H+m-{\rm jet}}\right)\\
    &\times \ x_a f_{f_a}(x_a, Q_a)\
    \cdot x_b  f_{f_b}(x_b, Q_b) \cdot (2\pi)^4\ \delta^2\!\!\left(\sum_{k=1}^n
      \mathbf{p}_{k\perp}+p_{\gamma_1\perp}+p_{\gamma_2\perp}\right )\
    \mathcal{O}_{2j}(\{p_i\}).
  \end{split}
\end{align}
The functions, $\mathcal{O}^e_{mj}(\{p_i\})$, are step-functions which determine
whether or not the given set of momenta cluster into exactly $m$ jets.  No
matching is performed for the high jet-multiplicity states, where the leading
order matrix element is very slow to evaluate, or not evaluated at all
(currently 4 jets and above).

Secondly, the momentum configurations which do not correspond to \HEJ
configurations are not described at all by
Eq.~\eqref{eq:sigma2jsum}.   We therefore add exclusive tree-level samples of these for
two and three jets, which gives a sum of terms like the following:
\begin{align}
   \label{eq:nonfklmatching}
   \begin{split}
     \sigma_{H+mj}^{\rm non-HEJ} =& \sum_{f_a,f_b}\  \sum_{\{f_i\}}\ \prod_{k=1}^m \left( \int_{p_{k\perp}=p_{\perp\mathrm{min}}}^{\infty} \frac{\mathrm{d}^2\mathbf{p}_{k\perp}}{(2\pi)^3}\
       \int_{y_\mathrm{min}}^{y_\mathrm{max}} \frac{\mathrm{d} y_k}{2}\right) \
    \int \frac{\mathrm{d}^3p_{\gamma_1}}{(2\pi)^3\ 2 E_{\gamma_1}}\
    \
    \int \frac{\mathrm{d}^3p_{\gamma_2}}{(2\pi)^3\ 2 E_{\gamma_2}} \\
     & \times \frac{\overline{|\mathcal{M}^{LO}_{f_af_b\to f_1 \ldots
           f_m H}\left(
           \left\{ p_i \right\} \right)|}^2}{\hat s^2} \cdot
\Theta_{mj}(\{f_i\}, \{p_i\}) \\ &\times x_a f_{f_a}(x_a, Q_a)\
    \cdot x_b  f_{f_b}(x_b, Q_b) \cdot (2\pi)^4\ \delta^2\!\!\left(\sum_{k=1}^n
      \mathbf{p}_{k\perp}+p_{\gamma_1\perp}+p_{\gamma_2\perp}\right).
   \end{split}
 \end{align}
The new function, $\Theta_{mj}(\{f_i\}, \{p_i\})$, returns 1 if the flavour
assignments and momenta correspond to a configuration not captured by the
all-order summation configuration (ie.~they are not FKL or one unordered emission) and zero
otherwise.

The full equation for the \HEJ cross section for the production of a Higgs
boson which decays to two photons in association with at least two jets,
including the two types of matching described above is therefore
\begin{align}
  \label{eq:fullsigma}
  \sigma_{H+2j}^{\rm HEJ} = \sigma_{H+2j}^{\rm resum,\ match} + \sum_{m=2}^{m_{max}}
  \sigma_{H+mj}^{\rm non-HEJ}.
\end{align}

The addition of tree-level events which do not correspond to \HEJ configurations is
important for the
description in regions of phase space which are far from the high-energy limit.
However, the description reached is obviously inferior to that reached by the
all-order treatment. The inclusion of momentum
configurations with one unordered gluon emission is the first important step in
reducing the influence of the tree-level samples in the overall description.  The theoretical developments described
in Section~\ref{sec:UnorderedandHiggsB} allow us to move these momentum
configurations from the ``non-HEJ'' terms to the ``resum, match'' term in
Eq.~\eqref{eq:fullsigma}.

\begin{table}[t]
  \centering
  \begin{tabular}{|l|c|c|c|}
    \hline
    & FKL-ordered & Unordered & non-FKL-ordered \\ \hline
    No unordered resummation & 1059~fb (85\%) & -- & 185~fb
                                                                  (15\%)\\
    With unordered resummation & 1059~fb (86\%) & 47~fb (4\%) & 120~fb (10\%)\\
    \hline
    $qg$-channel only &&& \\ \hline
    No unordered resummation & 452~fb (81\%) & -- & 103~fb (19\%)\\
    With unordered resummation & 452~fb (84\%) & 38~fb (7\%) & 48~fb
                                                                     (9\%) \\ \hline
    $qQ$-channel only &&& \\ \hline
    No unordered resummation & 84~fb (82\%) & -- & 18~fb (18\%)\\
    With unordered resummation & 84~fb (84\%) & 9~fb (9\%) & 7~fb
                                                                     (7\%) \\ \hline
  \end{tabular}
  \caption{The total inclusive 3-jet cross section split into different
    components when unordered emissions are and are not included in the
    description. The second and third sections show the same numbers for subprocesses
    with a single initial gluon (labelled ``$qg$'') and subprocesses with no gluons in
    the initial state (labelled ``$qQ$'').}
  \label{tab:sigmacomponents}
\end{table}
In order to illustrate this point, we give the components of the cross section
for inclusive $H+3j$ production
in Table~\ref{tab:sigmacomponents}, within simple cuts ($|y_H|<2.37$, $|p_{\perp
  j}|>30$~GeV, $|y_j|<4.4$).  
One can see that the effect of extending
the all-order summation to include next-to-leading order terms through the
unordered emissions has reduced the dependence on fixed-order matching (the
non-FKL-ordered component) from 15\% to 10\% overall. However, the total rate includes
the large $gg$-component which is unaffected by the description of unordered
emissions.  The equivalent numbers for ``$qg$''-channels (labelling all
channels with exactly one gluon in the initial state) show a much larger effect.
The cross section of the non-FKL-ordered component halves, and the relative
importance of this component drops from 19\% to 9\%.  There is a similarly
dramatic effect in the ``$qQ$''-channel (labelling all subprocesses with no
gluons in the initial state) where now the percentage significance of the
non-FKL-ordered component has dropped from 18\% down to 7\%.

\begin{figure}[btp]
  \centering
  \includegraphics[width=0.45\linewidth]{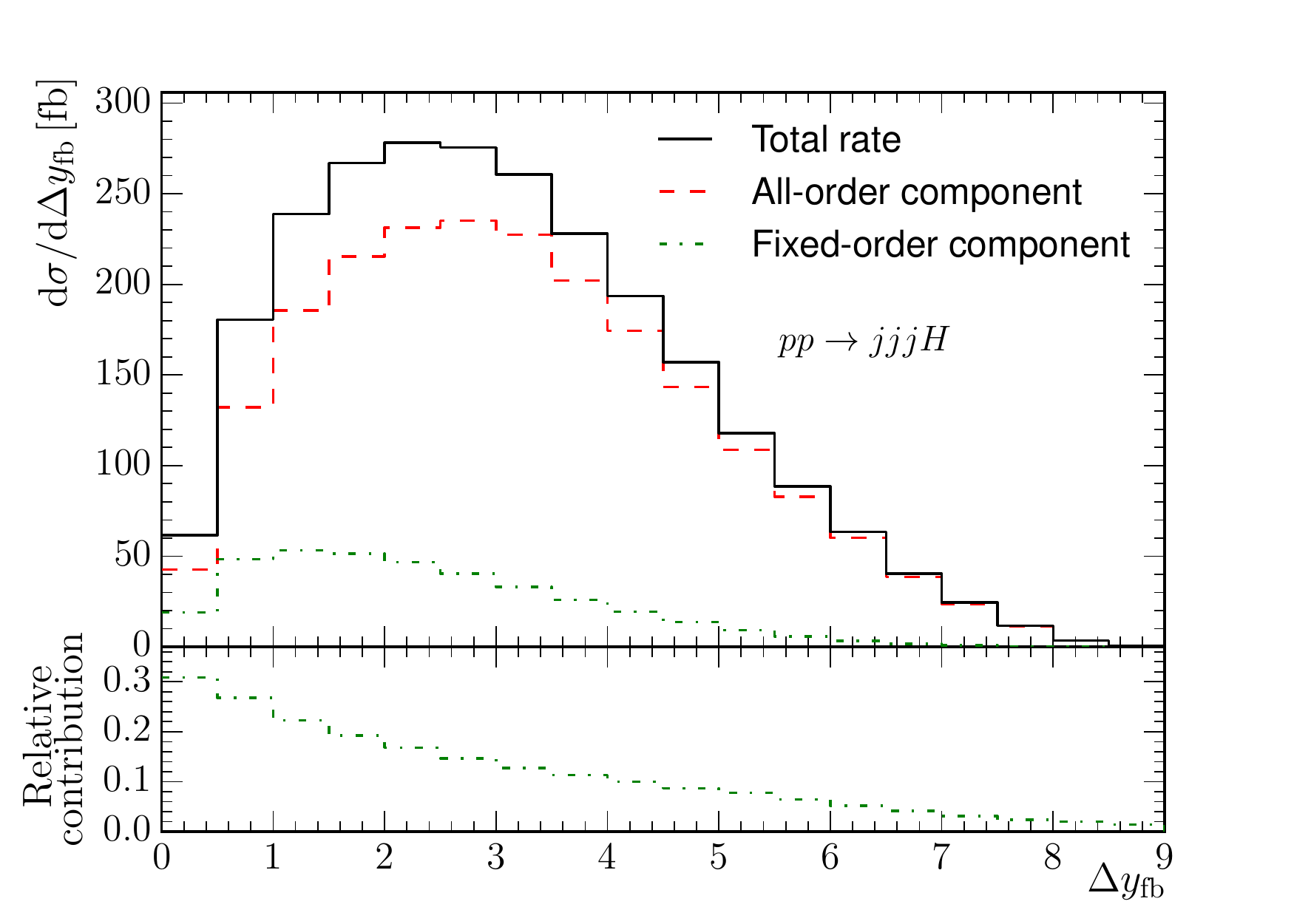}
  \includegraphics[width=0.45\linewidth]{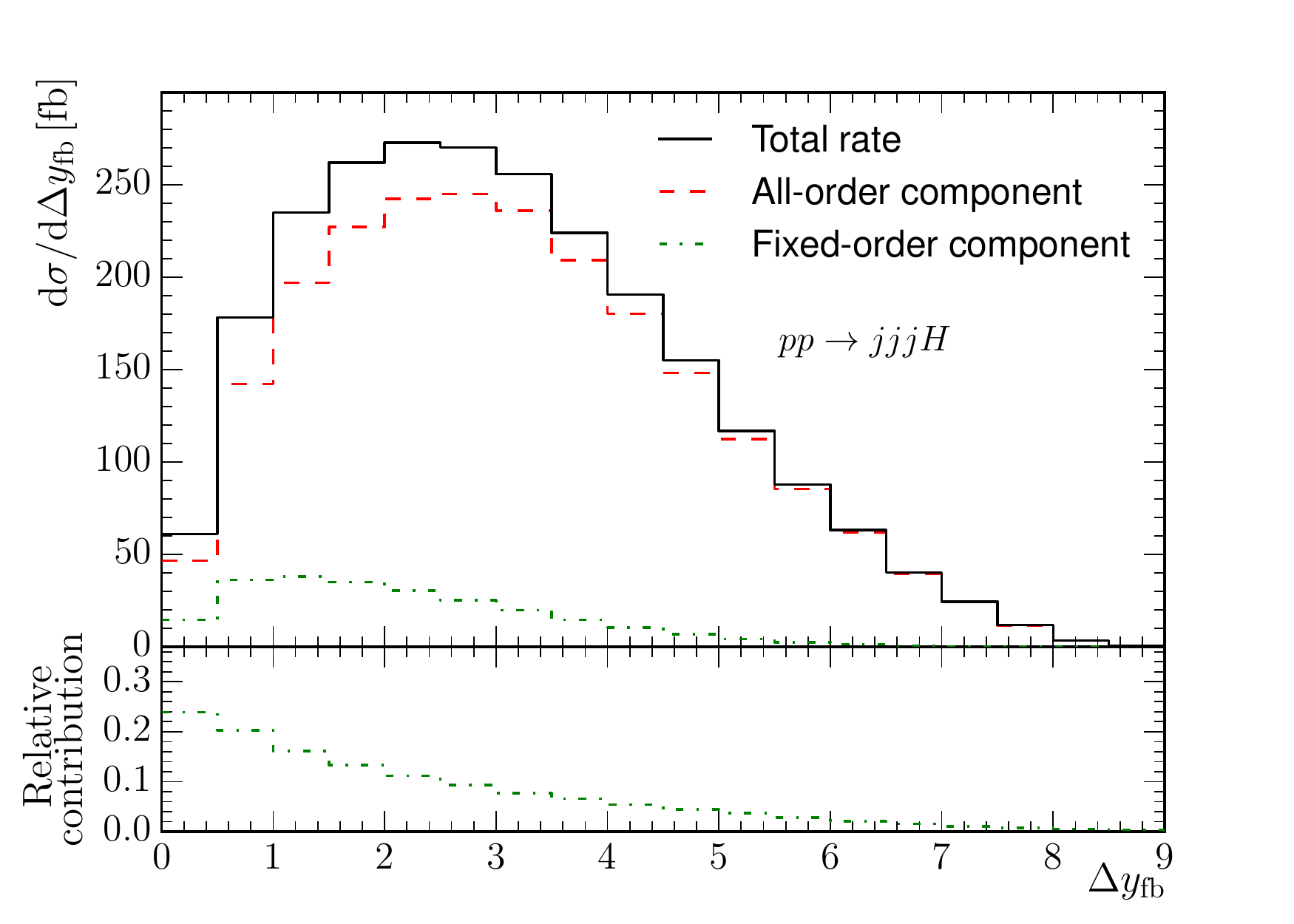}\\
  \includegraphics[width=0.45\linewidth]{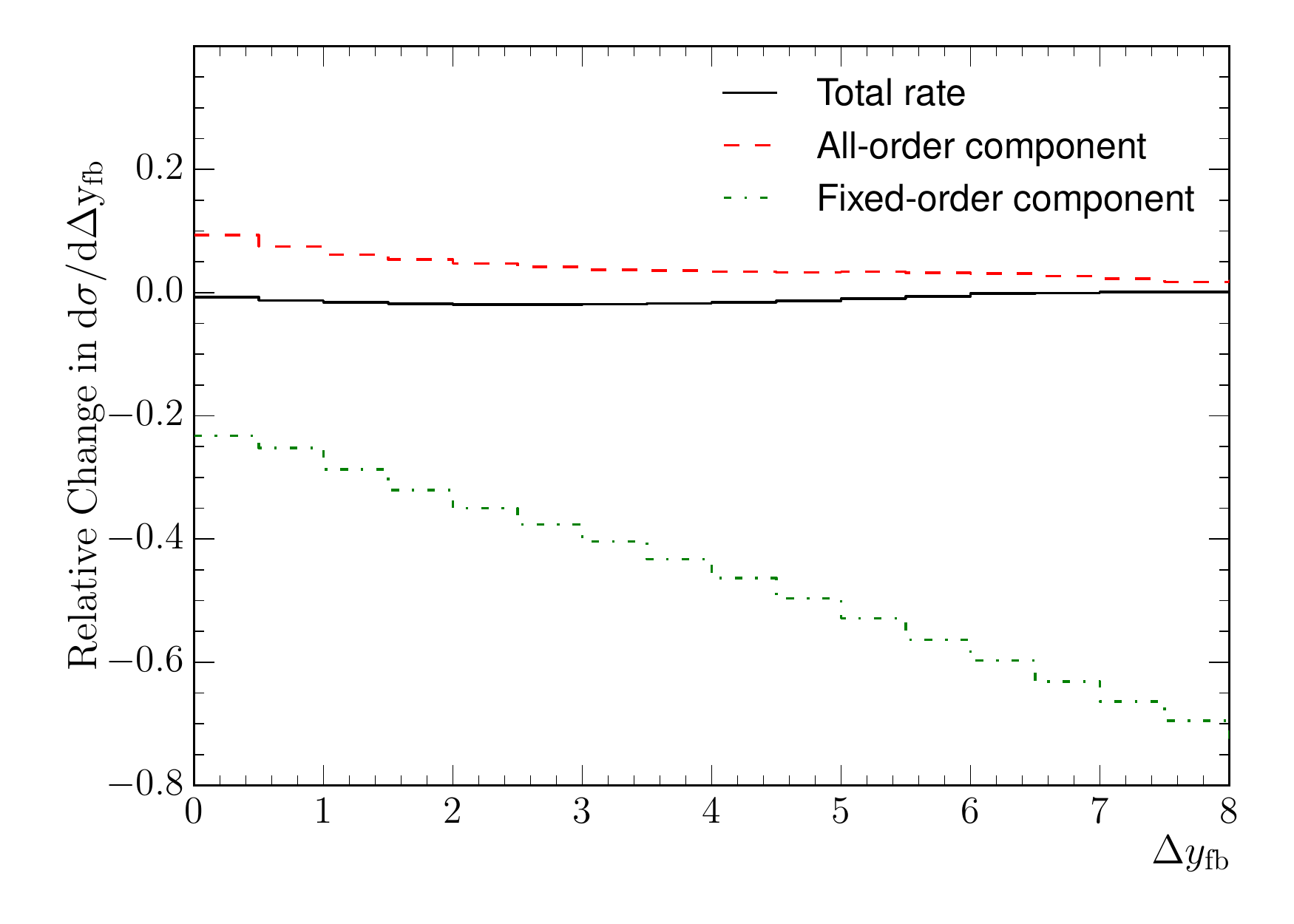}
  \caption{Plots showing the make-up of the cross-section as a function of the
    rapidity difference between the most forward and backward jets,
    $\Delta y_{fb}$.  The left-hand side shows the composition when the
    unordered emissions are included only through addition of fixed-order
    events. The green dotted line is the contribution from all such
    fixed-order events, the red dashed line is the contribution from
    the all-order summation, and the black solid line is the sum of the
    two. The right-hand side plot shows the same results, when the all-order
    summation is extended to included the unordered emissions. The bottom
    plot shows the relative change in the fixed-order, all-order and total
    rate after the extension of the all-order summation. The
    distributions are discussed further in the text.
  }
  \label{fig:ydifcpt}
\end{figure}

\begin{figure}[btp]
  \centering
  \includegraphics[width=0.45\linewidth]{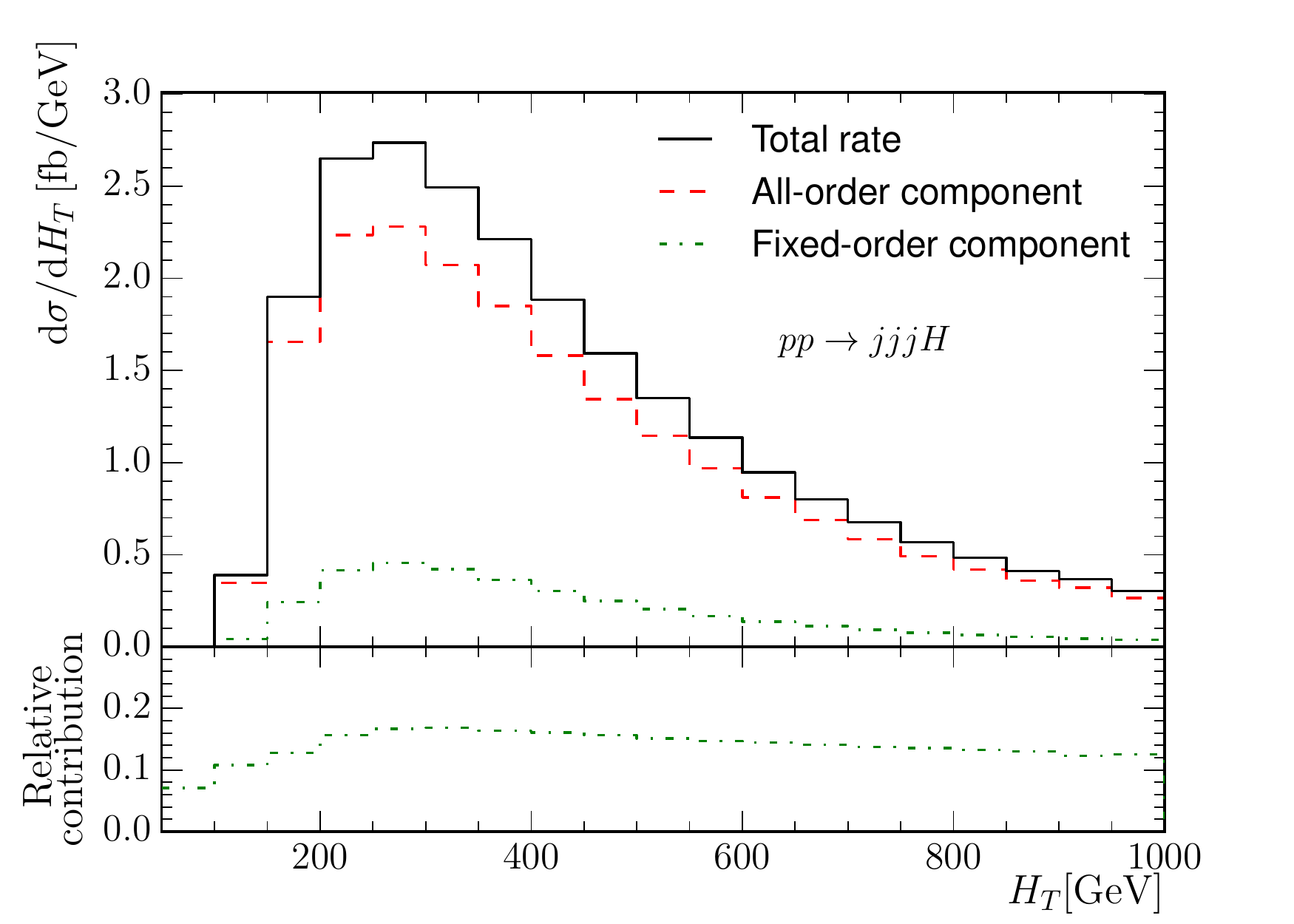}
  \includegraphics[width=0.45\linewidth]{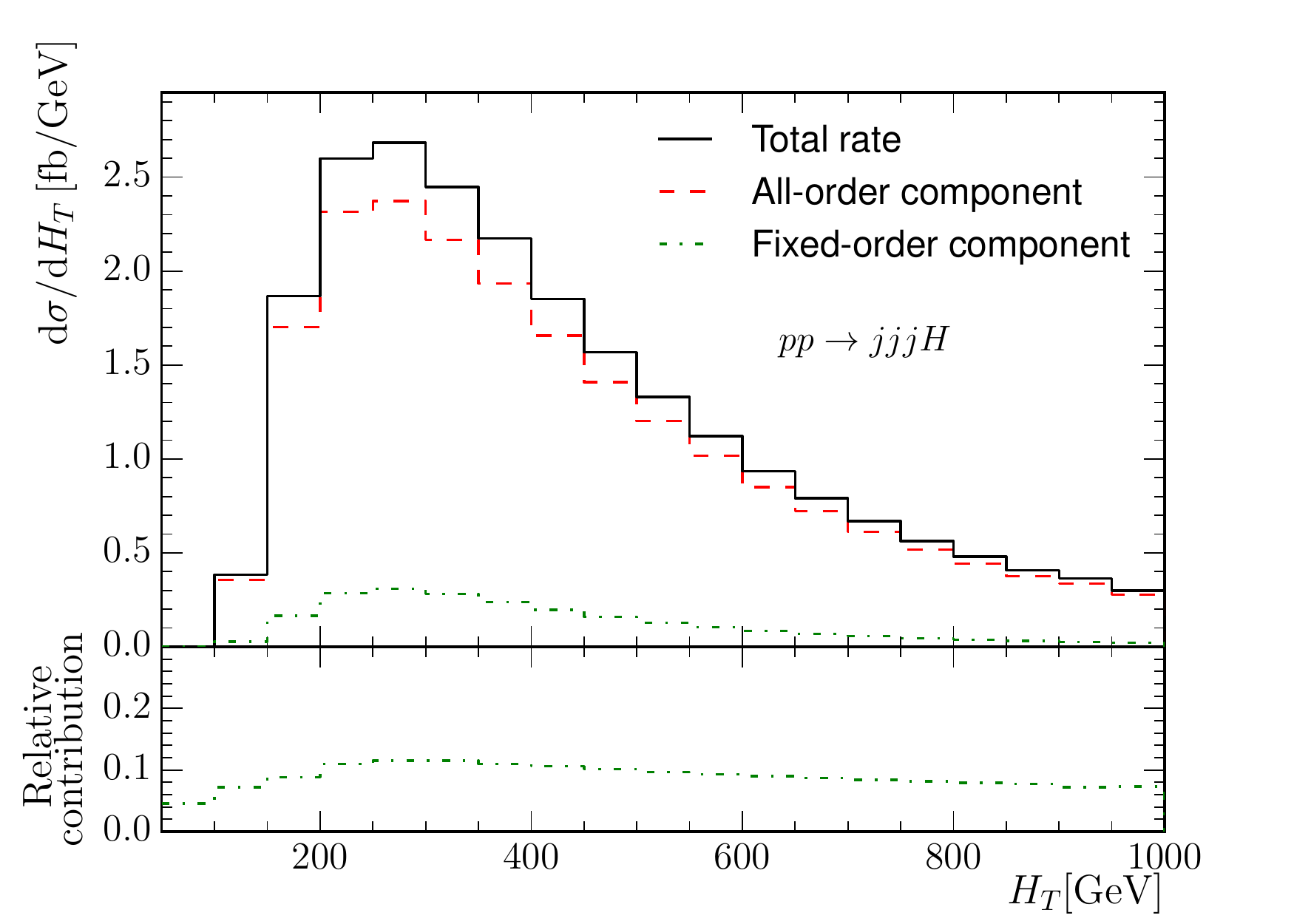}\\
  \includegraphics[width=0.45\linewidth]{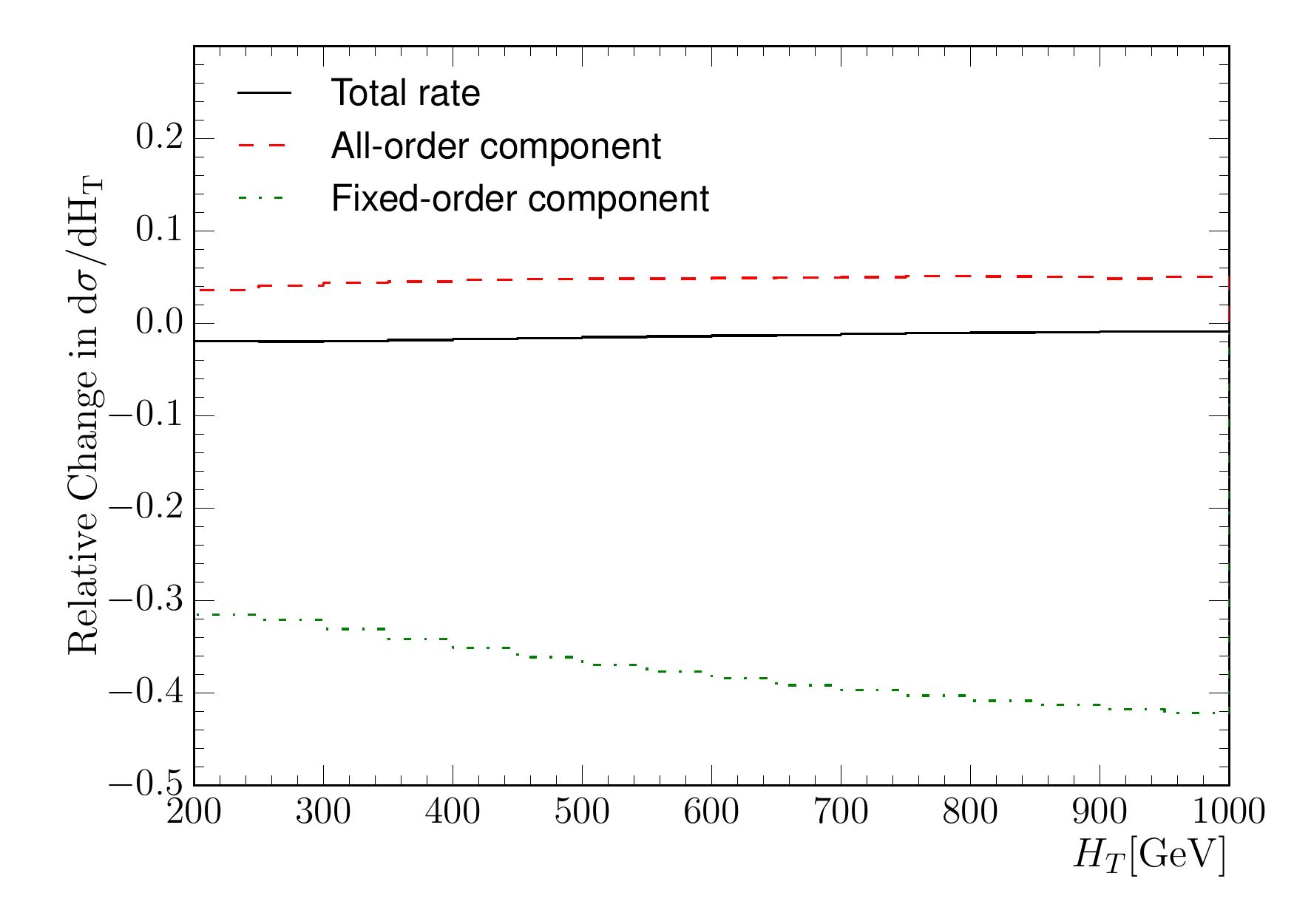}
  \caption{Plots showing the make-up of the cross-section as a function of the
    total transverse momentum of the event, $H_T$.  The left-hand side shows the composition when the
    unordered emissions are included only through addition of fixed-order
    events. The green dash-dotted line is the contribution from all such
    fixed-order events, the red dashed line is the contribution from
    the all-order summation, and the black solid line is the sum of the
    two. The right-hand side plot shows the same results, when the all-order
    summation is extended to included the unordered emissions. The
    distributions are discussed further in the text.}
  \label{fig:htcpt}
\end{figure}

Figures \ref{fig:ydifcpt} and \ref{fig:htcpt} show the composition of the
Higgs-boson plus three-jet cross-section in terms of the all-order and
fixed-order components as a function of the rapidity span of the event,
$\Delta y_{fb}$, and the scalar sum of transverse momenta, $H_T$. The top plot on
the left-hand side shows the composition when the unordered emissions are
included only through the addition of fixed-order events. The green dash-dotted line
is the contribution from all such fixed-order events, the red dashed line is
the contribution from the all-order summation, and the black solid line is
the sum of the two. The top right-hand side plots shows the same results, once
the all-order summation is extended to include the unordered emissions.

The
first thing to note on Fig.~\ref{fig:ydifcpt}~(top left) is that the relative
contribution of the fixed-order component is uniformly decreasing from 30\%
to 0\% for
increasing rapidity-spans $\Delta y_{fb}$. This is because the FKL-ordered
contributions dominate for large $\Delta y_{fb}$. Secondly, we note that including the
unordered emissions in the all-order treatment reduces the impact of the
fixed-order matching significantly (as seen by comparing the lower panels of
Fig.~\ref{fig:ydifcpt}), specifically from roughly 30\% to 24\% in the bin of
lowest rapidity span (where it peaks), and that the approach to 0\% is much
faster, since the largest sub-leading logarithmic contribution is now included in
the all-order approach. Lastly, we note that the sum of the fixed-order and
all-order results are largely unchanged after the inclusion of the unordered
emissions in the all-order summation: this is seen by the black lines being
largely unchanged between the left and right plots. This is made clearer in the bottom plot on
Fig.~\ref{fig:ydifcpt}, which shows the relative change in the
differential cross section for the two components and for the total
rate. We see that the total rate is almost unchanged for all $\Delta
y_{fb}$. This is in line with the rough expectation, since NLL corrections
should amount to a correction of order $\alpha_s$ compared to the LL in the
relevant channels (and the unordered emissions lead to corrections to only
the channels with incoming quarks, not the $gg$-channel). However, the dramatic reduction in the fixed-order component of the cross
section starts at about 25\% and rises linearly to 70\% over the same interval. The
increase in the reduction of the fixed-order component is driven by the
leading logarithms in the unordered $H+3j$ cross section, which as discussed
earlier constitutes part of the sub-leading corrections to $H+3j$. The fact
that the reduction is linear in $\Delta y$ is a nice illustration of the
dependence on $\log(s/t)\approx\Delta y$ of the component moved from the
fixed-order treatment to the all-order component. 

Fig.~\ref{fig:htcpt} shows the same information versus $H_T$. While $H_T$ is
not systematically connected with the all-order summation, a large value of $H_T$
limits the range of $\Delta y_{fb}$; so as seen on the top left plot (the
results when the unordered emissions are left in the fixed-order component),
the contribution from the fixed-order component of the cross section
increases from 8\% to 16\% and decreases to around 12\% as $H_T$ increases from 200~GeV to 1~TeV. The plot
at the top right shows the results when the unordered, NLL emissions are
included in the all-order treatment, and here the contribution from the
fixed-order component is reduced to 4\%-12\% throughout the range of
$H_T$, and is below 8\% by $H_T=1$~TeV.. This shows again that the first NLL terms of the unordered emissions
amount to a large portion of the $\mathcal{O}(\alpha_s^5)$-contribution not
accounted for by the FKL configurations. As
seen on the lower plot, the change in the all-order rate increases slightly from 4\%
to 5\%, while the fixed-order contribution decreases from 32\% to 42\%,
leading to an overall decrease in the total differential rate of a few percent.

As demonstrated in Figs.~\ref{fig:ydifcpt}--\ref{fig:htcpt}, the inclusion of
the first NLL terms through the unordered emissions leads to a large
systematic reduction in the dependence of the cross section on the
matching through the fixed-order component. The inclusion of the unordered
emissions and reduction in the dependence on the fixed-order matching is
particularly important for studies of the average number of jets versus the
rapidity span, as discussed later.


\section{Analysis of Results}
\label{sec:analysis-results}

In this section we will present the predictions which arise from the formalism
described in the previous section.  The study of the gluon-fusion component
of Higgs-boson
production in association with dijets is interesting for two separate
reasons: firstly, it is a background to the extraction of the measurement of
the weak-boson-fusion component, and while both production mechanisms manifest themselves in
the Higgs boson+dijet channel, several kinematic distributions and in particular their
higher-order corrections differ, and a thorough understanding of these can
aid in the suppression of the gluon-fusion-component when the aim is a study of
VBF. Secondly, the gluon-fusion component in Higgs boson+dijets can be studied on its own and as
such e.g.~the azimuthal correlation between the jets can be used for an
extraction of the $CP$-structure of the Higgs boson to gluon coupling, even in
the case of direct $CP$-violation and mixing in extended
Higgs sectors\cite{Klamke:2007cu,Andersen:2010zx,Dolan:2014upa}. These two
studies would evidently need separate cuts and approaches for event
selection, in order to enhance or suppress the gluon-fusion component. For
both purposes, the region of phase space with large rapidity span and large
dijet invariant mass is of interest.

\subsection{Setup and Parameters}
\label{sec:analysis-setup}

In the
current investigation we will focus on a few variables from the first
experimental analyses\cite{Aad:2014lwa}, except that the predictions presented here
will be for the LHC@13TeV. Furthermore, we require that the
events contain at least two jets
(anti-$k_T$ algorithm, $R=0.4$) which satisfy
\begin{align}
  \label{eq:atlascuts}
  \begin{split}
  &p_{\perp, j} > 30~{\rm GeV}, \quad |y_j|<4.4.
\end{split}
\end{align}

Since the weak-boson fusion process is initiated by two quarks, which
often carry a large part of the proton momenta, and receive only a modest
transverse momentum in the $t$-channel exchange of a weak boson, such events
will frequently result in a pair of jets separated by a large invariant mass
and rapidity.  Following the early analysis of the ATLAS collaboration~\cite{Aad:2014lwa}, we
will also investigate the gluon-fusion contribution within the VBF-selection
cuts applied to the two hardest jets in the event
\begin{align}
  \label{eq:vbfcuts}
  |y_1-y_2| > 2.8, \qquad m_{j_1 j_2} > 400~\textrm{GeV}.
\end{align}
As already discussed, the radiative corrections for the weak-boson fusion
process are significantly smaller than those for the gluon-fusion process. In
particular, the contribution from the 3-jet rate is small, and so for the VBF
process it is less
relevant to distinguish whether the two jets which are asked to fulfil the VBF
cuts are also the two hardest jets, the forward-backward jets (which always
have the largest rapidity separation, and often the largest invariant mass),
or whether one merely requires the existence of at least two jets which fulfil the VBF
cuts.

As in~\cite{Aad:2014lwa}, we consider Higgs boson decays into two photons with
\begin{align}
  \label{eq:photon_cuts}
  |y_{\gamma}| &< 2.37, \qquad 105~\textrm{GeV} <
  m_{\gamma_1 \gamma_2} < 160~\textrm{GeV},\notag\\
  p_{\perp, \gamma_1} &> 0.35\,m_{\gamma_1 \gamma_2}, \qquad p_{\perp, \gamma_2} > 0.25\,m_{\gamma_1 \gamma_2},
\end{align}
and require the photons to be separated from the jets and each other by
$\Delta R(\gamma, j), \Delta R(\gamma_1, \gamma_2)  > 0.4$.

We use the CT14nlo pdf set~\cite{Dulat:2015mca} as provided by
LHAPDF6~\cite{Buckley:2014ana}, choosing central renormalisation and
factorisation scales of $\mu_r = \mu_f = H_T/2$. To estimate the
perturbative uncertainty we also consider all combinations of $\mu_r,
\mu_f \in \{H_T/4, \sqrt{2} H_T/4, H_T/2, H_T/\sqrt{2}, H_T\}$ that
fulfil $1/2 < \mu_r/\mu_f < 2$. Larger ratios of the scales are excluded
in order to avoid artificially large logarithms. In the effective $ggH$
coupling in both calculations, we take the limit of an infinite top mass and set the
renormalisation scale to the Higgs boson mass.

\subsection{Differential Distributions for Higgs Boson Plus Dijets}
\label{sec:shape-higgs-dijets}
This subsection will present a comparison of results for the
gluon-fusion component of Higgs-boson-plus-dijets from \HEJ and from a NLO QCD
calculation facilitated by
MCFM~\cite{Campbell:2006xx,Campbell:2010cz}. We also show leading-order
results in order to demonstrate the higher-order effects in both
schemes. To avoid visual clutter, we refrain from including the
scale-variation uncertainties for the leading-order curves. We start by
discussing distributions obtained within the inclusive cuts of
Eq.~\eqref{eq:atlascuts} and Eq.~\eqref{eq:photon_cuts} which will be important for understanding the impact of the
VBF cuts in Eq.~\eqref{eq:vbfcuts}.

Firstly, we find that with the scales choices made, the inclusive cross
section for Higgs-boson-plus-dijets at NLO is $6.48^{+0.08}_{-0.57}$~fb, while the result
obtained in \HEJ is $4.06^{+1.15}_{-0.87}$~fb. The central value found at leading
order is $4.41$~fb, and so the result for \HEJ for the inclusive cross
section for Higgs-boson-plus-dijets is slightly less than the LO value, and
the correction compared to LO is in the opposite direction to the result
found at NLO. The cross-section obtained at LO is not within the scale variation of the NLO result,
and the higher-order corrections are therefore expected to be
large\footnote{The explanation for this is different to that for the case of
  inclusive Higgs-boson production, since all possible combinations of
  incoming partons are allowed even at LO.}.

\begin{figure}[btp]
  \centering
  \includegraphics[width=0.75\linewidth]{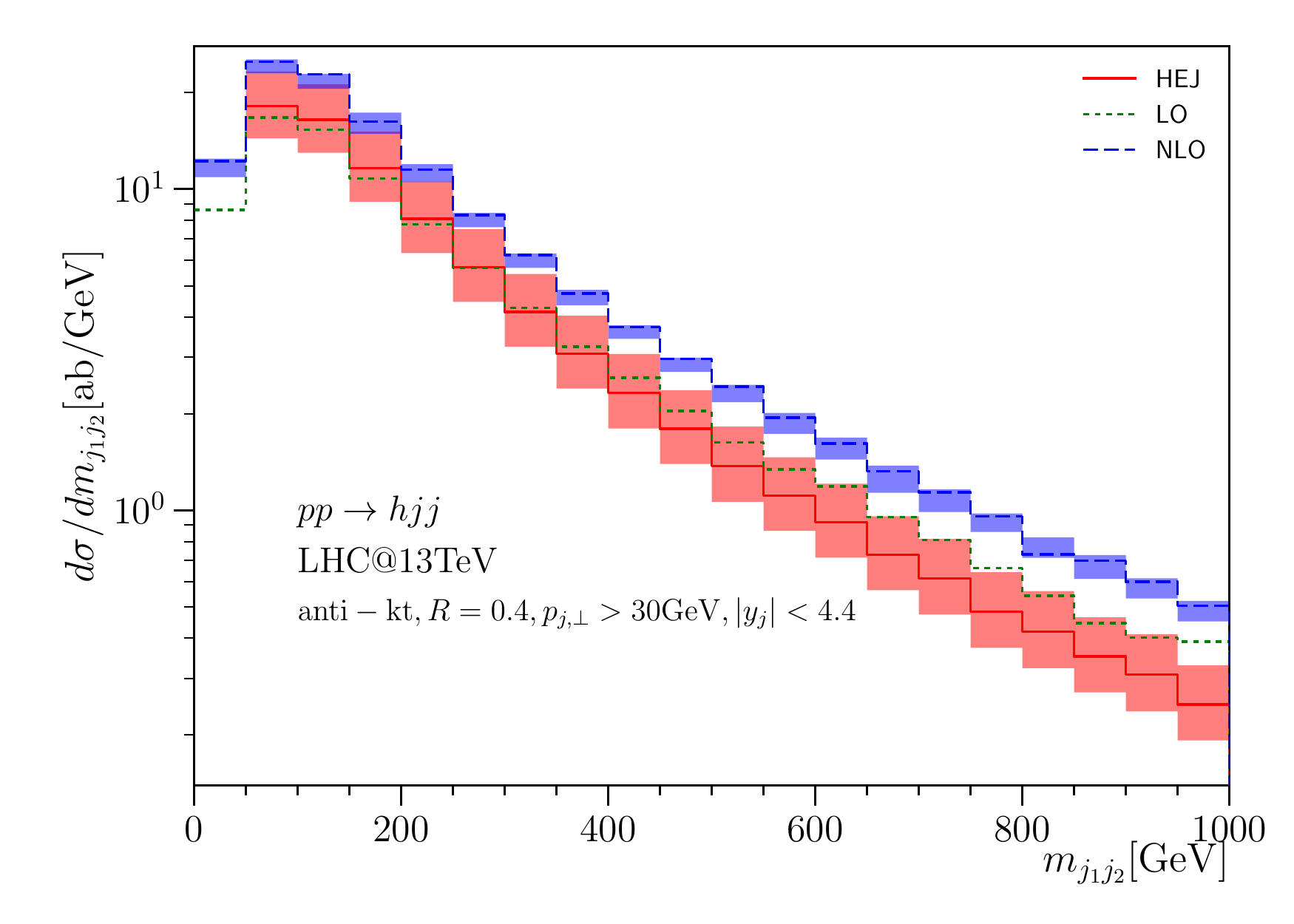}
  \caption{The distribution of the invariant mass between the two hardest
    jets. Predictions from \HEJ are shown in red (full line) while the NLO
    result is shown in blue (dashed line). The distributions are discussed
    further in the text.}
  \label{fig:minv}
\end{figure}
Fig.~\ref{fig:minv} shows the distribution in the invariant mass between the
two hardest (in transverse momentum) jets within the inclusive cuts. The
distribution obtained with \HEJ is slightly steeper than that at NLO; we
will see below that this is because \HEJ allows for more jet radiation than a
NLO-calculation, and the samples with more than just two jets carry more
relative weight. This in turn means the hardest two jets on average are
closer in rapidity and therefore have a smaller invariant mass.

\begin{figure}[btp]
  \centering
  \begin{subfigure}[t]{0.495\textwidth}
    \includegraphics[width=\linewidth]{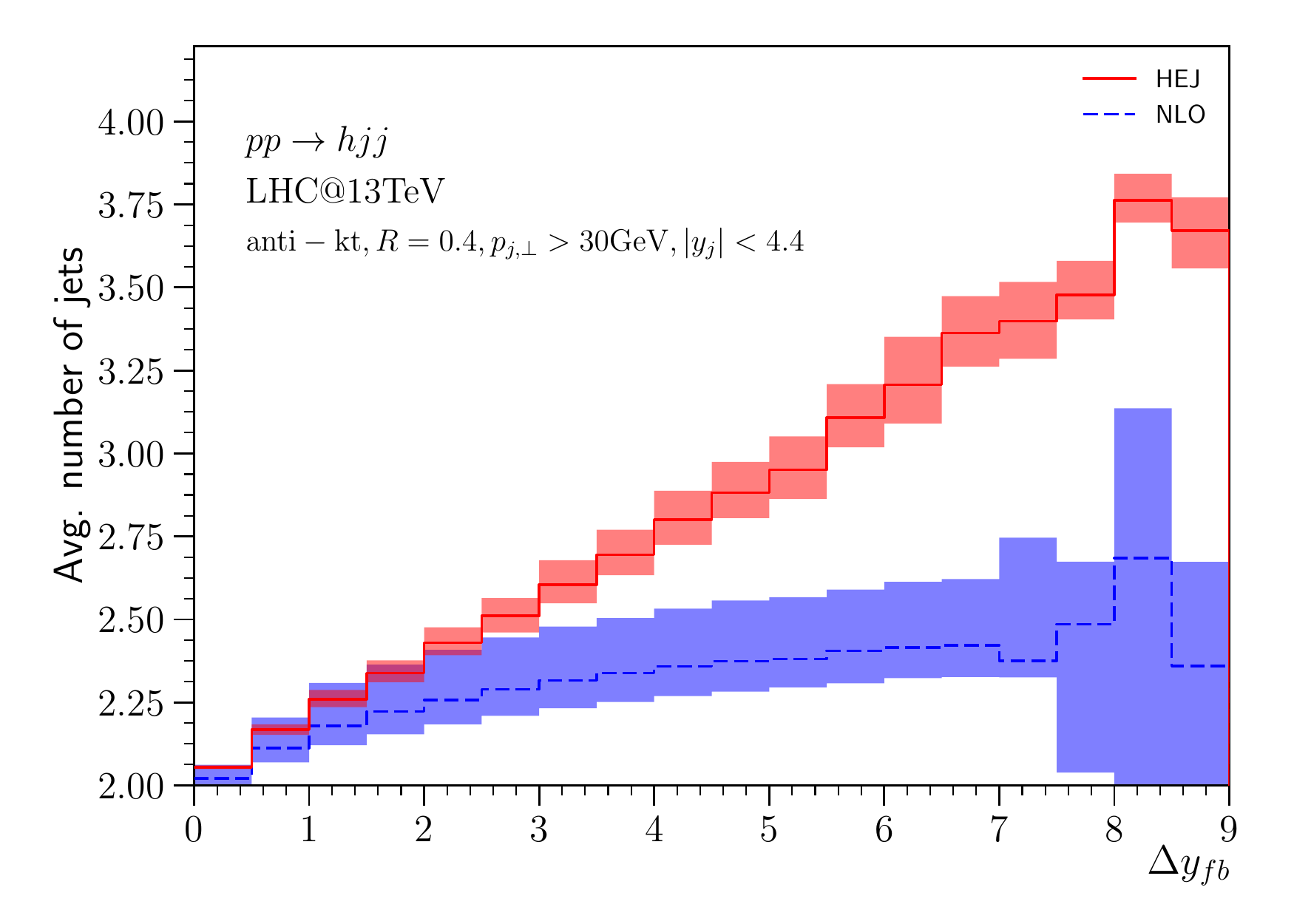}
    \caption{}
  \label{fig:nja}
  \end{subfigure}
  \begin{subfigure}[t]{0.495\textwidth}
    \includegraphics[width=\linewidth]{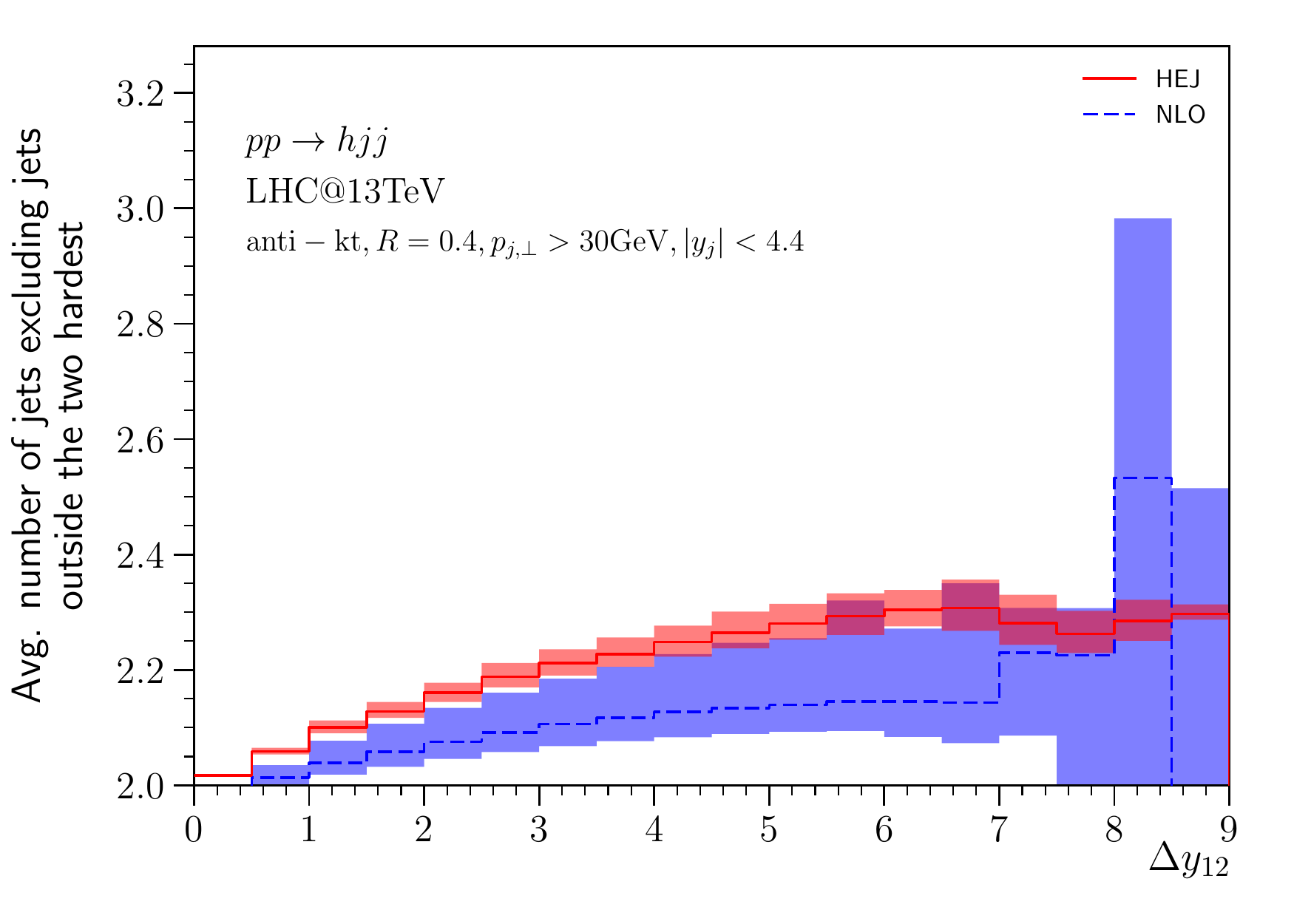}
    \caption{}
  \label{fig:njb}
  \end{subfigure}
  \caption{The average number of jets as a function of the rapidity
    difference between the most forward and backward jets (a)
    and between the hardest two jets (b). The \HEJ
    predictions are shown in red while the NLO results are shown in
    blue (dashed). Both distributions are discussed in the text.}
  \label{fig:njt}
\end{figure}
To investigate further the expected impact of the VBF cuts, we calculate the
average number of hard jets observed  versus the rapidity difference between
jets. A successful description of this radiation
pattern is necessary for the distinction of the VBF and GF process\cite{Dokshitzer:1991he}, and in
particular for a correct description of the effect of the
VBF cuts on the GF component. It is well-known that in all descriptions of dijet processes, and
indeed data, the average number of hard jets increases with the rapidity difference
between the most forward and backward hard jets, $y_{fb}$.  This is clearly
seen in the results for both NLO and \HEJ in Fig.~\ref{fig:nja}.  The prediction from \HEJ rises more
steeply than the equivalent prediction from the NLO calculation, which
plateaus at a value of 2.4 already for $\Delta y_{fb}=5$, where the
prediction for the exclusive
hard 3-jet rate is nearly as large as the exclusive hard 2-jet rate (obviously the
NLO calculation for Higgs boson plus dijet production gives an NLO estimate for the dijet rate, but only a
LO estimate for the trijet rate). It is indeed expected that \HEJ should rise higher than NLO, as the NLO calculation contains only contributions from 2- and 3-jet
events and does not contain the all-order evolution in rapidity which is present
in \HEJ.  This steeper dependence obtained in \HEJ has been seen to give a good description of
data in other dijet processes, where data has already allowed detailed
analyses, see e.g.~Ref.~\cite{Abazov:2013gpa}. Since the
contribution from higher jet counts is small in the VBF-process, a large
number of jets from the gluon-fusion process would make it easier to
distinguish the two. This will be the source of the difference between the
prediction of \HEJ and NLO for the GF contribution within the VBF-cuts.

Fig.~\ref{fig:njb} shows the average number of hard jets within the same
phase space as Fig.~\ref{fig:nja}, but as a function of the rapidity
separation between the two \emph{hardest} jets, and not counting jets with
rapidities outside the two hardest jets. \HEJ has been shown to also give a
good description of this observable for other dijet
processes\cite{Abazov:2013gpa}. When the jets outside the two hardest ones
are excluded, the rise in the average number of hard jets counted is far less
for both NLO and \HEJ. Indeed, both predictions plateau with a value of
roughly 2.2 at around $\Delta y_{12}=6$. The difference between
Fig.~\ref{fig:nja} and Fig.~\ref{fig:njb} is caused only by events with three
or more jets (since if there are just two, there is no difference between the
two hardest jets, and the two furthest apart in rapidity), and thus no large
difference between the two observables is expected for the VBF process. The
large contribution from the component with 3-jets and higher in the
gluon-fusion process means that significant differences can arise in
superficially similarly defined quantities as illustrated in
Fig.~\ref{fig:njt}. This is important for the use of cuts to suppress the
gluon-fusion component in VBF analysis, and separately for the focus on the
gluon-fusion component e.g.~for the extraction of the $CP$-structure of the
$ggH$-coupling.

We will now discuss kinematic distributions of the Higgs-boson and the jets,
both for the inclusive and the VBF cuts. The prediction obtained with NLO for
the cross-section within the VBF-cuts is $0.87^{+0.02}_{-0.09}$~fb (with LO
it is 0.62~fb), and with \HEJ it is
$0.38^{+0.11}_{-0.08}$~fb. We argue that for the VBF-cuts the results obtained with
\HEJ are more reliable than those obtained with NLO. This is because a
successful description of the VBF-cuts relies on the description of the
emission of further hard jets from the production of Higgs-boson plus
dijets. Even at the Tevatron centre-of-mass energy of 1.96~TeV, the pure
NLO-calculation gives an insufficient description of the average number of
jets in other dijet-processes such as W+dijets. This deficiency of the
NLO-calculation will be even larger at the LHC, whereas \HEJ gives a good
description of the hard jet-production in other processes with similar
jet-cuts as those applied in this study of Higgs-boson production with
dijets.

\begin{figure}[btp]
  \centering
  \includegraphics[width=0.49\linewidth]{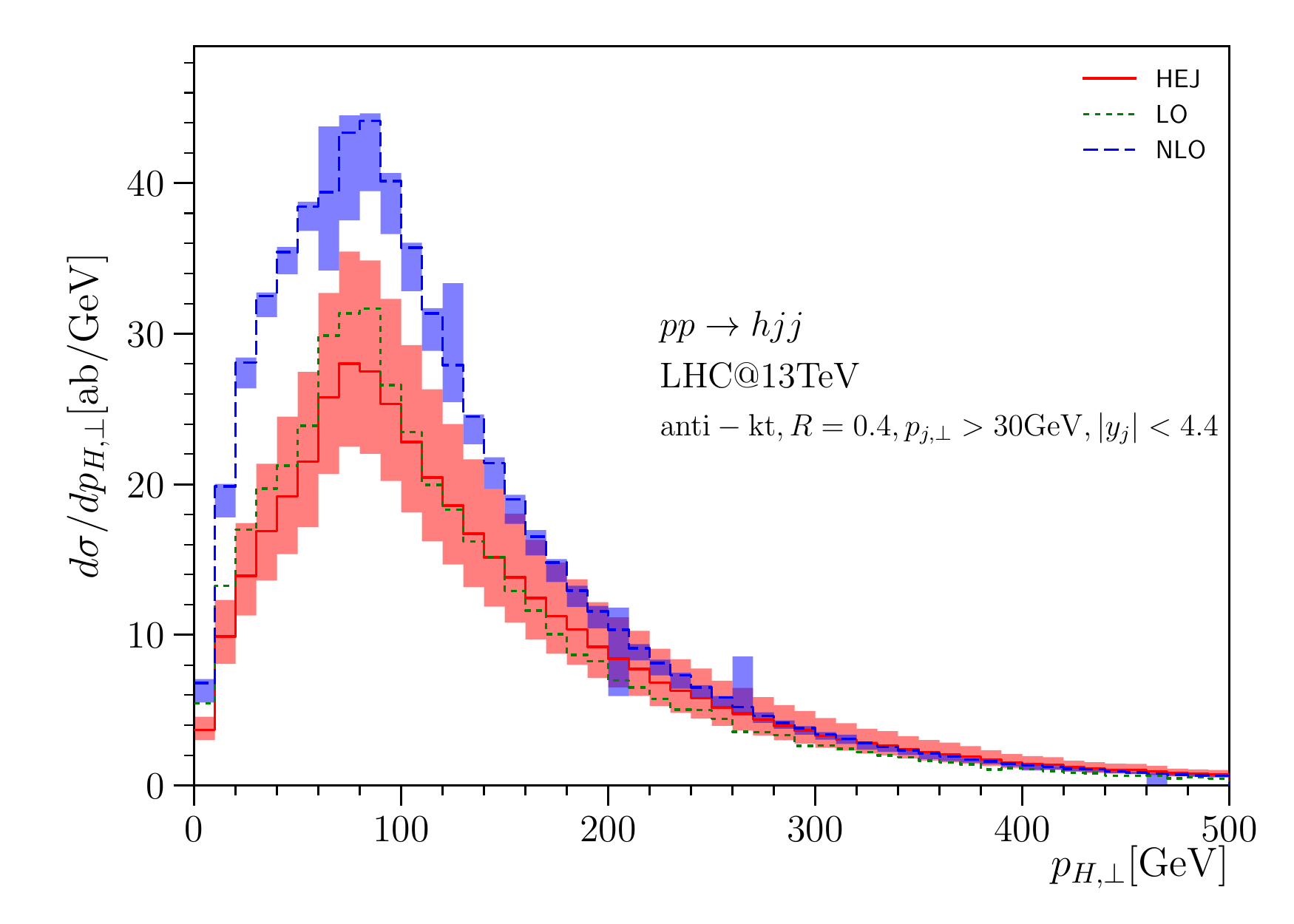}
  \includegraphics[width=0.49\linewidth]{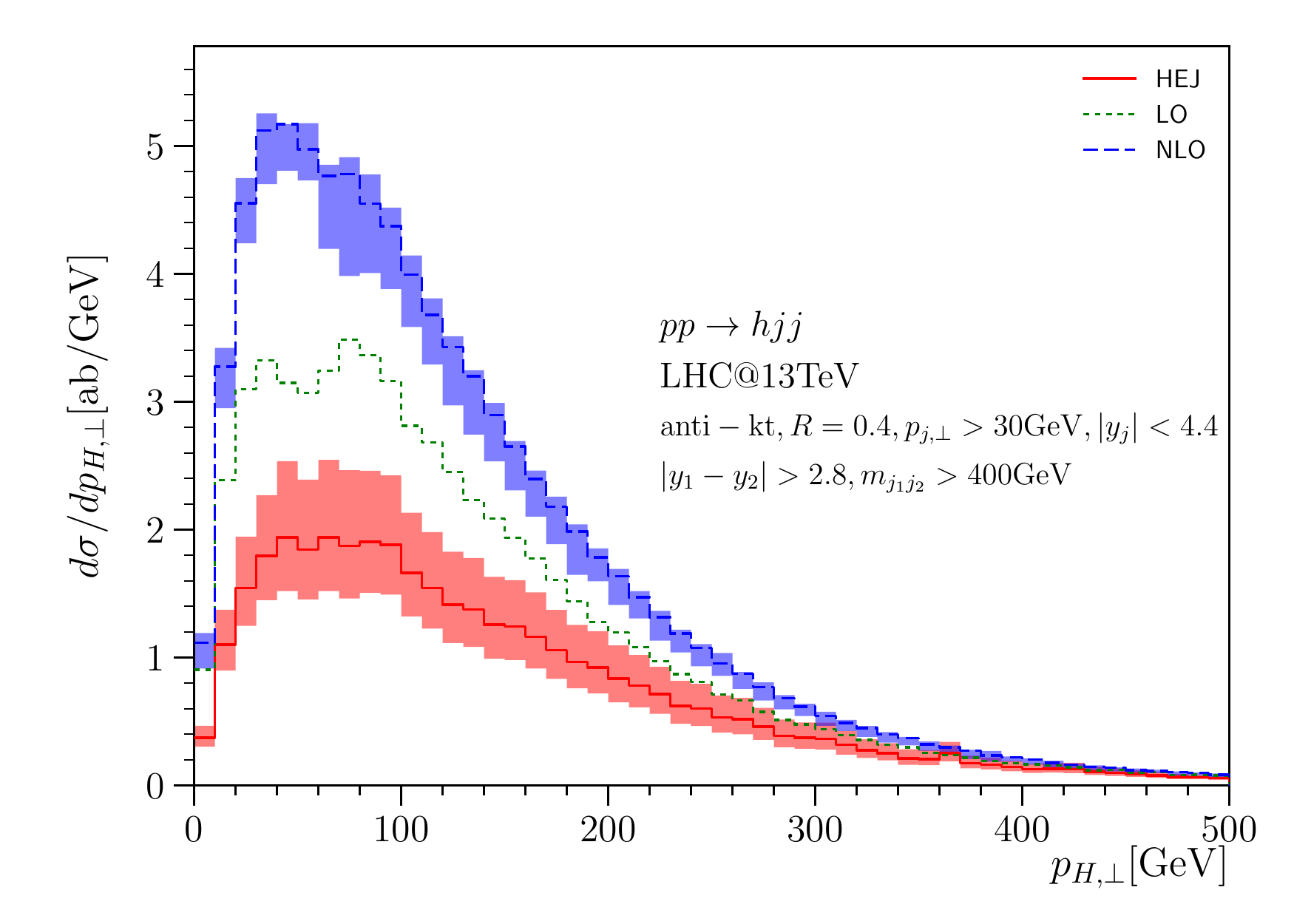}\\
  (a) \hspace{6.5cm} (b)
  \caption{(a) The transverse momentum distribution of the Higgs boson in
    inclusive dijet events, and (b) when the hardest two jets are required to pass the VBF-cuts.
    Predictions from \HEJ are shown in red while the NLO result is shown in
    blue.  Both distributions are discussed further in the text.}
  \label{fig:ptHphi}
\end{figure}

In Fig.~\ref{fig:ptHphi}(a), we show the Higgs transverse momentum
distribution within the cuts of Eq.~\eqref{eq:atlascuts}, while
Fig.~\ref{fig:ptHphi}(b) is the same distribution when also the VBF
cuts of Eq.~\eqref{eq:vbfcuts} are fulfilled. We observe the understood
reduction in cross-section obtained with \HEJ compared to NLO. The two peaks visible in
the LO obtained within the VBF cuts are caused by the azimuthal
structure of the $ggH$ coupling. As we will see later, the cross-section
peaks when the jets are back-to-back and has another local maximum when
they are collinear. This induces the two features in the LO curves,
which become broader and indistinguishable when further radiation is
included through either the NLO corrections or the all-order summation.
 The
radiative corrections at NLO are found to be large for the gluon-fusion
component of Higgs-plus-dijets; in particular, the 3-jet component forms a significant part
of the 2-jet cross section at NLO --- contrary to the situation for
the VBF component.  Furthermore the one-loop interference between the QCD
and EW component is negligible\cite{Andersen:2007mp}. Requiring that the two
hardest jets are separated in rapidity and invariant mass according to
Eq.~\eqref{eq:vbfcuts} reduces the gluon fusion component more compared to
just requiring the existence of two jets which satisfy the
requirement. Other selection processes may be of significance for the study
of the gluon-fusion component alone, and will be the focus of further
studies. We note again here that the application of the further VBF cuts
reduces the gluon-fusion cross section from $6.48$~fb (inclusive) to $0.87$~fb
(VBF cuts) at NLO and from $4.06$ fb to $0.38$~fb in the \HEJ
resummation. This corresponds to a severe reduction of the \HEJ
cross section to $9.4\%$, whereas NLO QCD predicts a reduction to
$13.4\%$ of the inclusive cross section, and the difference is explained by
the deficiency of a NLO-calculation in describing the number of hard jets
produced by the gluon-fusion process in the VBF-region of phase
space.\footnote{The NLO calculation of the inclusive rate does of course not
  answer the question of the number of hard jets produced at NLO accuracy.}

We also note that the transverse momentum distribution for
the Higgs boson is relatively hard such that the effective theory
derived from $m_t\to\infty$ will obviously not apply in all the relevant
region, but the results presented here are still relevant for inspecting
the impact of the high-energy summation. Furthermore, the
$m_t\to\infty$-limit and the high-energy limits commute, and the leading
high-energy effects can be calculated with full top-mass
dependence. This is the focus of ongoing work within \HEJ.

\begin{figure}[btp]
  \centering
  \includegraphics[width=0.49\linewidth]{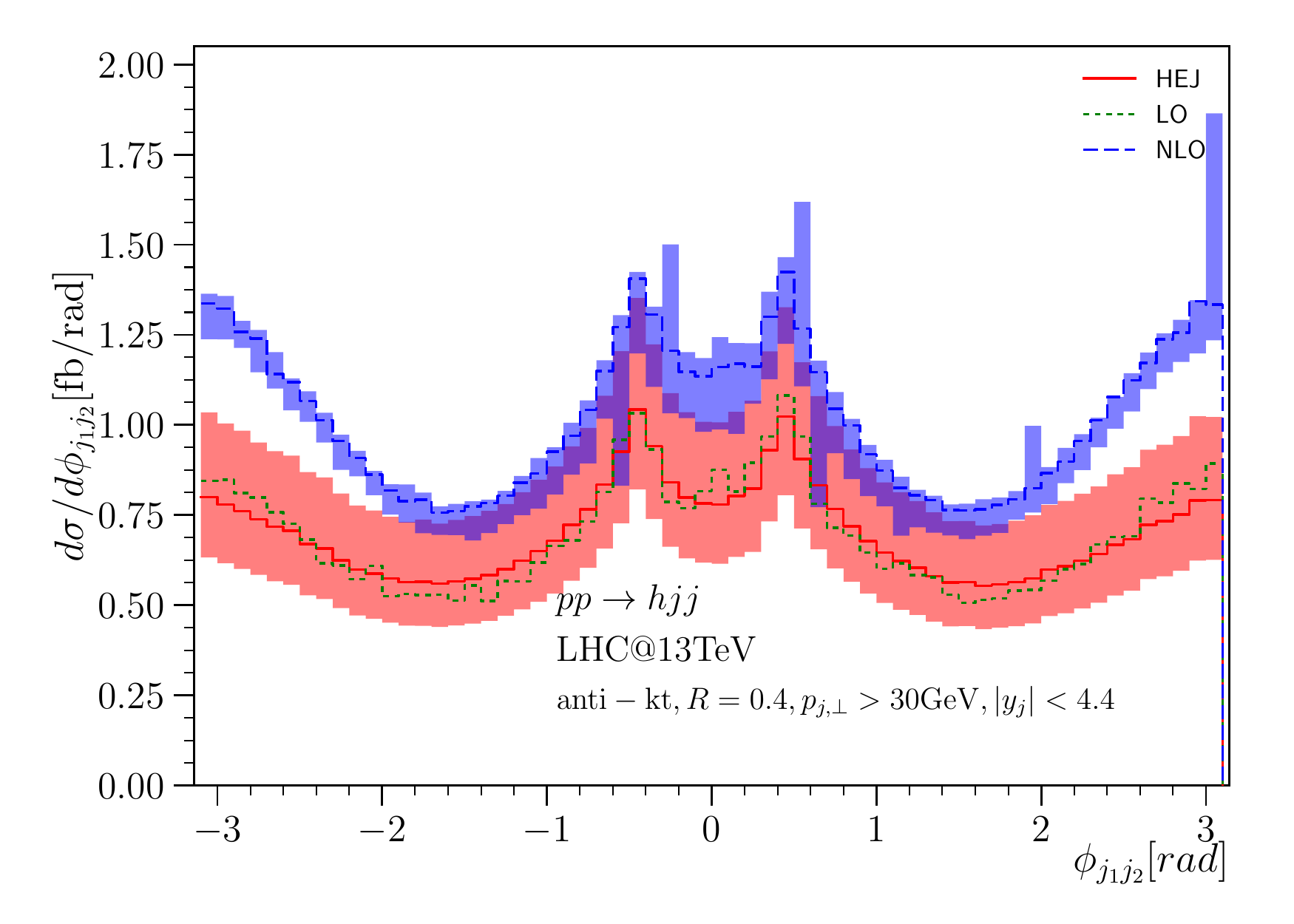}
  \includegraphics[width=0.49\linewidth]{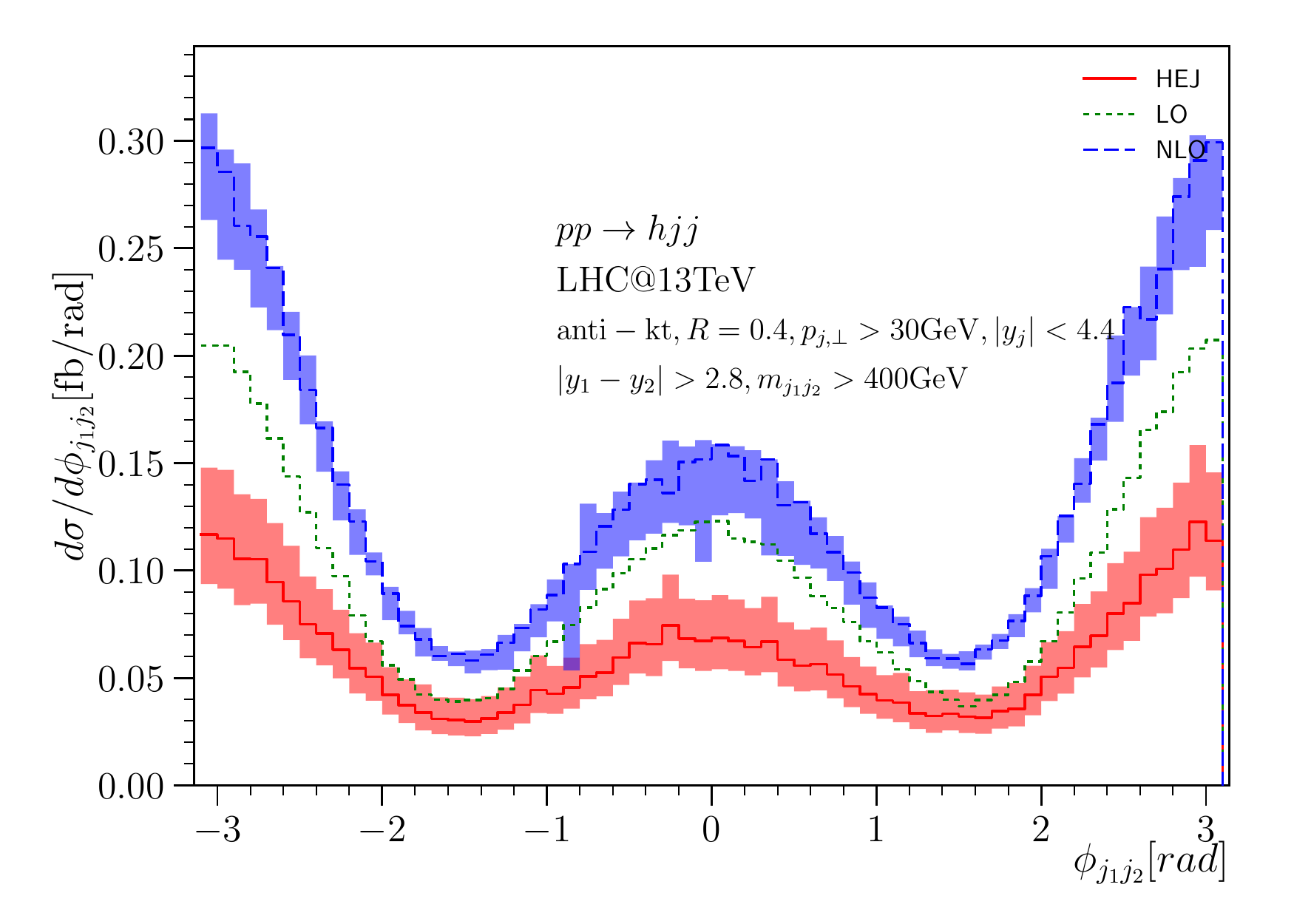}\\
  (a) \hspace{6.5cm} (b)
  \caption{(a) The distribution of the azimuthal angle between the two
    hardest jets, $\phi_{j_1j_2}$, and (b) ditto within the VBF cuts.  The
    cosine-like even distribution is a finger-print of the $CP$-even structure
    of the $ggH$-vertex. Predictions from \HEJ are shown in red (solid line) while the NLO
    result is shown in blue (dashed).}
  \label{fig:phi}
\end{figure}

A tree-level analysis indicates that the $CP$ structure of the Higgs coupling
can be cleanly studied using the azimuthal angle between the two
jets\cite{Klamke:2007cu}, with the definition of the azimuthal angle extended
to the full range $[-\pi;\pi]$ by e.g.~always measuring it counter-clockwise
relative to a predefined forward direction. The Born-level analysis of the Standard Model couplings predicts an even,
cosine-like behaviour, and the extension of the azimuthal angle to the full
range of $-\pi$ to $\pi$ allows for a probe of $CP$
admixtures\cite{Klamke:2007cu}.
In Fig.~\ref{fig:phi}, we
show the distribution of the angle between the hardest two jets,
$\phi_{j_1j_2}$ with (a) inclusive and (b) VBF cuts.  We again see the same
reduction in cross section of \HEJ compared to the NLO prediction.  The shape around
$\phi_{j_1j_2}=0$ in Fig.~\ref{fig:phi}(a) is caused by the removal of tree-level
three-parton events which appear in two-jet configurations --- the extension
of the dip is determined by the $R$-parameter in the jet-clustering, which
removes the contribution arising from collinear splittings of 2-jet events.

We have therefore seen in this section that higher-order corrections in
Higgs-boson-plus-dijet production are large and have a significant impact on the results of imposing
VBF event selection cuts.  


\section{Conclusions}
\label{sec:conclusions}

In this paper we have described the production of a Higgs boson with at least
two jets within the High Energy Jets (\HEJ) formalism.  This key process will be
central to efforts to pin down properties of the Higgs couplings to vector
bosons. We implemented the process of Higgs-boson production in association with
at least two jets within the framework of \HEJ. Furthermore, we calculated the
first next-to-leading logarithmic corrections to the framework by including the
un-ordered emission of a gluon (i.e.~the emission of gluon outside of the
rapidity region contributing to the leading-logarithmic behaviour of the cross
section) in the all-order treatment for the first time. Such regions were
previously accounted for only through matching to fixed-order
matrix-elements. The new results increases the fraction of the total cross
section which is controlled by \HEJ and subject to resummation, while also
reducing our dependence on fixed-order matching.

We have then studied the predicted jet radiation patterns for various
distributions within typical experimental cuts, and compared these to the
corresponding results for a fixed-order NLO calculation.  The
inclusion of higher-order corrections beyond NLO are  clearly observed in the
average number of jets as a function of rapidity, where other variables show
less pronounced differences. This result can be used to distinguish the
gluon-fusion and vector-boson fusion component of the Higgs boson+dijet cross section.

We have also seen that imposing topological ``VBF'' cuts has a significant
impact on the cross section beyond that predicted at NLO (for the particular
choice here, the cross section was reduced to $9.4$\% of the original).  This is
understood as a combination of increased jet activity in any event with a
reasonable rapidity separation and the impact of the all-order virtual
corrections included in the \HEJ description.

\section*{Acknowledgements}
This project has received funding from the European Union's Horizon 2020 research and innovation programme under the Marie Sk\l{}odowska-Curie grant
agreement No 722104. JRA is
supported by the UK Science and Technology Facilities Council (STFC). AM
is supported by a European Union COFUND/Durham Junior Research
Fellowship under EU grant agreement number 267209. JMS is supported by a
Royal Society University Research Fellowship and the ERC Starting Grant
715049 ``QCDforfuture''. The conclusion of this project was made possible by
sustained efforts of the Penrith, Keswick and RAF Mountain Rescue Teams, the
Great North Air Ambulance and
the kind hospitality of the RVI, Newcastle.


\appendix
\boldmath
\section{Tree-Level Amplitudes for $qg\to qg$}
\label{sec:tree-level-ampl}
\unboldmath

A short calculation gives the amplitude for the
process $qg\to qg$.  We use the following notation for spinors:
\begin{equation}
  \label{eq:spinor}
  \begin{split}
    u_\pm(p)=\vert p\pm \rangle, \qquad \overline{u}_\pm(p)=\langle p\pm \vert, \\
    \asp{pk} = \langle p- \vert k+ \rangle=\overline{u}_-(p)u_+(k), \\
    \ssp{pk} = \langle p+ \vert k- \rangle=\overline{u}_+(p)u_-(k),
  \end{split}
\end{equation}
and then find for $q^-(p_a) + g^-(p_b) \to q^-(p_1) + g^-(p_2)$
\begin{align}
  \label{eq:qgPT}
  i\mathcal{M}_{qg\to qg} = 2ig^2 \left( t^2_{1e}t^b_{ea} 
  \frac{\langle 2a \rangle \langle 12 \rangle^2}{\langle a1\rangle \langle
  2b\rangle \langle ba \rangle} + t^b_{1e}t^2_{ea}  \frac{[ab]^3}{[1a][a2][2b]}
  \right).
\end{align}
The factors of $t^X_{MN}$ are fundamental colour matrices; where an index is one
of $\{a,b,1,2\}$, it represents the index associated with that particle.
Repeated indices are summed over.  

We now wish to consider the behaviour of this expression in the HE limit.
Without loss of generality, we take $p_a$ to be in the incoming positive
direction and $p_b$ to be in the incoming negative direction.  We consider first
the configuration that is consistent with FKL-ordering such that $y_1\gg y_2$.
The magnitude of each spinor product $\langle ij \rangle$ or $[ij]$ is given
by the square root of the magnitude of the corresponding invariant:
\begin{align}
  \label{eq:sigs}
  |\langle ij \rangle| = \sqrt{|s_{ij}|} = | [ij] |.
\end{align}
Therefore in this configuration, for example, $|[ab]|=\sqrt{s}\to\infty$ in the
HE limit and $|[b2]| = \sqrt{-t}$ remains finite.  We therefore find that both
terms in Eq.~\eqref{eq:qgPT} scale as $s/t$, and in particular that the
$s$-dependence is $s^1$ in agreement with Regge theory.

Alternatively, if we take $y_2\gg y_1$, this means that $t=(p_2-p_a)^2$ such
that $\langle b1 \rangle$ scales like $\sqrt{-t}$ while $\langle b2\rangle$ now
scales like $\sqrt{s}$.  Therefore the terms in Eq.~\eqref{eq:qgPT} now scale as
$\sqrt{t/s}$ and $\sqrt{s/t}$ respectively.  The dominant behaviour in the HE
limit is therefore $\sqrt{s/t}$, again in agreement with Regge theory.

We have chosen a particular helicity
assignment here. The analogous expression for $q^-(p_a) + g^+(p_b) \to q^-(p_1)
+ g^+(p_2)$ is
\begin{align}
  \label{eq:subhel}
    i\mathcal{M}_{qg\to qg} = -2ig^2 \left( t^2_{1e}t^b_{ea}
  \frac{[a2]^3}{[1a][ab][b2]} + t^b_{1e}t^2_{ea}  
  \frac{\langle ba \rangle \langle 1b \rangle^2}{\langle 1a\rangle \langle
  a2\rangle \langle 2b \rangle} 
  \right).
\end{align}
Again, in the FKL configuration both terms scale as $s/t$.  However, in this
case in the non-FKL configuration neither term contributes a leading
$\sqrt{s/t}$ term, and instead yield $\sqrt{t^3/s^3}$ and $\sqrt{t/s}$
respectively.  The other two non-zero helicity configurations may be obtained by
complex conjugation.


\bibliographystyle{JHEP}
\bibliography{Hpapers}

\end{document}